\documentclass[12pt]{article}

\usepackage{amsmath}
\usepackage{amssymb}
\usepackage{graphicx}
\usepackage{enumerate}
\usepackage[round,authoryear]{natbib}
\setcitestyle{aysep={}} 
\usepackage{xurl}
\usepackage{multirow}
\usepackage{subfig}
\usepackage{algorithm}
\usepackage{algorithmicx}
\usepackage{xcolor}
\usepackage{cancel}
\usepackage{ulem}

\newcommand{\blind}{0}

\newcommand{\real}{\ensuremath{\mathbb{R}}}

\newcommand{\ltwo}{\ensuremath{\mathbb{L}^2}}

\newtheorem{defn}{Definition}
\newtheorem{lemma}{Lemma}

\begin{document}
\tabcolsep 1pt

\def\spacingset#1{\renewcommand{\baselinestretch}%
{#1}\small\normalsize} \spacingset{1}


\if0\blind
{
  \title{\bf Data-Driven, Soft Alignment of Functional Data Using Shapes and Landmarks}
  \date{}
  \author{Xiaoyang Guo \\
    Wei Wu \\
    Anuj Srivastava \footnote{xiaoyang.guo.fl@gmail.com, \{wwu, anuj\}@stat.fsu.edu} \\
    Department of Statistics, Florida State University}
  \maketitle
} \fi

\if1\blind
{
  \bigskip
  \bigskip
  \bigskip
  \begin{center}
    {\LARGE\bf Title}
\end{center}
  \medskip
} \fi

\bigskip
\begin{abstract}
Alignment or registration of functions is a fundamental problem in statistical analysis of functions and shapes. While there are several approaches available, a more recent approach based on Fisher-Rao metric and square-root velocity functions (SRVFs) has been shown to have good performance. However, this SRVF method has two limitations: (1) it is susceptible to over alignment, i.e., alignment of noise as well as the signal, and (2) in case there is additional information in form of landmarks, the original formulation does not prescribe a way to incorporate that information. In this paper we propose an extension that allows for incorporation of landmark information to seek a compromise between matching curves and landmarks. This results in a soft landmark alignment that pushes landmarks closer, without requiring their exact overlays to finds a compromise between contributions from functions and landmarks. The proposed method is demonstrated to be superior in certain practical scenarios.
\end{abstract}

\noindent%
{\it Keywords:}  Data-driven alignment, Functional Data Analysis, Phase-Amplitude Separation, Landmark Registration, Square-root Velocity Function
\vfill

\newpage
\spacingset{1.5} 

\section{Introduction}

Functional and shape data analysis is a branch of statistics that seeks tools for statistical analysis of signals, curves, surfaces, or even more
complex objects while being invariant to certain shape-preserving transformations. 
Among the different kinds of objects that one comes across in shape analysis, the simplest types are real-valued functions on a fixed interval.
One of the most important problems in functional and shape data analysis is the registration of points across objects.   
In case of real-valued
 functions, registration boils down to warping the temporal domain of functions so that geometric features (peaks and valleys) of 
 functions are well-aligned. 
 This task,  also called {\it curve registration}~\citep{ramsay1998curve,srivastava2011registration}, 
 or {\it functional alignment}~\citep{ramsay2006functional},
 or {\it phase-amplitude separation}~\citep{marron2014,marron2015functional} is 
 omnipresent in the functional data analysis and its applications. For example,  in analysis of data collected by 
wearable devices, it is critically important to register time-series data across experiments and observations~\citep{choi2018temporal}. 
Similarly, in analysis of biological growth data, including the famous
 Berkely growth data \citep{tuddenham1954physical}, the growth bursts (peaks) are located at different times for 
 different subjects and need to be registered across subjects for analysis. 
 In all these cases, a simple cross-sectional statistical analysis based on the $\ltwo$ norm, and without any alignment, leads to 
 a loss of geometric features and, thus, incorrect inferences~\citep{marron2015functional,srivastava2016functional}.
If one does not account for misalignment in the given data, this can artificially inflate variance and can defeat any statistical 
 analysis, especially the one based on analysis of variance. Registration or alignment is very important for preserving geometric features and to 
 obtain statistics inferences that are more meaningful and interpretable.

\begin{figure}[H]
\begin{center}
\includegraphics[height=1.in]{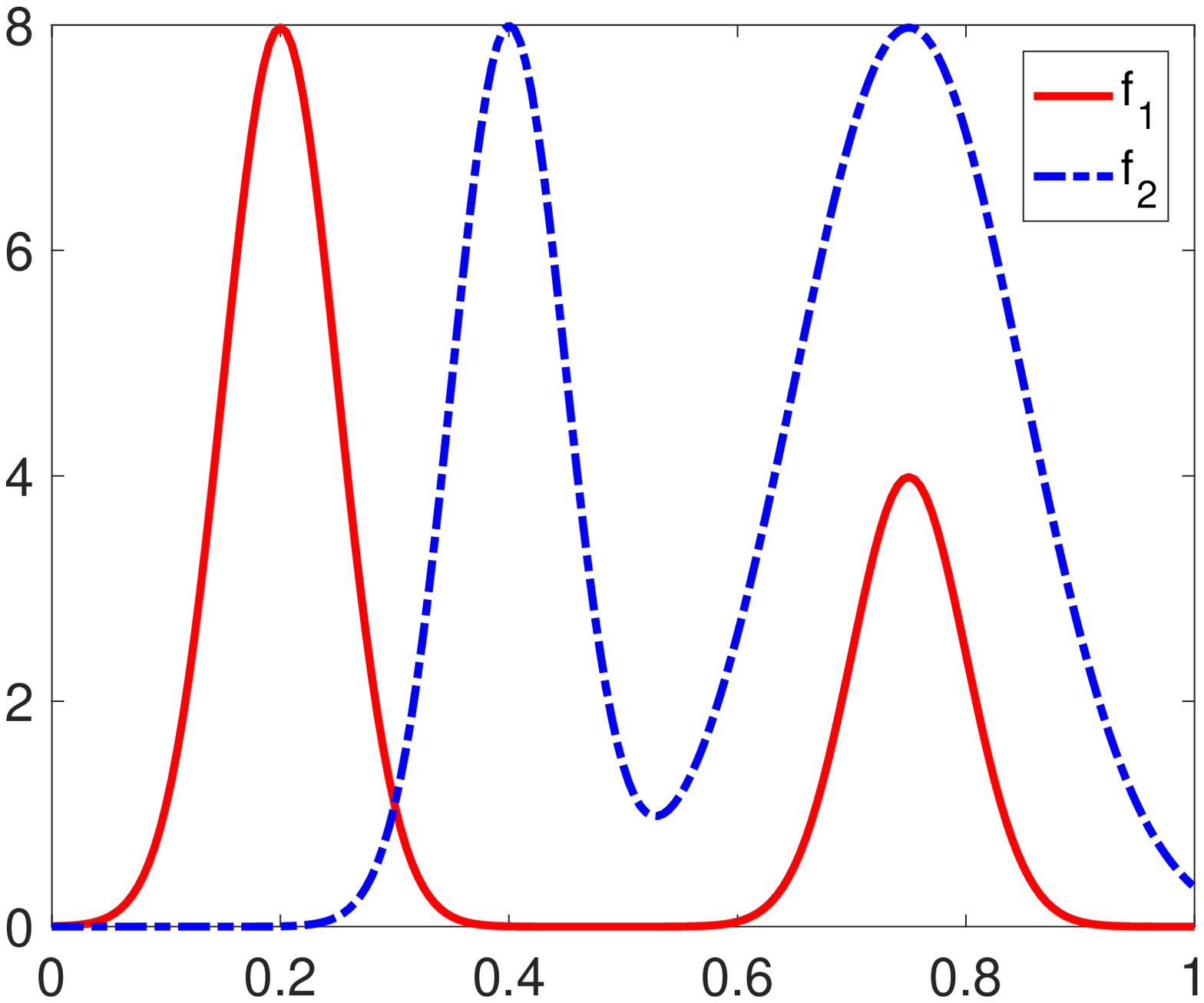} 
\\[30pt]
\begin{tabular}{lcccc}
$\lambda = 0$ \\[-50pt] &\includegraphics[height=1.in]{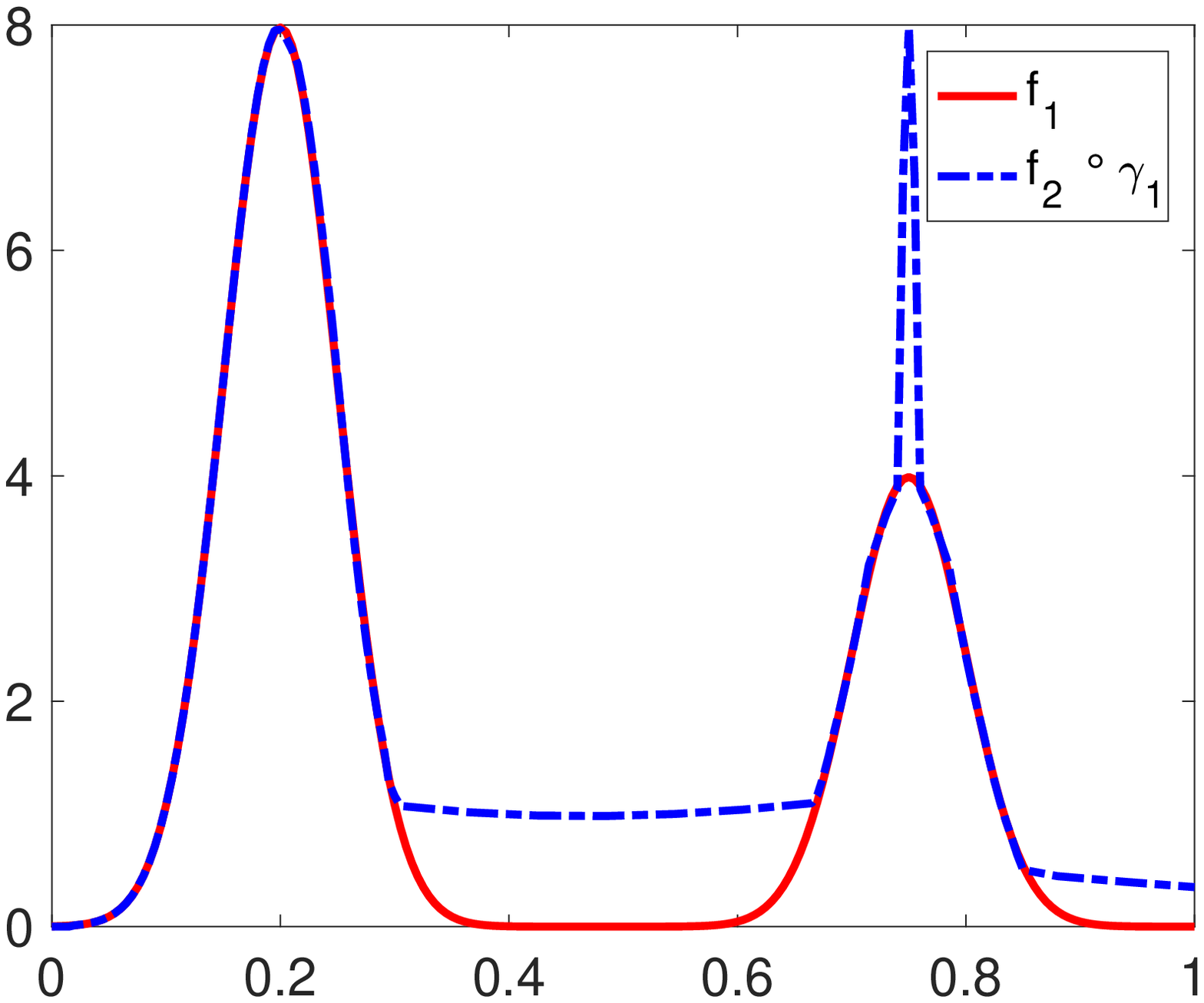}&
  \includegraphics[height=1.in]{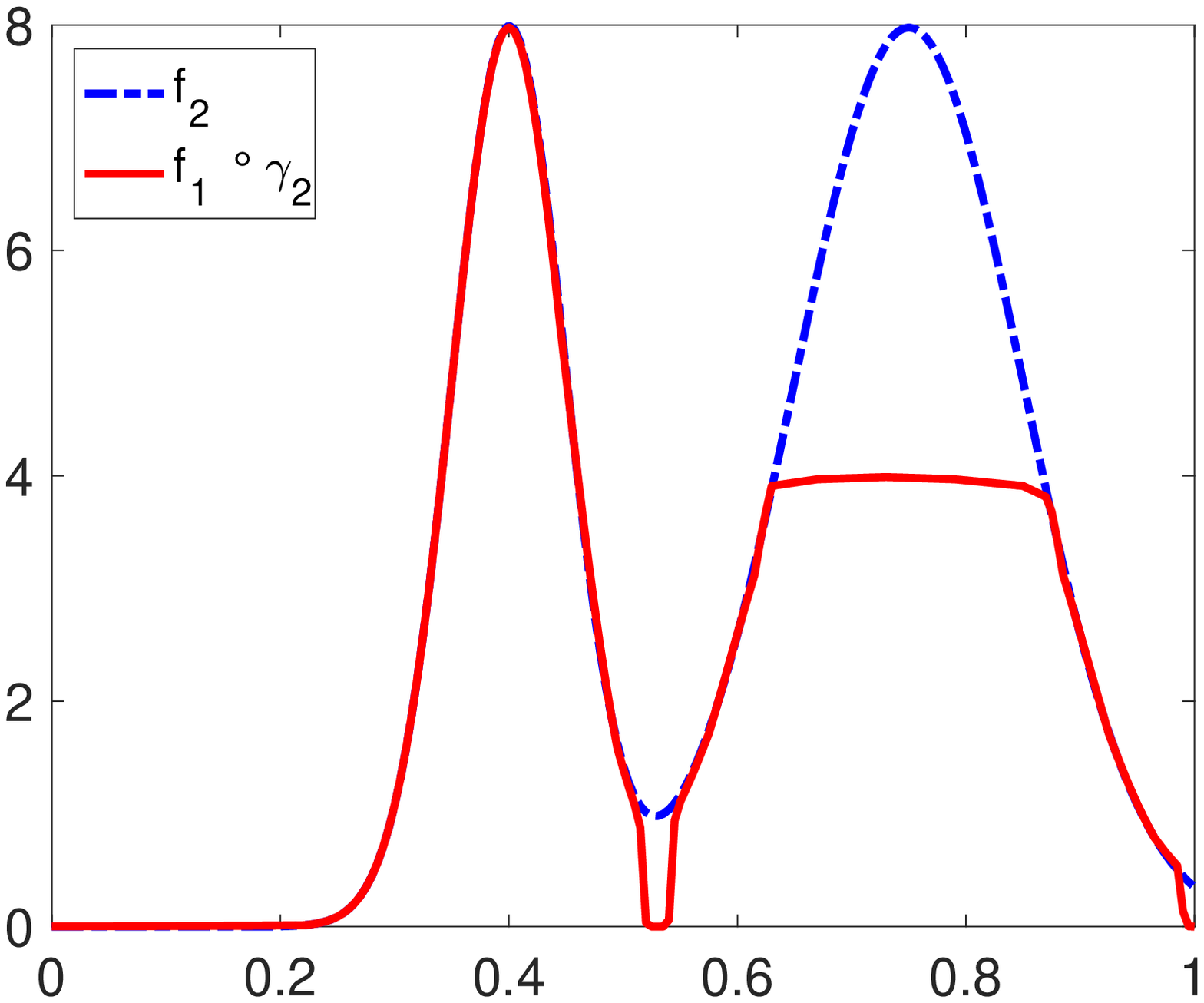} &
   \includegraphics[height=1.in]{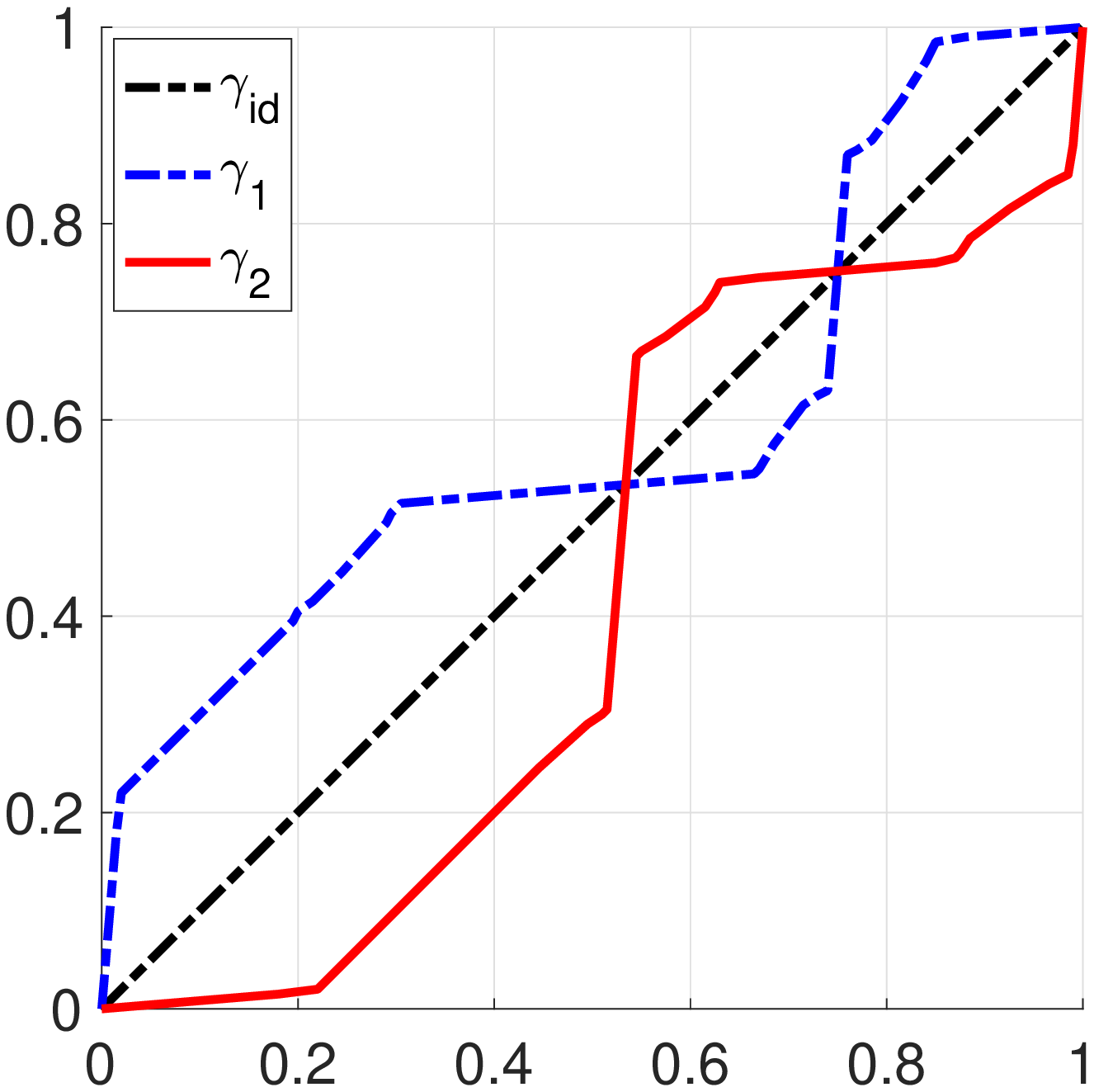} &
   \includegraphics[height=1.in]{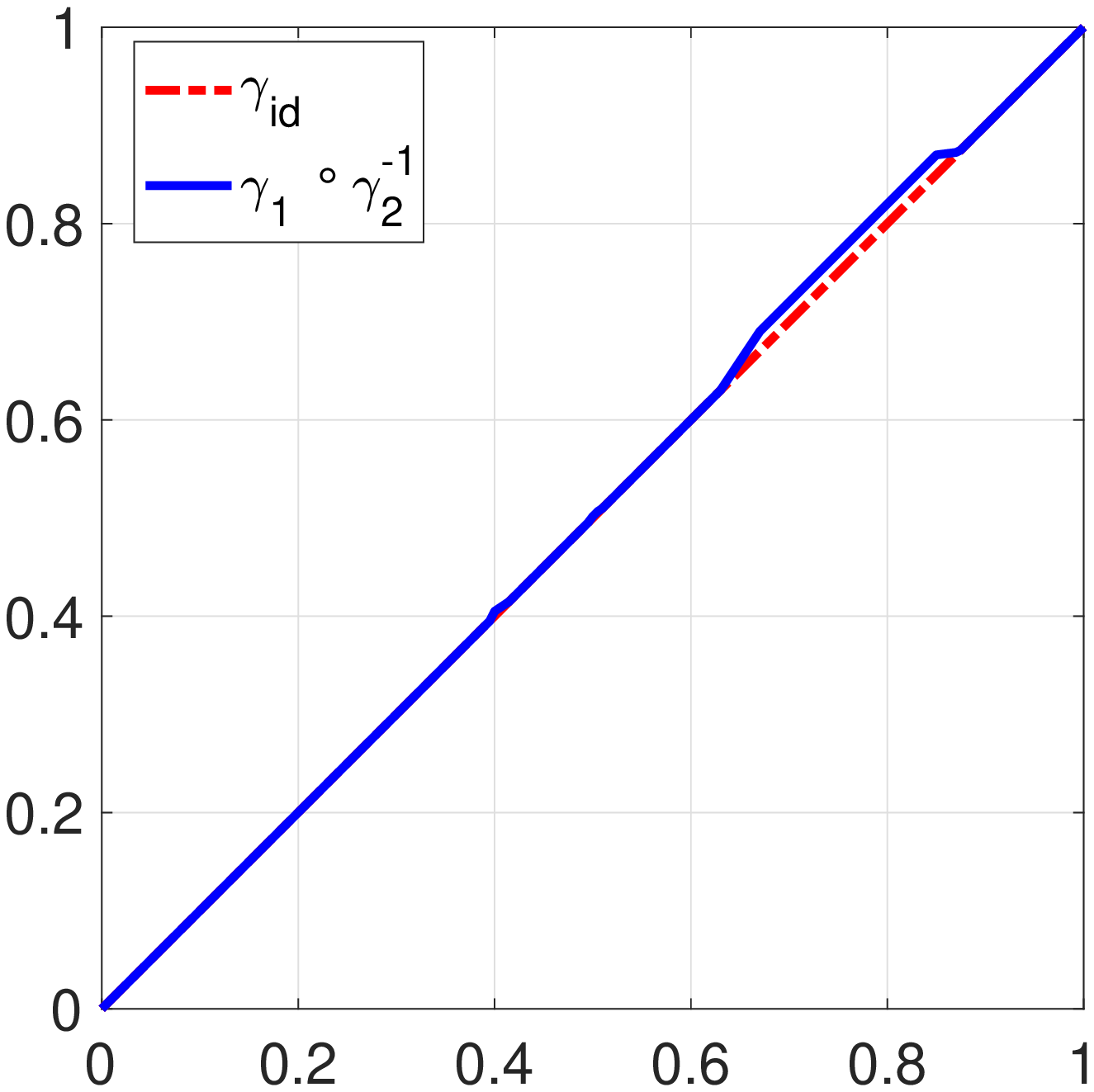}\\
\hspace{1.5cm} &{$f_1$ and $f_2 \circ \gamma_1$} &
       {$f_2$ and $f_1 \circ \gamma_2$} &
       {$\gamma_1$ and $\gamma_2$} &
       {$\gamma_1 \circ \gamma_2$}
\\[30pt] 
$\lambda = 0.1$ \\[-50pt]&\includegraphics[height=1.in]{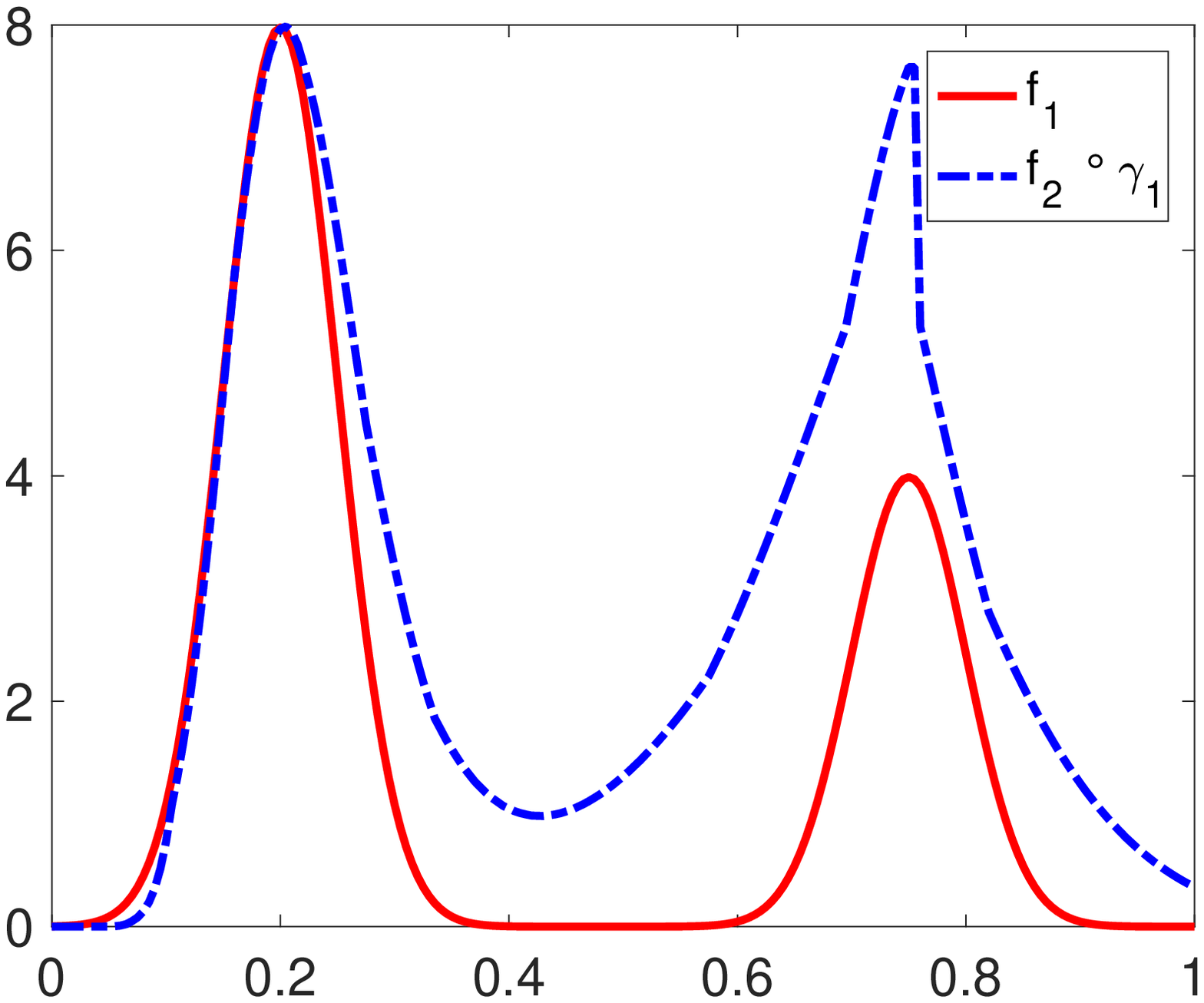}&
  \includegraphics[height=1.in]{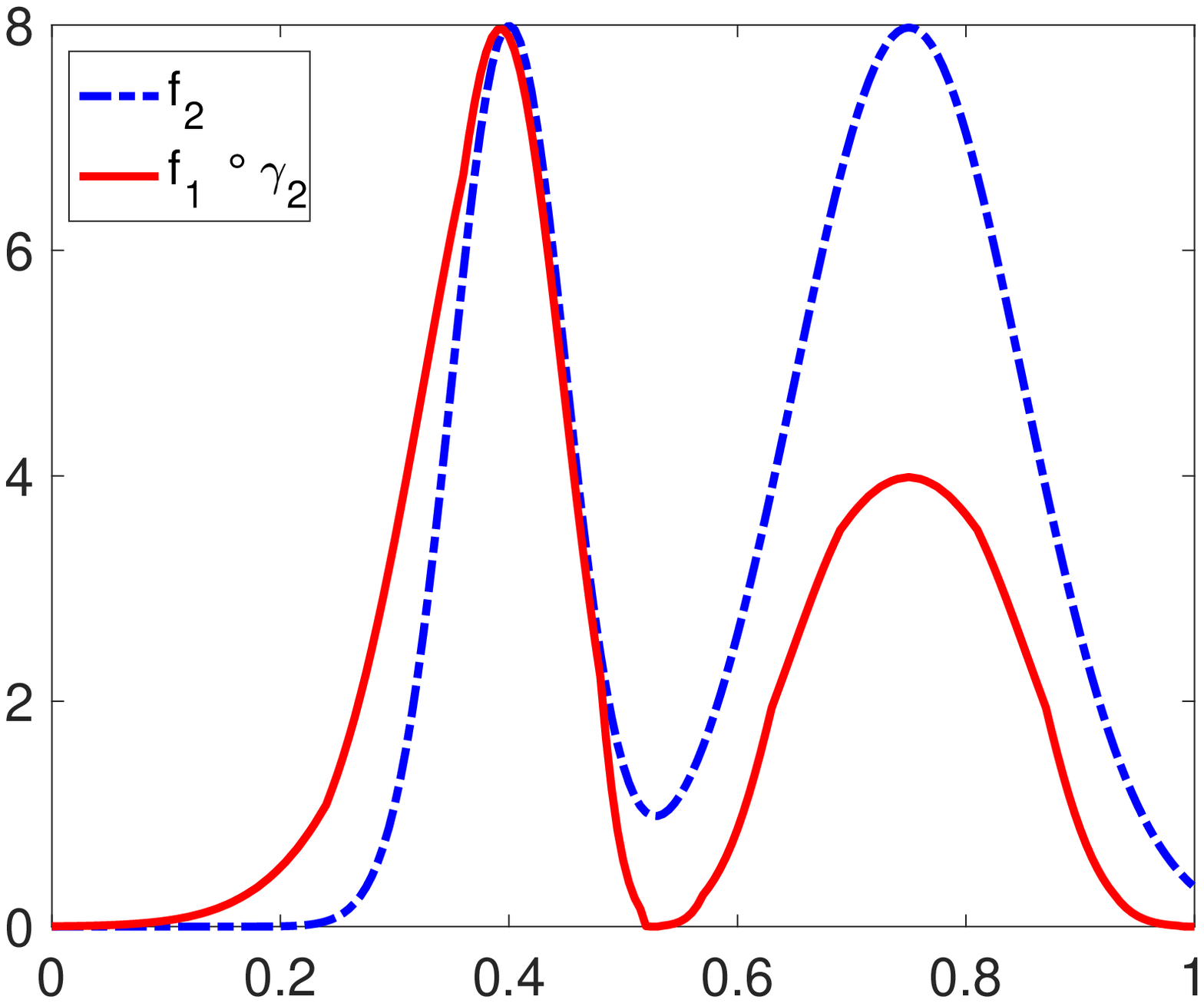} &
   \includegraphics[height=1.in]{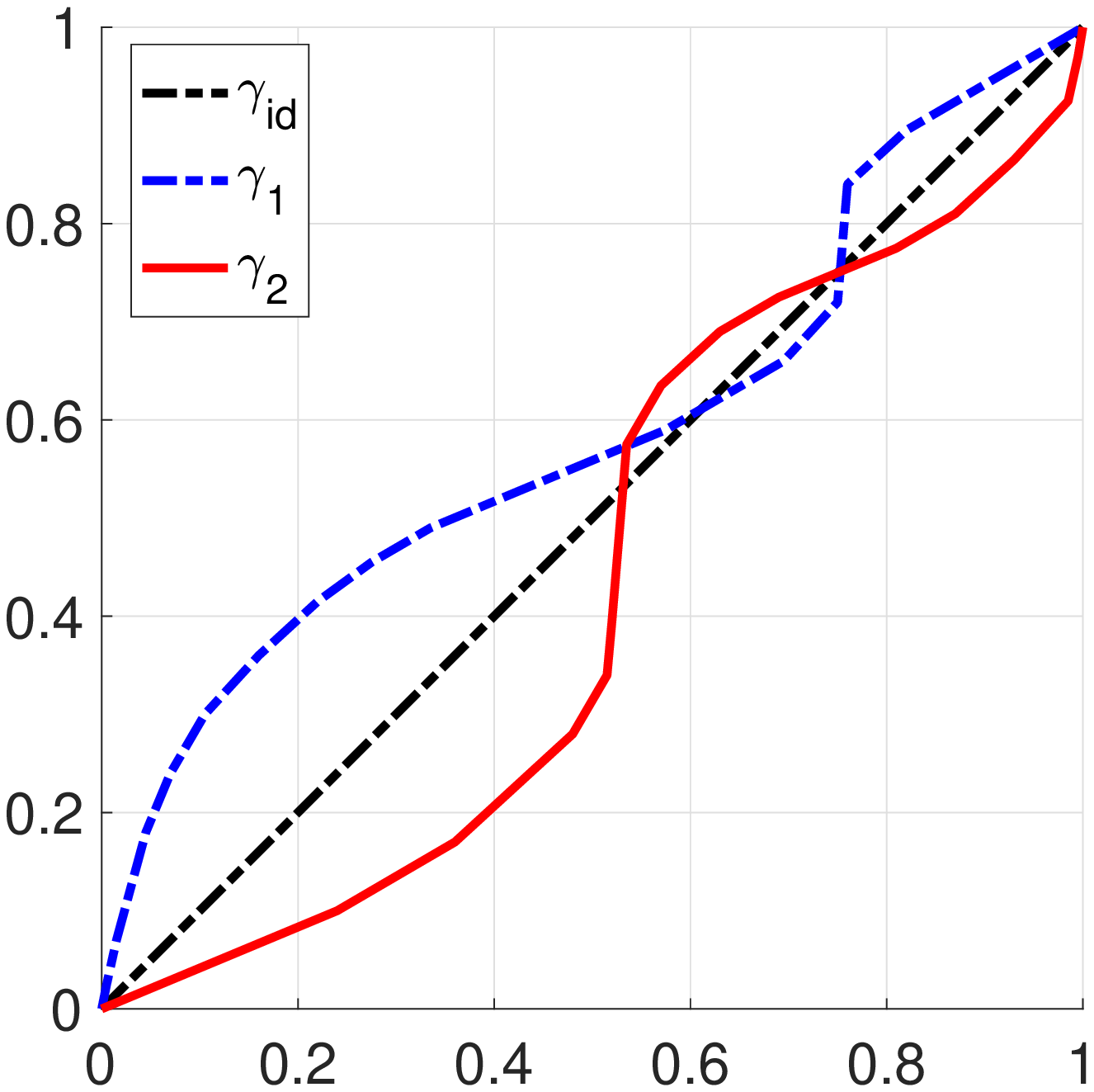} &
   \includegraphics[height=1.in]{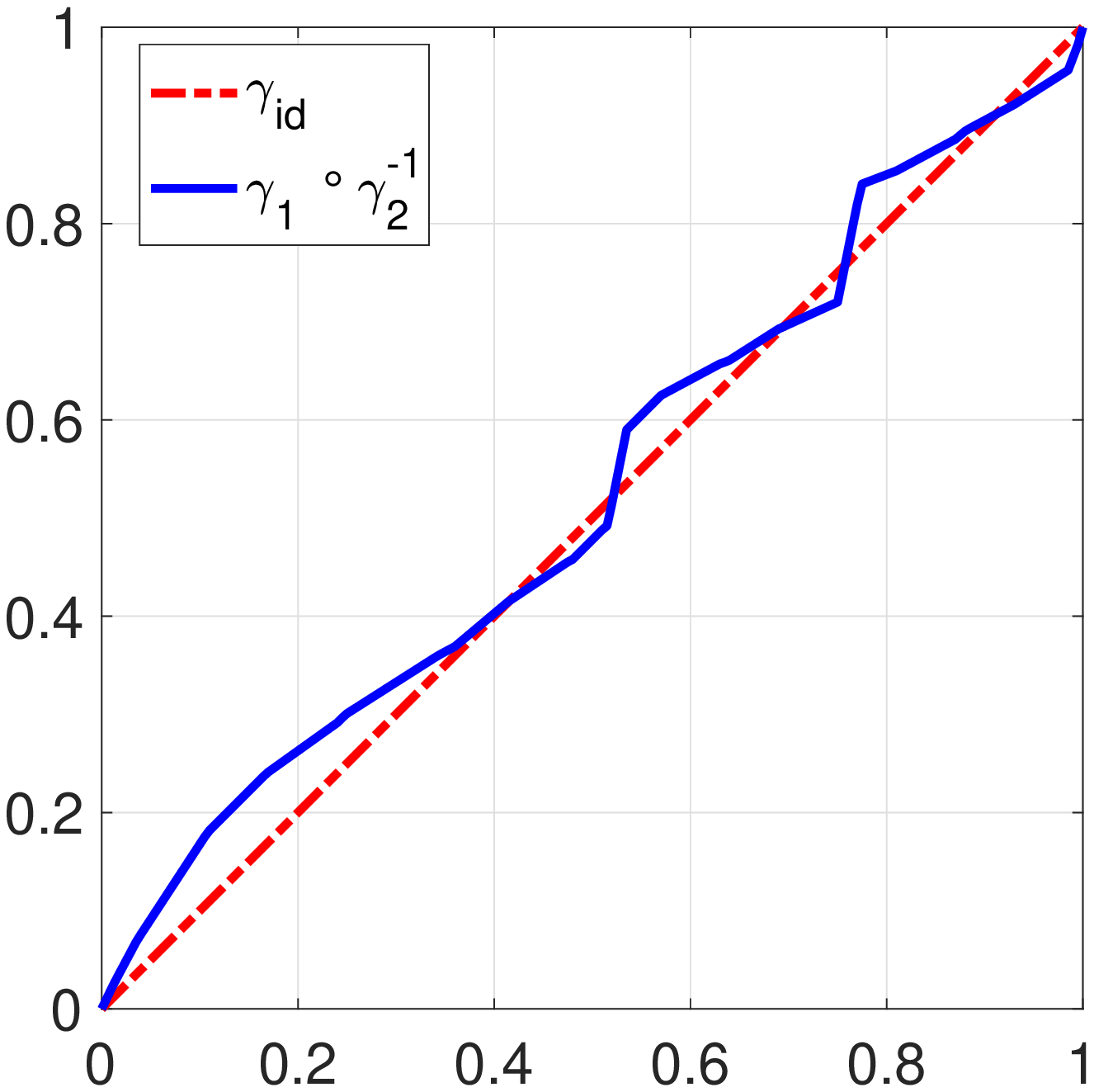}\\
\hspace{1.5cm} &{$f_1$ and $f_2 \circ \gamma_1$} &
       {$f_2$ and $f_1 \circ \gamma_2$} &
       {$\gamma_1$ and $\gamma_2$} &
       {$\gamma_1 \circ \gamma_2 $}
 \\[30pt]
 $\lambda = 0.5$ \\[-50pt]&\includegraphics[height=1.in]{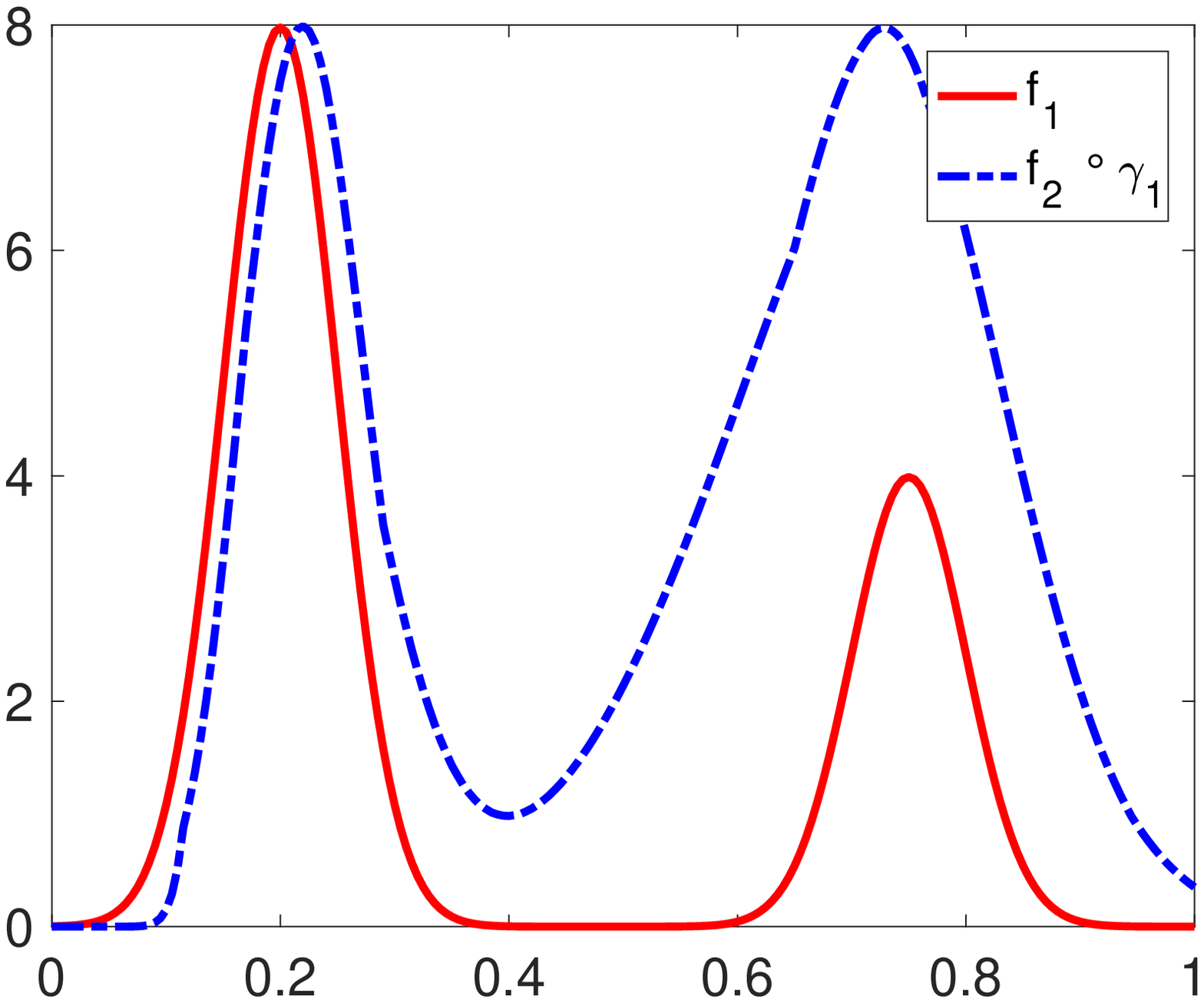}&
  \includegraphics[height=1.in]{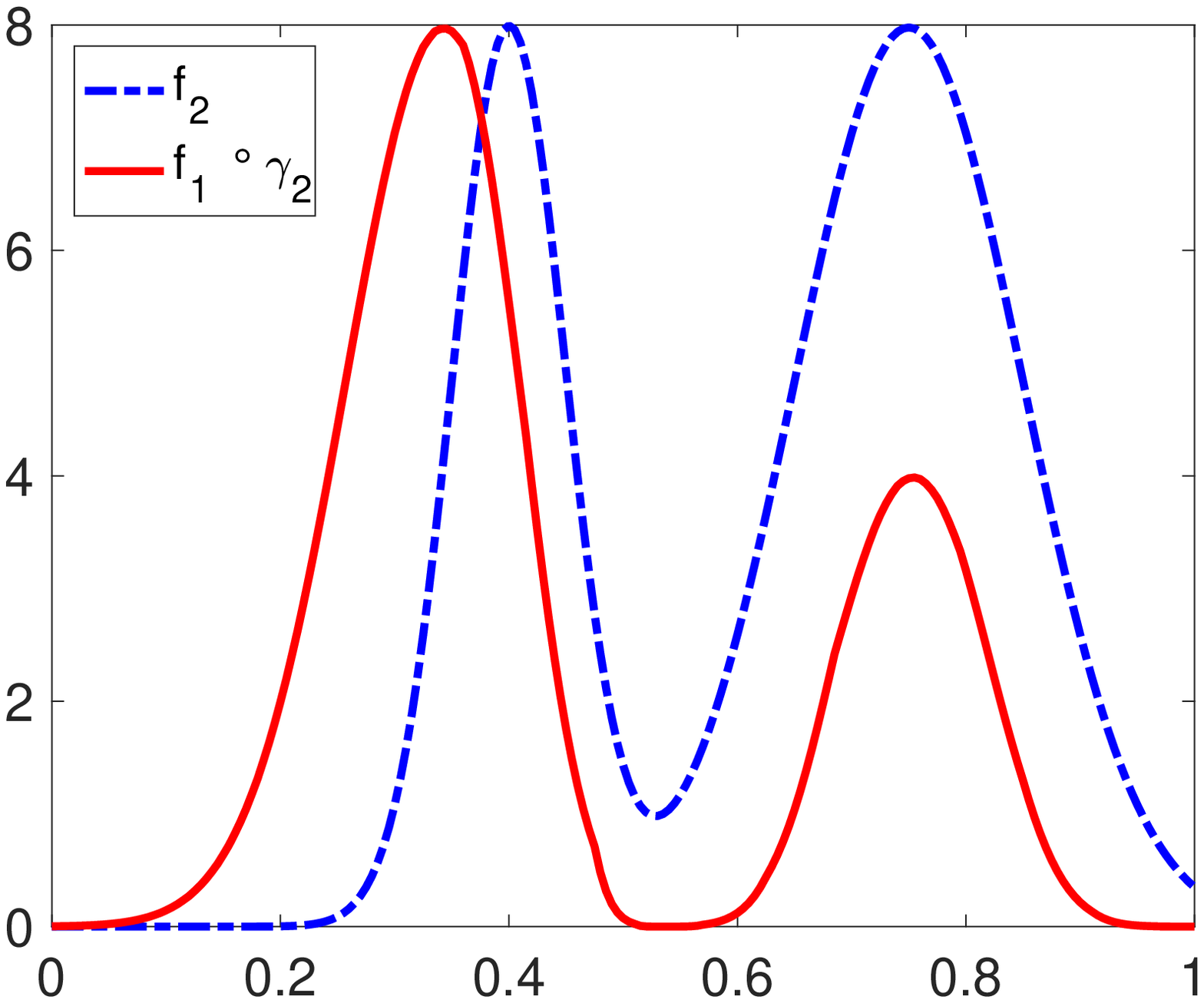} &
   \includegraphics[height=1.in]{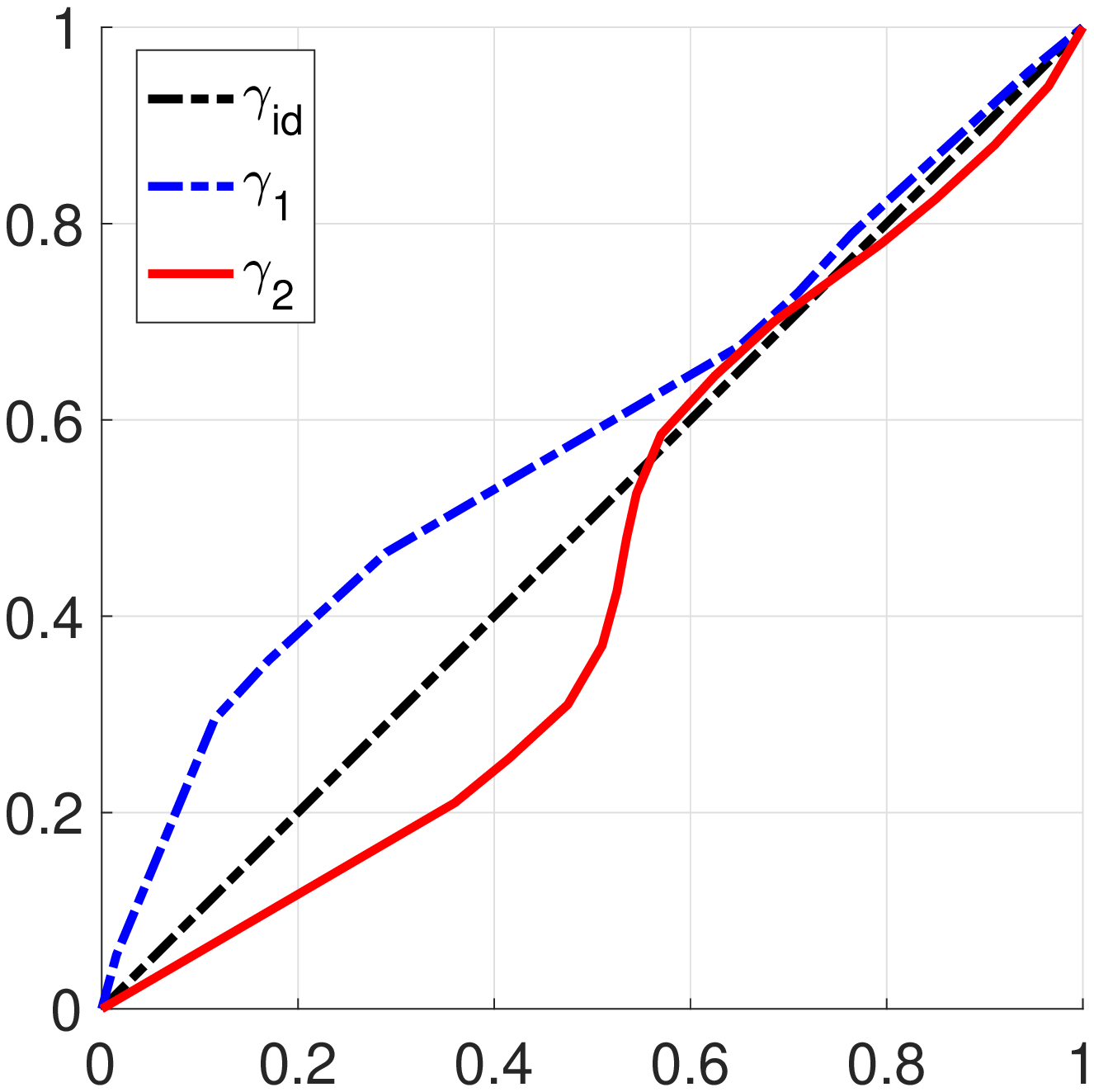} &
   \includegraphics[height=1.in]{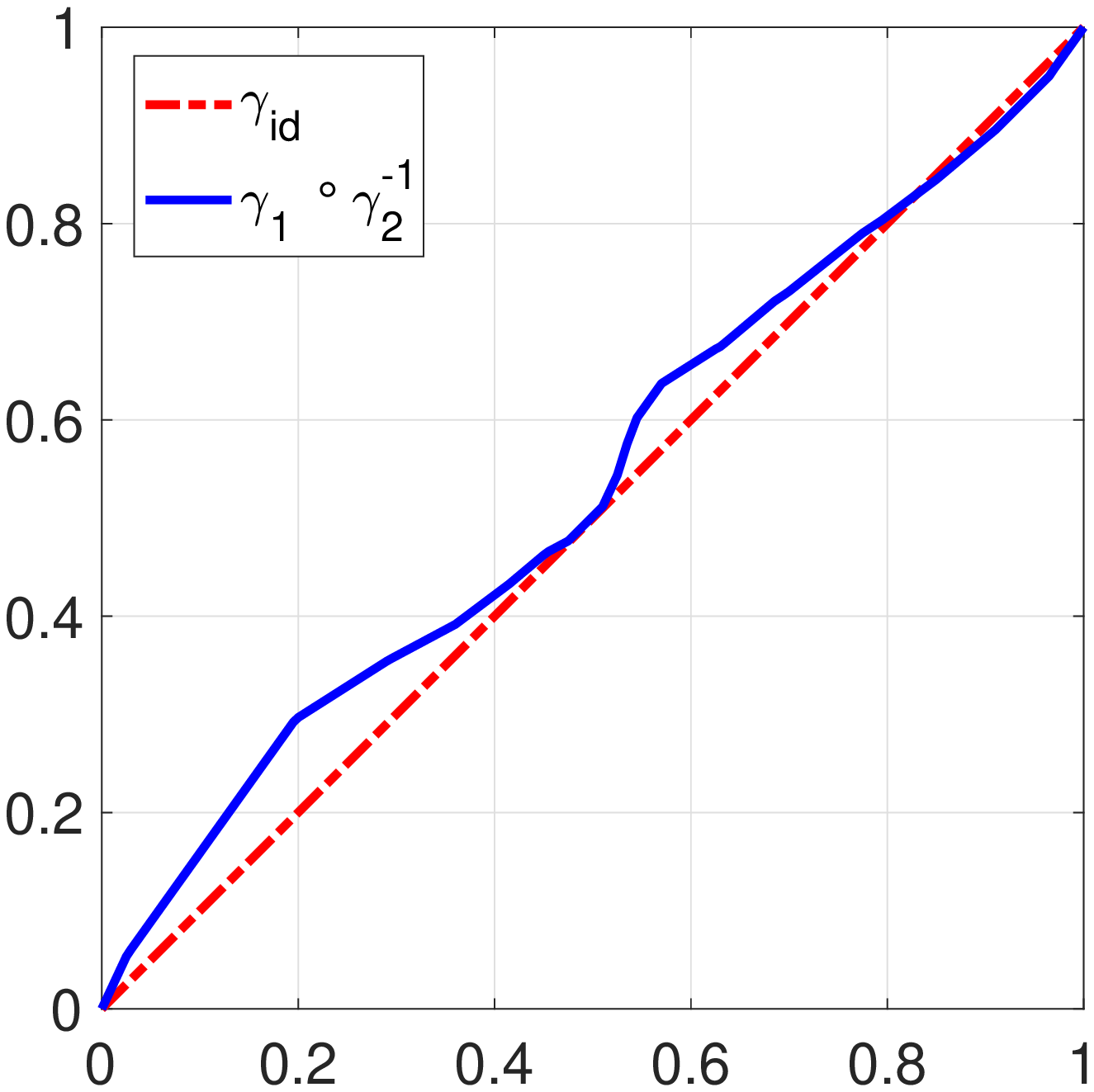}\\
\hspace{1.5cm} &{$f_1$ and $f_2 \circ \gamma_1$} &
       {$f_2$ and $f_1 \circ \gamma_2$} &
       {$\gamma_1$ and $\gamma_2$} &
       {$\gamma_1 \circ \gamma_2 $}
\end{tabular}
   \caption{Issues in penalized-$\mathbb{L}^2$ based alignment of Eqn. \ref{eqn:pen-L2}. Top row: original functions. 
Second-Fourth row: alignment results for tuning parameter $\lambda = 0, 0.1, 0.5$, respectively.}
   \label{fig:invf}
   \end{center}
\end{figure}

In order to formulate the problem, consider a 
given set of functions: $\{f_i: [0, 1] \rightarrow \mathbb{R}|i=1,2,...,m\}$ and the goal is to find a set of time warping functions, $\{\gamma_i\}$, such that the warped functions $\{f_i\circ \gamma_i\}$ are 
as well aligned as possible. We will use the terms {\it alignment} and {\it registration} interchangeably throughout the paper. 
 For simplicity, we restrict the domain to be $[0, 1]$ but any finite interval can be transformed into $[0, 1]$ by a simple translation and scaling.
The time warping functions $\{\gamma_i\}$ are typically boundary-preserving, 
positive diffeomorphisms of $[0,1]$ to itself.  
The quantification of how well a set of functions are aligned depends, of course, on the application and is open to 
subjectivity of an observer. However, one does need an objective function for automating the alignment process, especially for handling 
large volumes of data. 
There are several objective functions defined in the literature for such alignment. 
The classical solution is the {\it dynamic time warping} (DTW) that has been used extensively in 
signal, especially speech processing. 
DTW and related methods are based on minimizing a penalized-$\mathbb{L}^2$ distance between aligned functions \citep{sakoe1978dynamic, ramsay1998curve}. 
That is, for any pair $f_i$, $f_j$, they solve for the optimization problem: 
\begin{equation}
\hat{\gamma} = \arg\inf_{\gamma} \left( \| f_i - f_j \circ \gamma\|^2 + \lambda {\cal R}(\gamma) \right)\ ,
\label{eqn:pen-L2}
\end{equation}
where ${\cal R}$ is a roughness penalty on $\gamma$. For instance, the first-order penalty is $(\int \dot{\gamma}^2(t)~dt)$ and 
the second-order penalty is $(\int \ddot{\gamma}^2(t)~dt)$. 
Despite its popularity, this method has some major 
limitations: (1) The solution is not inverse symmetric in $f_i$ and $f_j$; (2) In fact, the objective function is not a proper metric
and cannot be used for ensuing statistical analysis; (3) Finally, it maybe difficult to find a balance between the penalty and the data term.
This is because the penalty term does not carry any additional information which relates the warping function to the observed data.; it simply imposes arbitrary 
constraints on the warping function. (This is in contrast to the proposed method where the extra term comes from 
the additional information present in form of landmarks.) 
These issues are illustrated using an example in Figure \ref{fig:invf} for the two functions $f_1$, $f_2$ shown in the top panel. 
The next three rows show the result of alignments under the first order penalty, each obtained using Eqn. \ref{eqn:pen-L2} but for different values of $\lambda$.
In order to study the symmetry of the solution, we perform alignment both ways --
$f_1$ to $f_2$ and $f_2$ to $f_1$ -- and compose the resulting $\gamma$s. Ideally, if the solution is symmetric, this composition 
should be identity, the $45\deg$ line.  We see that when $\lambda = 0$, the solution is symmetric but  we have the
{\it pinching effect}{~\citep{marron2015functional}} in alignment (second row of Figure \ref{fig:invf}). As $\lambda$ increases, the pinching effect is gone but two things happen: 
(1) there is a decrease in the level of alignment, and (2) the solution is less and less symmetric. 
Despite its multiple limitations, this is the most commonly-used technique for registration of functional data.

\begin{figure}[H]
\begin{center}
\begin{tabular}{cc}
   \includegraphics[width=0.33\linewidth]{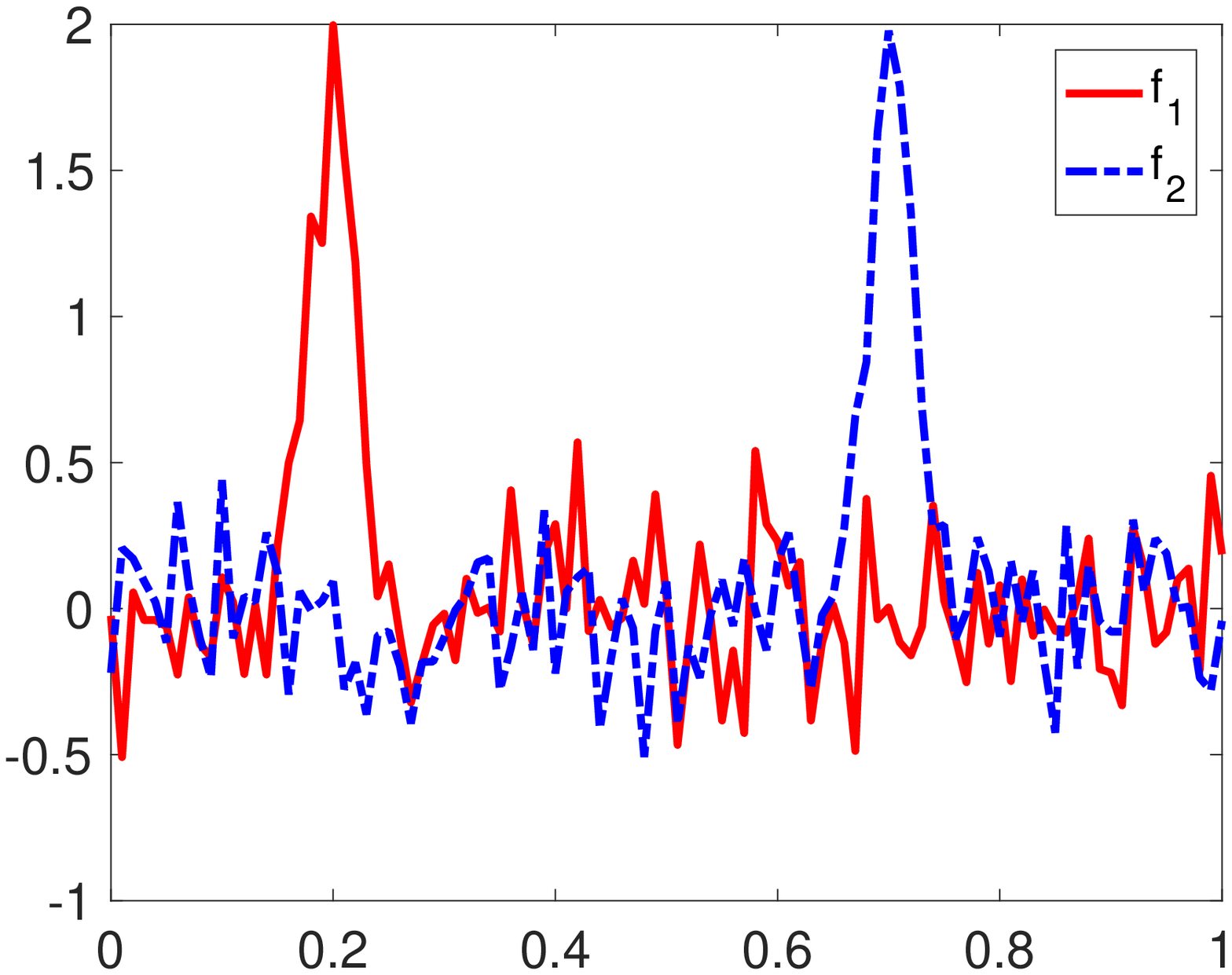}
   \includegraphics[width=0.33\linewidth]{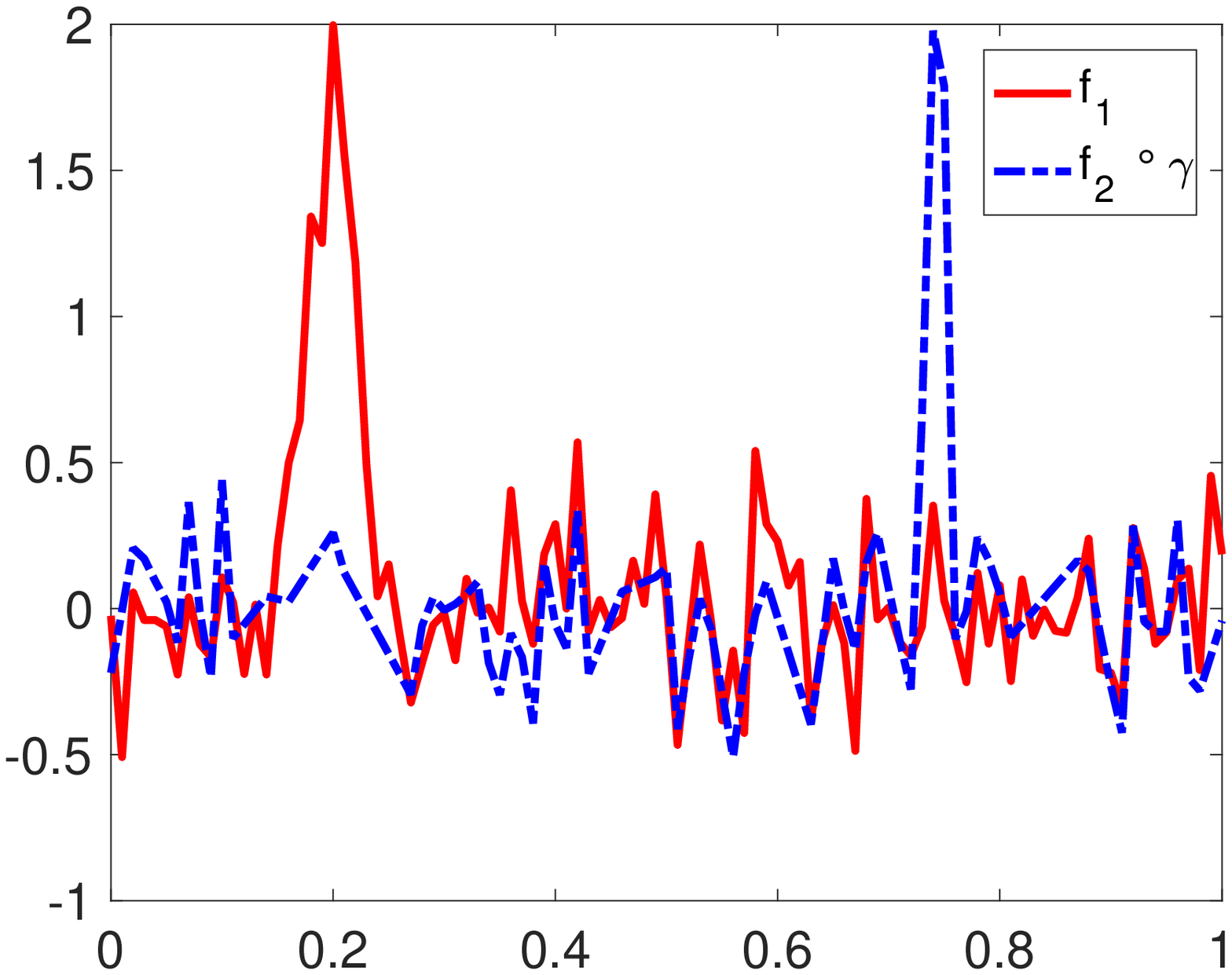}&
   \includegraphics[width=0.33\linewidth]{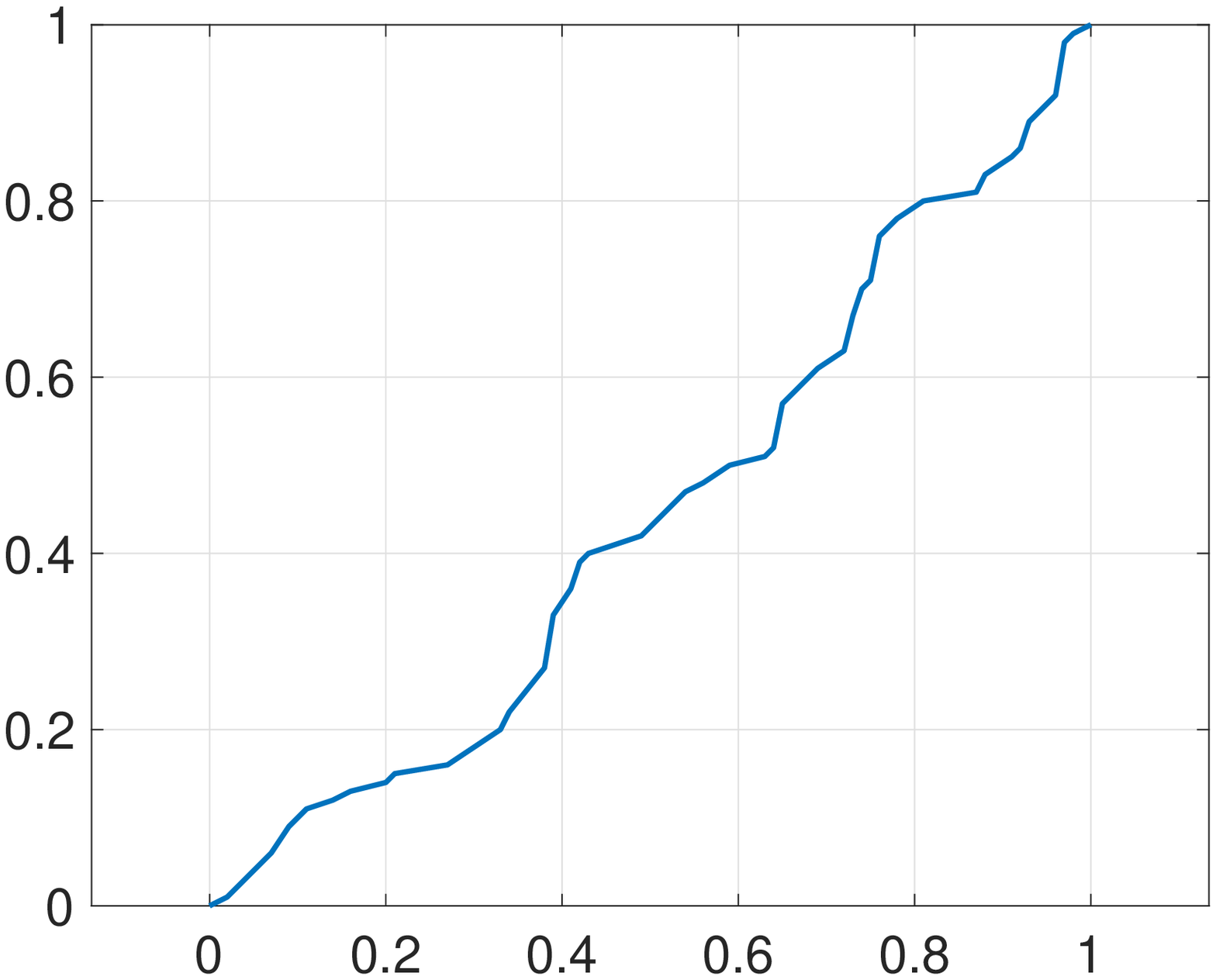}\\
\end{tabular}
\end{center}
   \caption{Example of over alignment using SRVF-based framework. 
   Left: the original functions; middle: aligned functions; right: 
   optimal time-warping.}
\label{fig:noise}
\end{figure}

	A recently developed theory \citep{srivastava2011registration,srivastava2016functional} addresses these  issues by defining a new mathematical
representation of functions, called 
{\it square-root velocity function} (SRVF). 
For an absolutely-continuous function $f: [0,1] \to \real$, its SRVF 
is defined to be $q(t) = \mbox{sign}(\dot{f}(t)) \sqrt{|\dot{f}(t)|} \in \ltwo([0,1],\real)$. 
If $f$ is time warped using a warping $\gamma$, resulting in $f \circ \gamma$, the corresponding effect on its 
SRVF is given by $(q \circ \gamma) \sqrt{\dot{\gamma}}$, henceforth denoted by the expression $(q * \gamma)$. This constitutes
the right action of $\Gamma$, the group of all time warping functions, on $\ltwo$, the set of all SRVFs. 
For any two functions $f_i$, $f_j$, with the associated SRVFs $q_i$, $q_j$, the registration problem is 
then given by $\arg\inf_{\gamma} \| q_i - (q_j * \gamma)\|^2$. No additional penalty term is needed here making it 
a fully automatic method. 
As described later, this formulation: (1) avoids
the pinching effect, (2) results in a symmetric solution in terms of $f_i$ and $f_j$, and (3) the infimum is a proper 
metric that can be used for statistical analysis. However, this solution has some limitations. 
Despite its good performance in raw peak alignment, sometimes it is susceptible to overalignment. In case the data is noisy and contains 
spurious peaks, the solution results in alignment of these spurious peaks with the signal peaks, as
shown by an example in Figure \ref{fig:noise}. This example studies two functions with one tall peak and each and numerous small 
peaks associated with the noise. The SRVF solution aligns a large peak in one function with a nearby smaller peak, 
and vice-versa. 

One can also introduce a roughness penalty 
on the time warping functions, similar to Eqn.~\ref{eqn:pen-L2}, 
according to: $\arg\inf_{\gamma} (\| q_i - (q_j * \gamma)\|^2 + \lambda ||\sqrt{\dot{\gamma}}-1||^2)$. 
The second term is the Fisher-Rao distance~\citep{srivastava2016functional} between $\gamma$ and 
identity warping function $\gamma_{id}$. 
However, a penalty simply curtails the amount of warping, irrespective of function shapes, 
and is seldom useful in handling the issue of over alignment. 
It will be better if there is some external information available to help overcome the effects of noise. 

%

\noindent {\bf Hard Landmark Alignment}: The external information can come 
in form of {\bf landmarks}. These are labeled points in the domain $[0,1]$ that represent locations of 
great importance in the functions, e.g., points of high curvature, anatomical landmarks, etc. 
They represent the same features across functions and it is meaningful to register them. 
Suppose we are given labeled landmarks
$\{\tau_{j}^{(i)}: 1 \leq i \leq m, 1 \leq j \leq n\}$ ($n$ time points for function $f_i$) that denote certain underlying temporal events that 
need to be aligned during registration. In other 
words,  one wants both $\{ f_i \circ \gamma_i\}$  and $\{ \gamma^{-1}_i(\tau_{j}^{(i)})\}$ to be aligned as well as possible. 
Examples of landmark-based registration are presented in \citep{kneip1992statistical,gasser1995searching,bigot2006landmark}. 
However, these approaches are essentially {\bf hard} registration methods in the sense that
they insist on matching the landmarks exactly, i.e., variance($\{ \gamma_i^{-1}(\tau_{j}^{(i)})\}) = 0$ for all $j$.
After registering all the landmarks using affine transformation actions, the remaining problem is to 
individually align segments separated by landmarks.
One of the drawbacks of any hard registration is that it ignores the global geometry of full functional objects
and relies only on local shapes.

\noindent {\bf Soft Landmark Alignment}: In contrast, if the landmarks are steered closer to 
each other, but without requiring them to overlay each other,  then 
it is called {\bf soft matching}, or  a {\bf landmark-guided} functional or shape data analysis 
\citep{doi:10.1111/cgf.12063,bauer2015landmark}. What is the motivation for soft registration?
Since one usually estimates the landmarks from the data, either manually or automatically, landmarks 
themselves may have errors, and the resulting hard registration can be erroneous.
Moreover, if one has multiple choices of landmarks, the hard registration has to 
select only one solution, while ignoring the other possibilities.
All this motivates the need for a data-driven soft registration, in order to reach  a more reasonable solution. 
It is termed data-driven because one determines the relative strengths of contributions from the 
function shape and landmarks from the data itself. 

{In recent years, there have also been some Bayesian framework for function alignment, see e.g., 
\citep{telesca2008bayesian,cheng2016bayesian,kurtek2017geometric,lu2017bayesian}. However, current Bayesian approaches do not consider incorporating extra landmark information but rather focus on global curve registration.} Moreover, the disadvantage of Bayesian methods is the computational cost and the dependence of the solution on the prior models. 

We propose an extension of the SRVF framework that incorporates landmark information in a soft fashion. 
This two-step procedure: (1) time warps the given functions or curves into hard alignment, and (2) performs further warping 
using a penalized elastic metric.
{We first introduce it for pairwise alignment and then naturally extend it for multiple alignment. }
The main 
strengths of this framework as as follows: 
\begin{enumerate}

\item {\bf Optimal Combination of Shapes and Landmarks}: 
It combines information from function shapes and landmark locations, 
and results in an optimal alignment solution. It provides a solution that lies between 
fully constrained (hard registration) and fully unconstrained. The choice of relative weights of the two terms is 
based on cross-validation and, hence, the whole framework is data driven. 

\item {\bf Mathematical Properties}: The cost function used for soft registration has nice mathematical properties -- it is non-negative, symmetric, 
and satisfies the triangle inequality, i.e., it is a pseudometric! Consequently, this 
pseudometric can be used for statistical analysis of functional data. This provides a consistency 
between the metric used for registration and metric used in statistical analysis. 

\item {\bf Computational Efficiency}: The solution to the proposed soft alignment is computationally efficient, as it uses the 
fast dynamic programming algorithm for alignment. {Code is available on Github: \url{https://github.com/xiaoyangstat/soft-alignment-of-functional-data-using-shapes-and-landmarks}}

\end{enumerate}

The rest of this paper is organized as follows. Section \ref{sec:sla} briefly summarizes the SRVF framework and 
introduces a penalized elastic pseudometric for soft landmark alignment. 
Section \ref{sec:multiple_sfa} extends this pairwise soft landmark alignment to multiple functions. In Section \ref{sec:exp_sfa}, we present a variety of results on 
both simulated data and real data to demonstrate the superiority. Finally, Section \ref{sec:con_sfa} provides concluding discussions and future directions. 

\section{Soft Landmark Alignment}\label{sec:sla}
In this section, we first introduce the SRVF framework for unconstrained alignment and then discuss the hard alignment, 
or the fully constrained approach. 
Then, we introduce the framework for intermediate constraint or soft alignment of two functions given their landmark data. 
Before we discuss different options, we outline some important desirable properties that a
soft framework should provide. 
\begin{enumerate}
\item {\bf Landmark Guidance}: The framework should obviously be able to 
incorporate both sets of information:  functional shapes and landmarks,
in performing time warping and registration. Ideally, the user should be able to estimate the 
relative weights of these components using the data itself. 

\item {\bf Metric or Pseudometric}: The objective function for pairwise registration of functions should be symmetric in the two 
input variables. In other words, the registration of $f_1$ to $f_2$ under this criterion will be compatible with the 
registration of $f_2$ to $f_1$.  Furthermore, if the objective function is non-negative and satisfies 
triangle inequality, then it can be used as a proper metric in the ensuing functional data analysis. Even if it is not 
a full metric, but just a pseudometric, it can still be useful in statistical analysis of registered functions.  

\item {\bf Invariance Condition}: It is important to note that an identical warping of any two functions preserves their registration. 
In other words, for any time warping function $\gamma$, the registration between $f_1$ and $f_2$ is the same as the registration 
between $f_1 \circ \gamma$ and $f_2 \circ \gamma$, for any $\gamma$. Therefore, the objective function used for registration should 
also be invariant to simultaneous warping. This requires that the associated metric or pseudo-metric preserves its value under the 
action of time warping. 
\end{enumerate}
As we introduce different options for soft registration, we will evaluate them using these desired properties.

\subsection{Unconstrained SRVF Alignment}
\begin{figure}[H]
\centering
\begin{tabular}{cccccc}
   \multirow{3}{*}[10pt]{\includegraphics[width=0.15\linewidth]{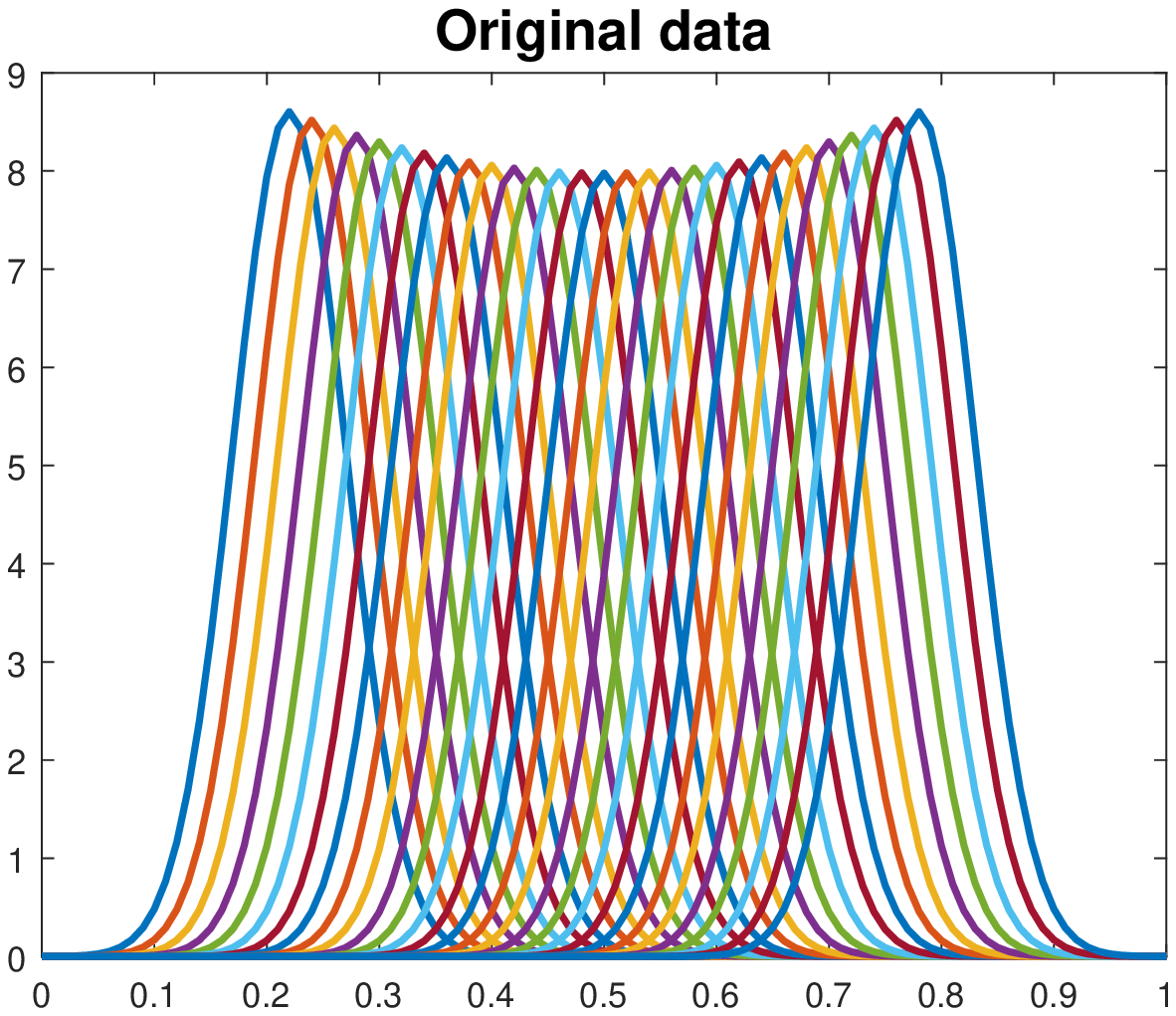}}
   &\includegraphics[width=0.18\linewidth]{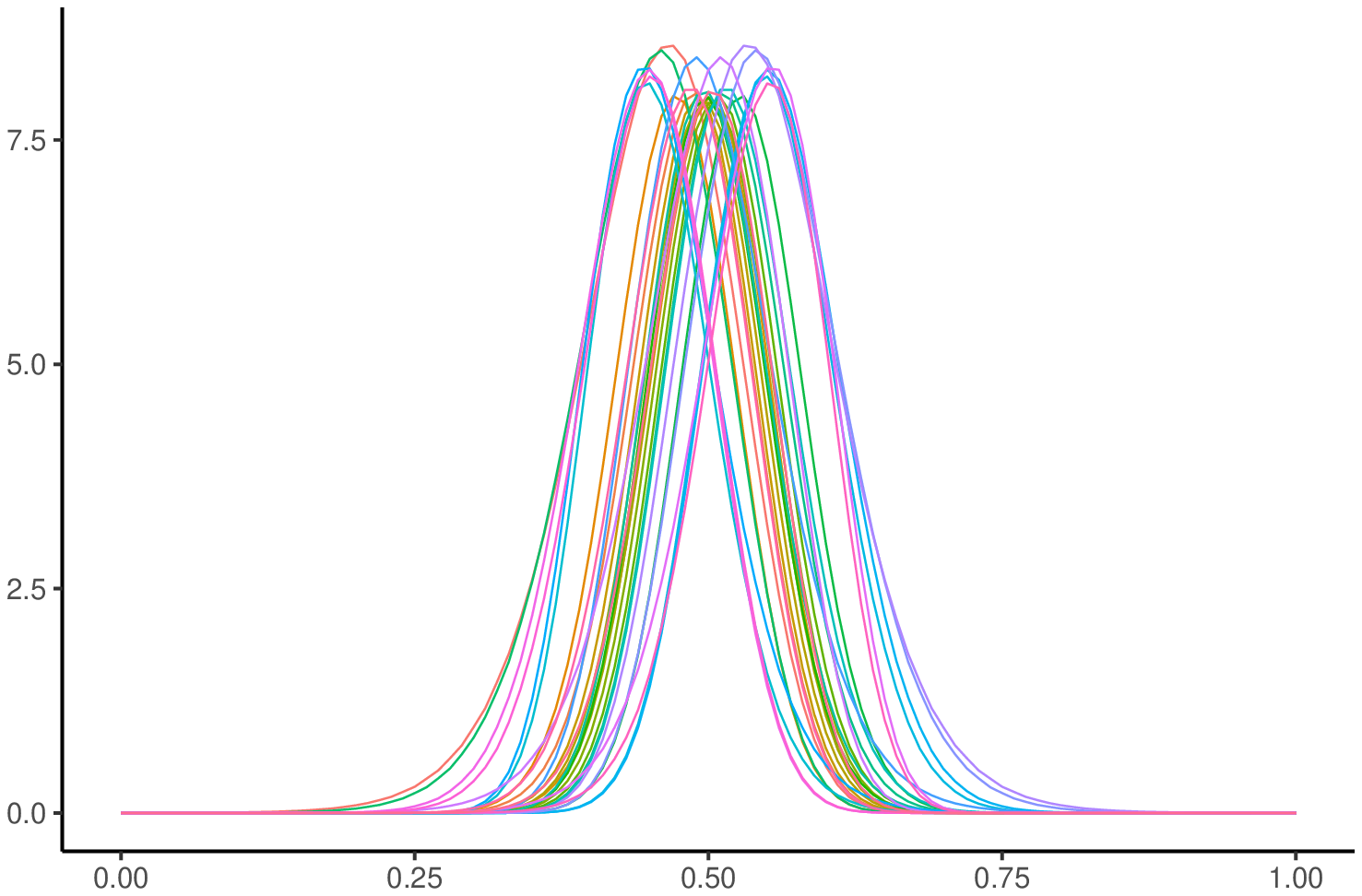}   
   &\includegraphics[width=0.15\linewidth]{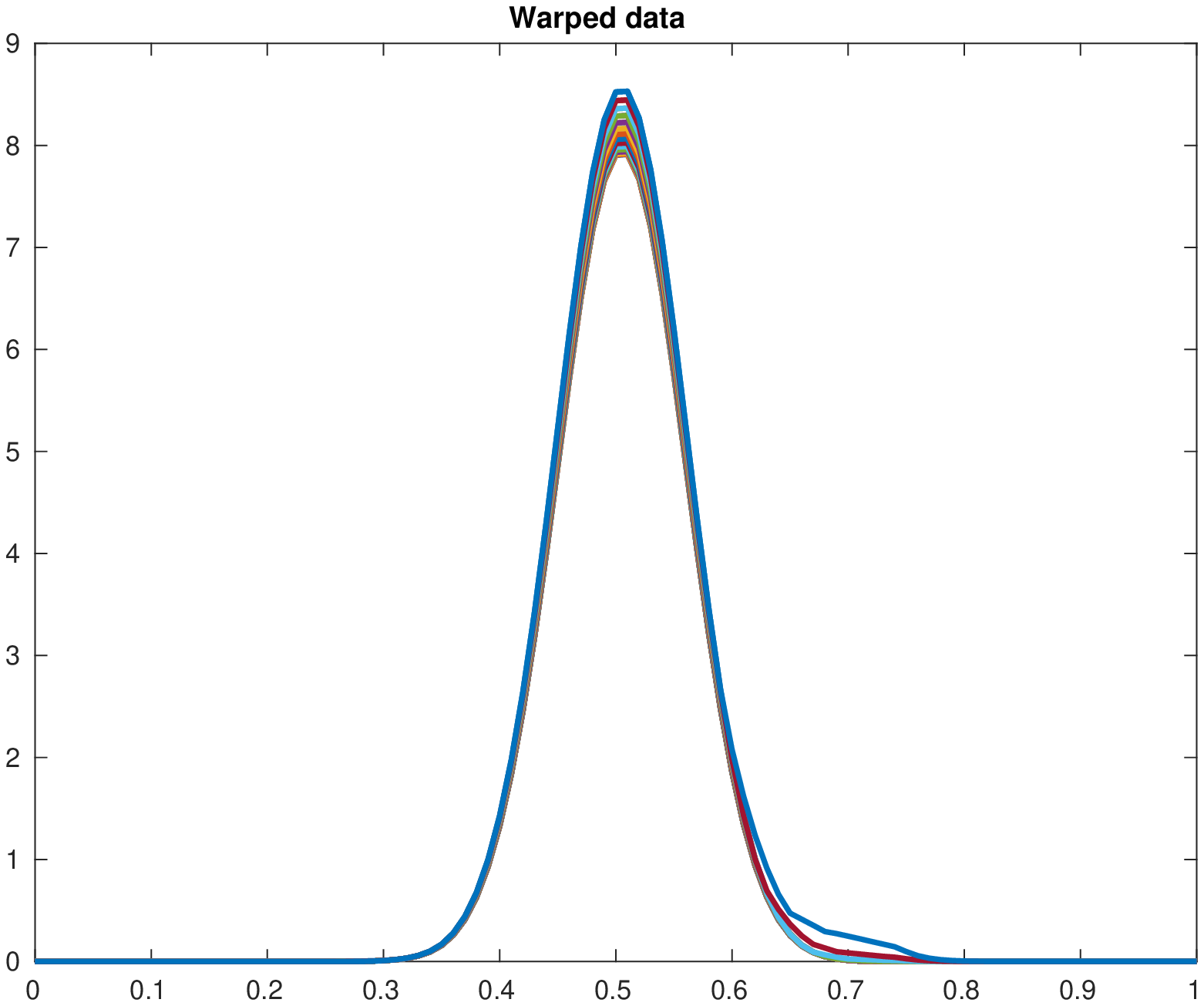}
   &\includegraphics[width=0.15\linewidth]{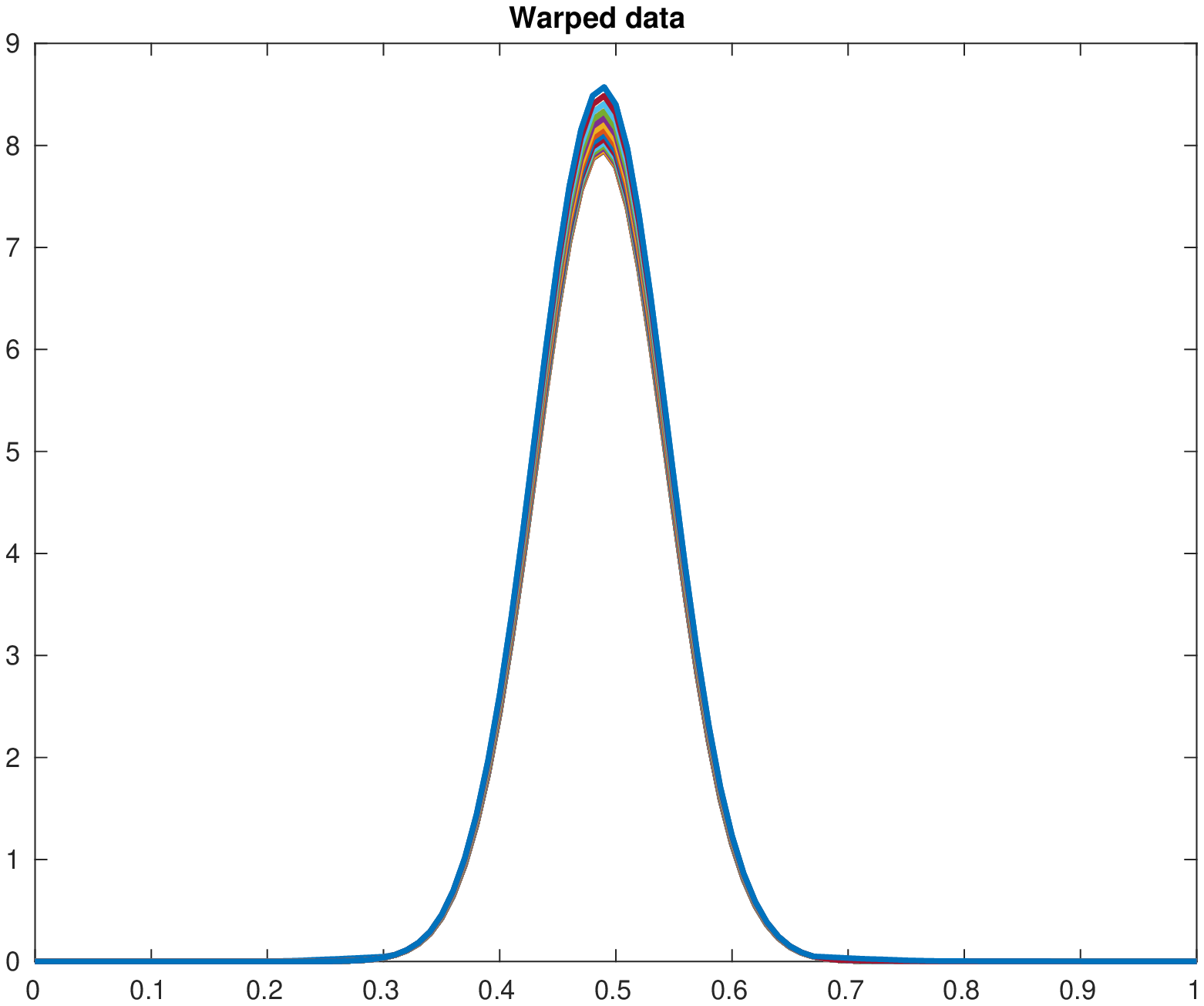}
   & \includegraphics[width=0.15\linewidth]{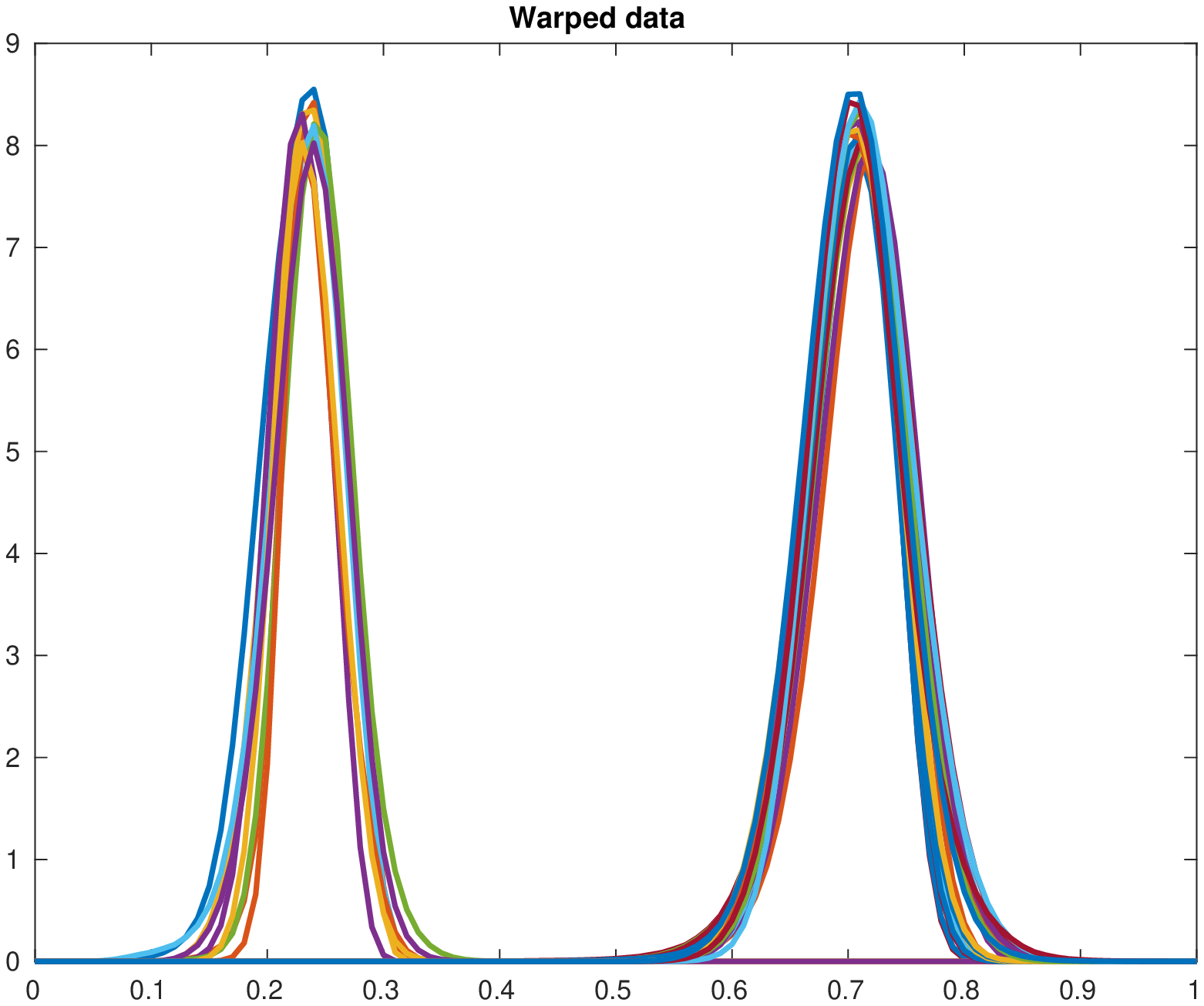}
   & \includegraphics[width=0.15\linewidth]{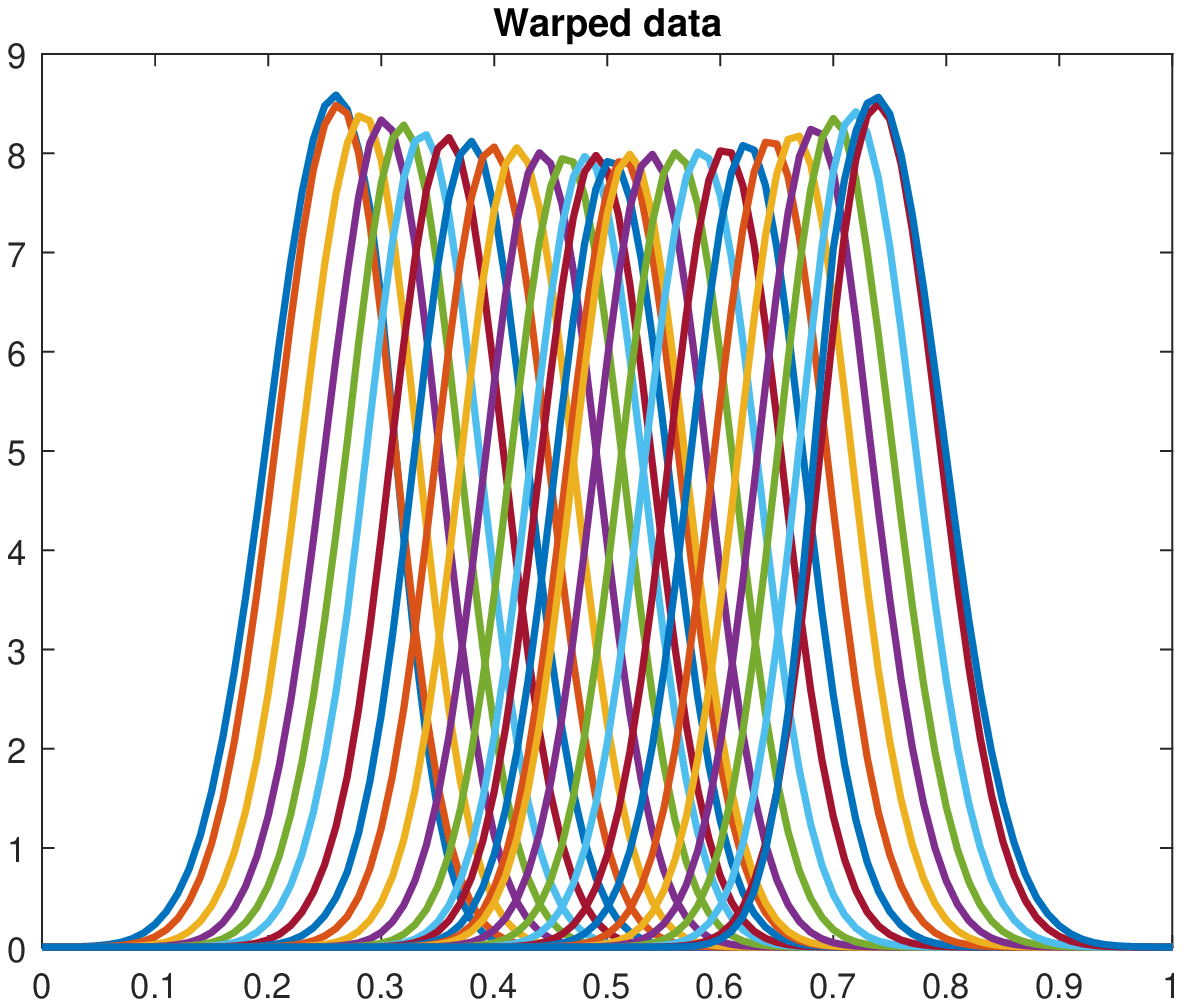} \\
   
    &\includegraphics[width=0.155\linewidth]{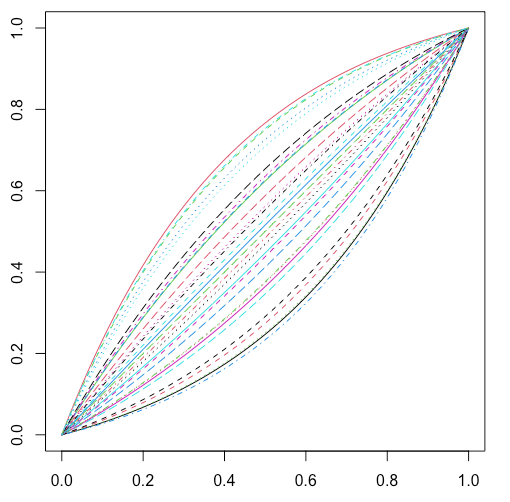}
   &\includegraphics[width=0.15\linewidth]{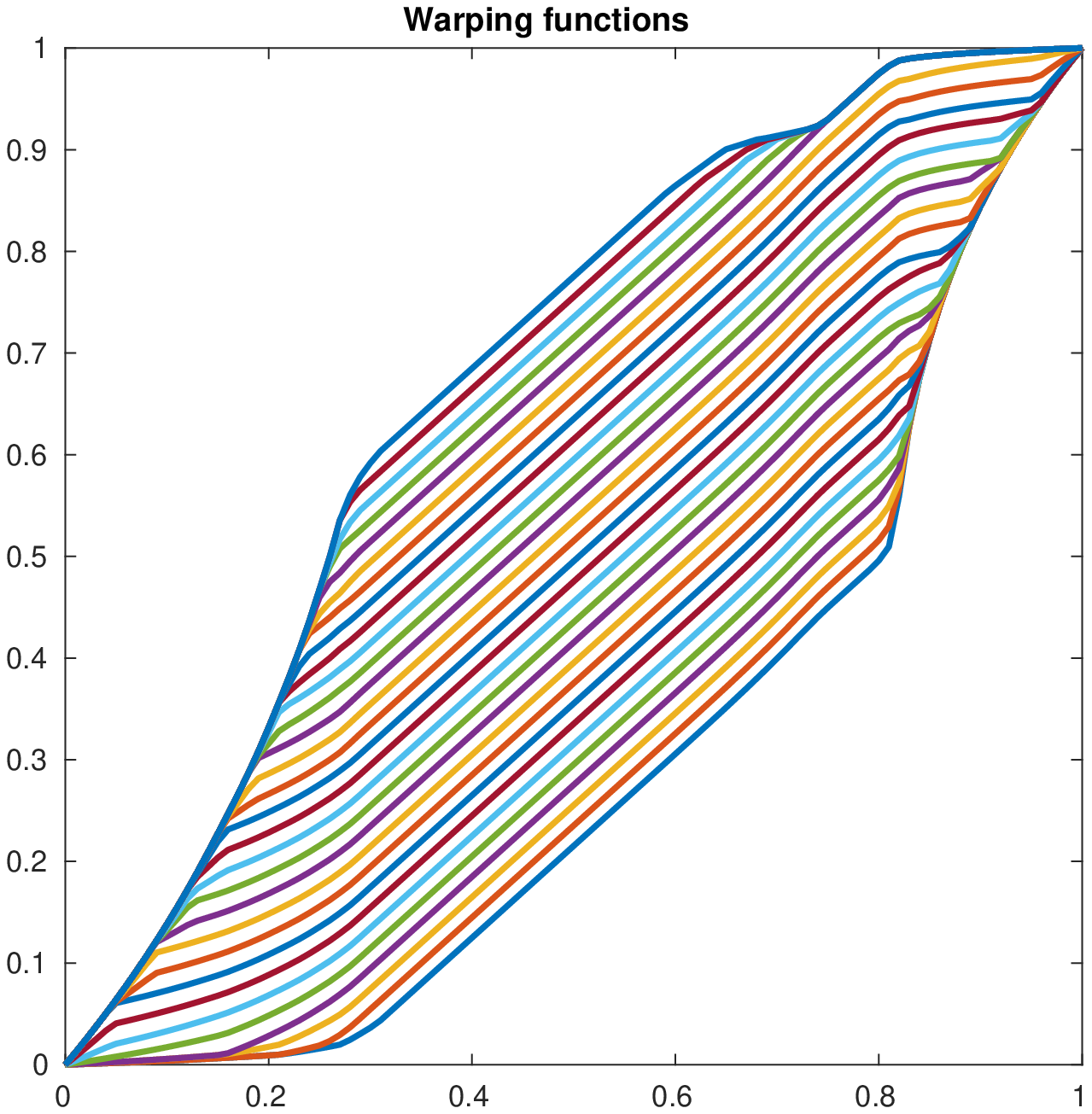}
   &\includegraphics[width=0.15\linewidth]{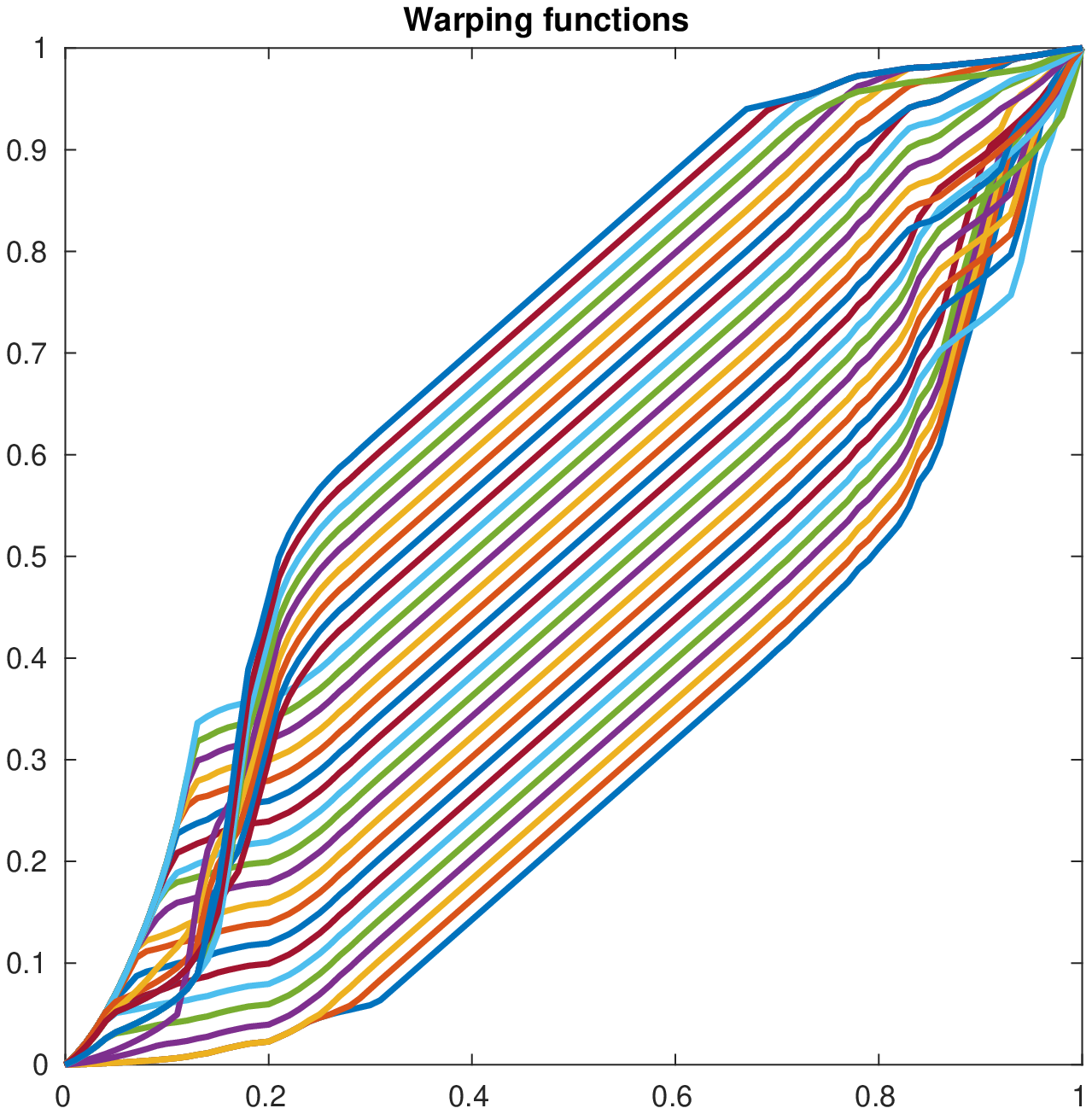}
   & \includegraphics[width=0.15\linewidth]{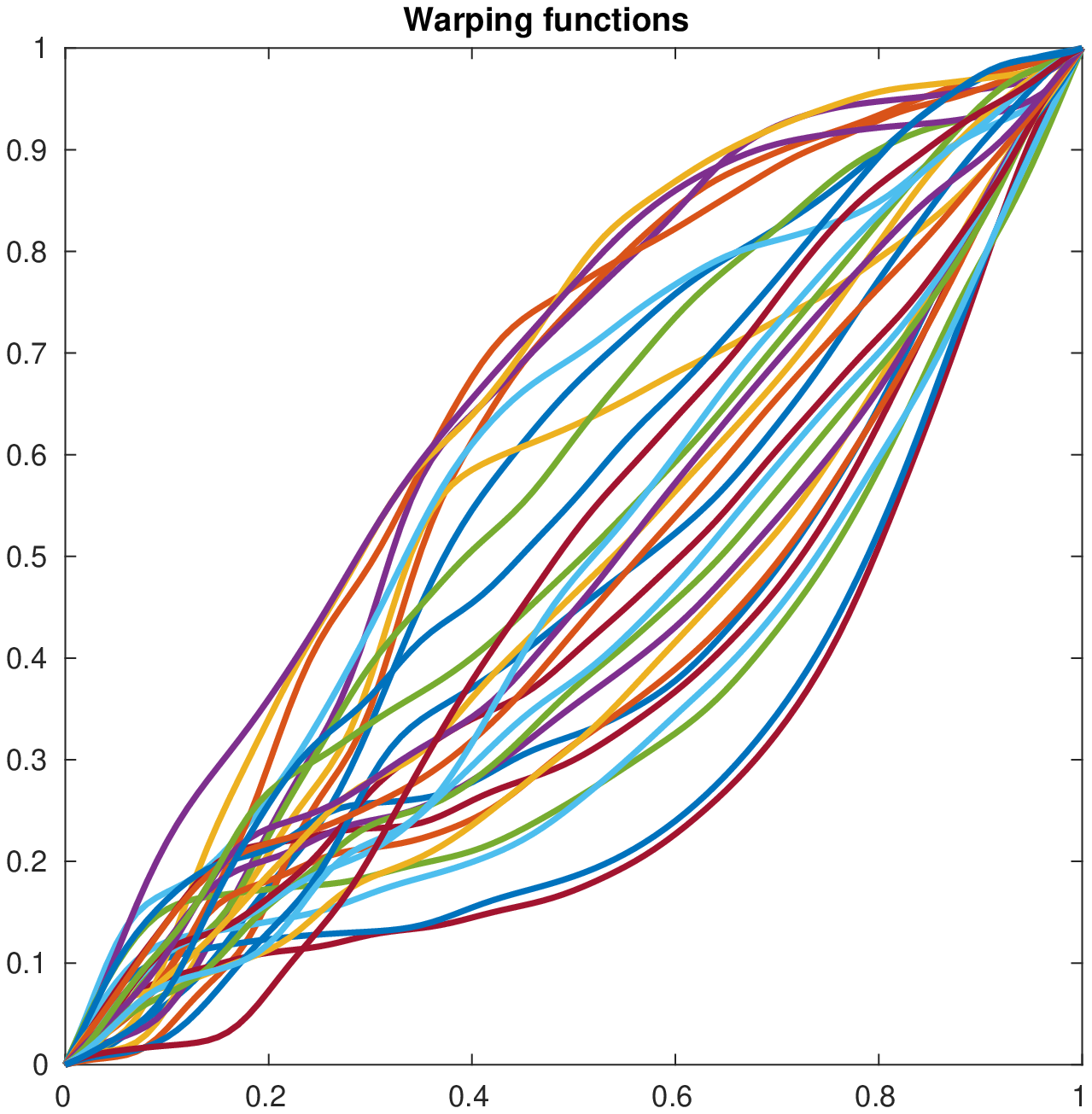}
   & \includegraphics[width=0.15\linewidth]{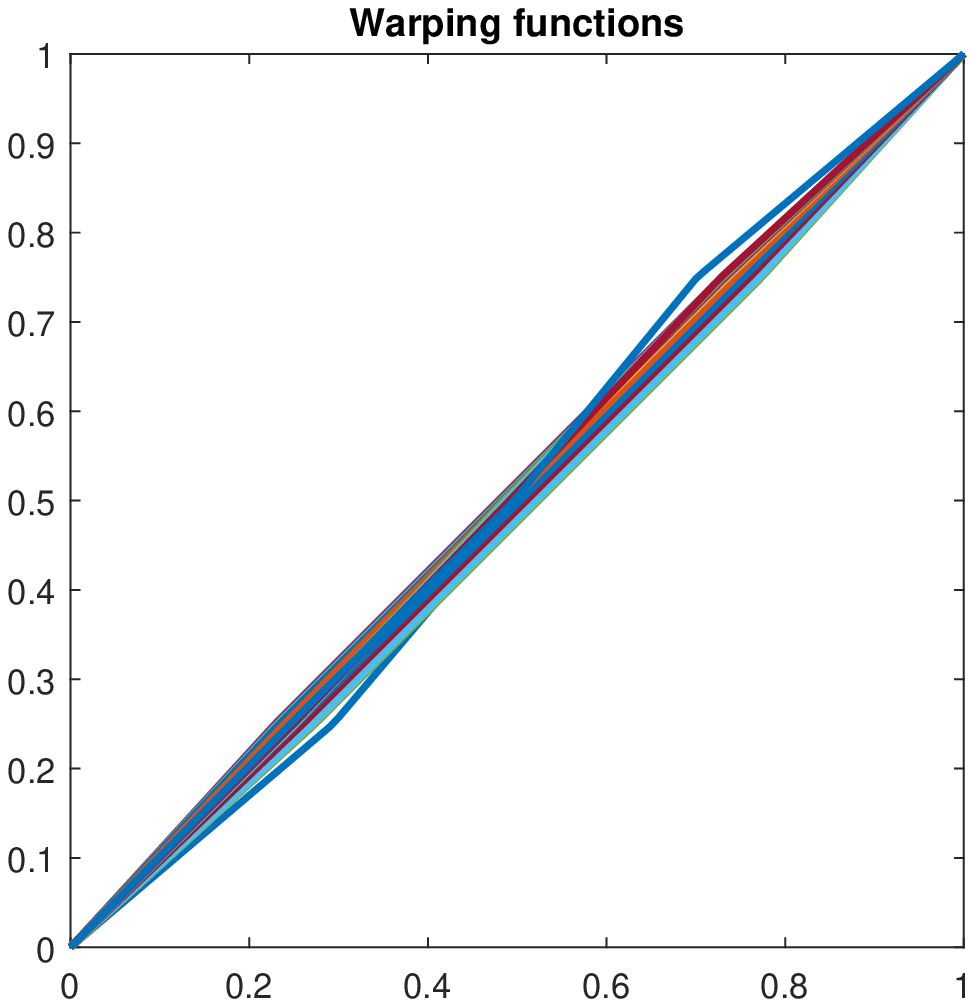} \\
   
   & \tiny{\cite{ramsay2021fda}}
   & SRVF
   & \ltwo
   & \tiny{\cite{lu2017bayesian}}
   & \tiny{\cite{tang2008pairwise}} \\

\end{tabular}
   \caption{{Example of unconstrained alignment. Leftmost shows the original functions;  Results from different methods are displayed on the right hand side. Top is aligned functions and bottom right is corresponding warping functions.}}
\label{fig:ex1}
\end{figure}

Our approach is to modify an SRVF-based technique for elastic registration and shape comparison. 
So we start by briefly introducing the SRVF framework. For details, please refer to the textbook 
~\citep{srivastava2016functional}.
Let $f: [0,1 ] \mapsto \mathbb{R}$ 
be an absolutely continuous function on $[0,1]$ and let ${\cal F}$ be the set of all such functions. 
Define the square-root velocity function (SRVF) of an $f \in {\cal F}$ to be the function 
$q: [0,1] \mapsto \mathbb{R}$, where $q(t)=\text{sign}(\dot{f}(t))\sqrt{|\dot{f}(t)|}$. It can be shown that this 
map provides a bijective mapping between ${\cal F}$ and $\ltwo$, up to an addition by a constant. 
In fact, one can reconstruct $f$ from the pair $(f(0),q)$ using $f(t) = f(0)+ \int_0^t q(s)|q(s)|ds$.
Let $\Gamma$ denote the set of all boundary-preserving positive diffeomorphisms of $[0,1]$ to itself; 
$\Gamma$ forms a group with the group operation given by composition, and the identity element being $\gamma_{id}(t) = t$. 
The group $\Gamma$ acts on ${\cal F}$ from right by composition $f \circ \gamma$, and the corresponding 
action of $\Gamma$ on the SRVF space ($\ltwo$) is given by 
 $(q \circ \gamma)\sqrt{\dot{\gamma}}$; we will denote it by $(q*\gamma)$. 
It can be shown that the $\mathbb{L}^2$ metric in the space of SRVFs is exactly the Fisher-Rao distance in ${\cal F}$~\citep{srivastava2016functional}. 
Furthermore, under this metric, the group action of $\gamma$ preserves the distance, i.e., 
for any $q_1, q_2 \in \ltwo$ and $\gamma \in \Gamma$, we have $\|(q_1*\gamma)-(q_2*\gamma)\|=
\|q_1-q_2\|$, where $\| \cdot \|$ denotes the $\mathbb{L}^2$ norm. 
Consequently, the problem of pairwise alignment can be formulated as 
finding the optimal warping according to: 
\begin{equation}
\gamma=\arg\inf\limits_{\gamma \in \Gamma}  \left( \| q_1-(q_2 \circ \gamma) \sqrt{\dot{\gamma}}\|^2 \right) \ .
\label{eq:SRVF-align}
\end{equation}
This minimization problem can be solved efficiently albeit approximately using
the Dynamic Programming algorithm (DPA). 
Figure \ref{fig:ex1} 
shows an example of this framework applied to the alignment of multiple functions. 
The left panel shows a set of given functions $\{f_i\}$. 
{
On the right hand side, we compare different alignment methods. 
The top row shows the aligned functions $\{\tilde{f}_i\}$ and the bottom row shows the warping functions 
$\{\gamma_i\}$ such that $\{\tilde{f}_i = f_i \circ \gamma_i\}$. 
From left to right, the results are from \ltwo-based~\citep{ramsay2021fda}, SRVF-based (Eqn. \ref{eqn:pen-L2}), \ltwo-based (Our implementation of Eqn. \ref{eq:SRVF-align}), Bayesian approach~\citep{lu2017bayesian} implemented in~\citep{Tucker2022}, \ltwo-based~\citep{tang2008pairwise}}. 
In addition to DPA for SRVF matching, an exact matching algorithm based on the change points of functions 
has also been developed for this setup~\citep{robinson2017exact}.

As mentioned earlier, while this technique is impressive in alignments of peaks and 
valleys on the given functions, it is also susceptible to noise and over-alignment. Since one can potentially use the 
landmark information, if available, to overcome this issue,
we turn our attention to the problem of incorporating landmark information
in the registration process.
The landmarks can be the locations of critical points or any other meaningful geometric features of given functions. 
Registered landmarks can often assist in alignment and lead to a more meaningful phase-amplitude separation, 
relative to the unconstrained problem. 
These landmarks can be obtained in different ways depending on the data
and the context~\citep{ramsay2006functional}, 
but here we shall assume that a set of sparse landmarks is given. 

\subsection{Pairwise Hard Landmark Alignment}
The SRVF framework, as defined originally,  is not designed to incorporate landmark information
in the matching problem but has been extended in recent years to a hard matching \citep{strait2017novel, strait2017landmark, bharath2017partition, strait2018automatic}.
We will first discuss the hard alignment, i.e., the solutions where the landmarks are to be matched precisely with each other. Later on, 
we will relax this constraint to reach the soft alignment. 

In the following discussion, 
we will assume that all functions come with the same number of ordered landmarks and are in one-to-one correspondence. 
Suppose a function $f \in {\cal F}$ has $n$ associated landmarks: 
$\boldsymbol{\tau}  \equiv (\tau_1, \tau_2, ..., \tau_n)$, where $0< \tau_1<\tau_2<...< \tau_n< 1$.  
A mathematical representation of $f$ with landmarks 
is given by : $(q, \boldsymbol{\tau}) \in \mathbb{L}^2 \times D_n$, 
where $D_n = \{(x_1, ..., x_n) \in (0,1)^n | x_1 < x_2 < ... < x_n\}$. 
Also, $q$ denotes the SRVF of $f$, as earlier. 
The action of the warping group $\Gamma$ on the pair $(q, \boldsymbol{\tau})$ is given by $((q*\gamma), \gamma^{-1}({\boldsymbol{\tau}}))$; note
that this action preserves the number and the ordering of the landmarks. 

Let $\boldsymbol{\tau}^{(1)} \in D_n$ be
the set of  $n$ landmarks of $f_1$ and $\boldsymbol{\tau}^{(2)} \in D_n$ be the landmarks of $f_2$. 
Rather than matching these landmarks to each other, we will choose a third set, called 
the {\it reference landmarks} $\bar{\boldsymbol{\tau}} = (\bar{\tau}_{1}, \bar{\tau}_{2},..., \bar{\tau}_{n})\in D_n$, and 
match $\boldsymbol{\tau}^{(1)}, \boldsymbol{\tau}^{(2)}$ to this reference set. 
The reference set $\bar{\boldsymbol{\tau}}$ is sometimes predefined from the problem context and is kept fixed through out the process. 
If not, we can choose the reference set from the given data.  For example, one can pick one of two given sets of landmarks, 
or compute arithmetic average of the two. 
(Note, if the reference landmarks depend on the data, as opposed to being 
pre-determined and fixed,  we will lose a nice theoretical property of invariance as explained later.)
Now, the goal of pairwise hard alignment is to find warping 
functions $\gamma_1, \gamma_2 \in \Gamma$ such that: (1)  $\gamma_1(\bar{\boldsymbol{\tau}}) = \boldsymbol{\tau}^{(1)}$ 
and $\gamma_2(\bar{\boldsymbol{\tau}}) = {\boldsymbol{\tau}}^{(2)}$, which means that one matches the landmarks exactly; 
and, (2) match features (peaks and valleys) of the two functions in the corresponding subintervals. 

\begin{figure}[H]
\begin{center}
\begin{tabular}{ccc}
   \includegraphics[width=0.33\linewidth]{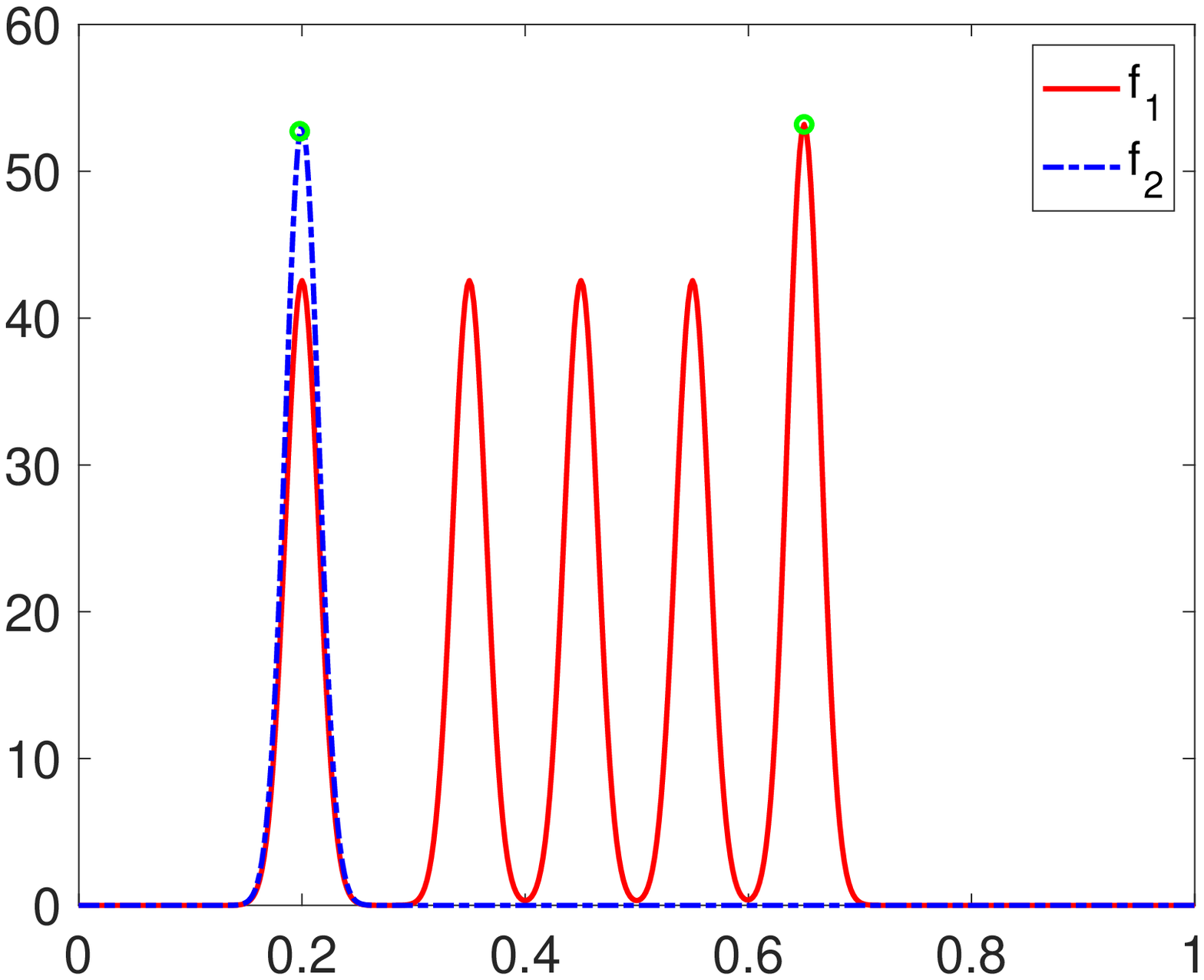}&
   \includegraphics[width=0.33\linewidth]{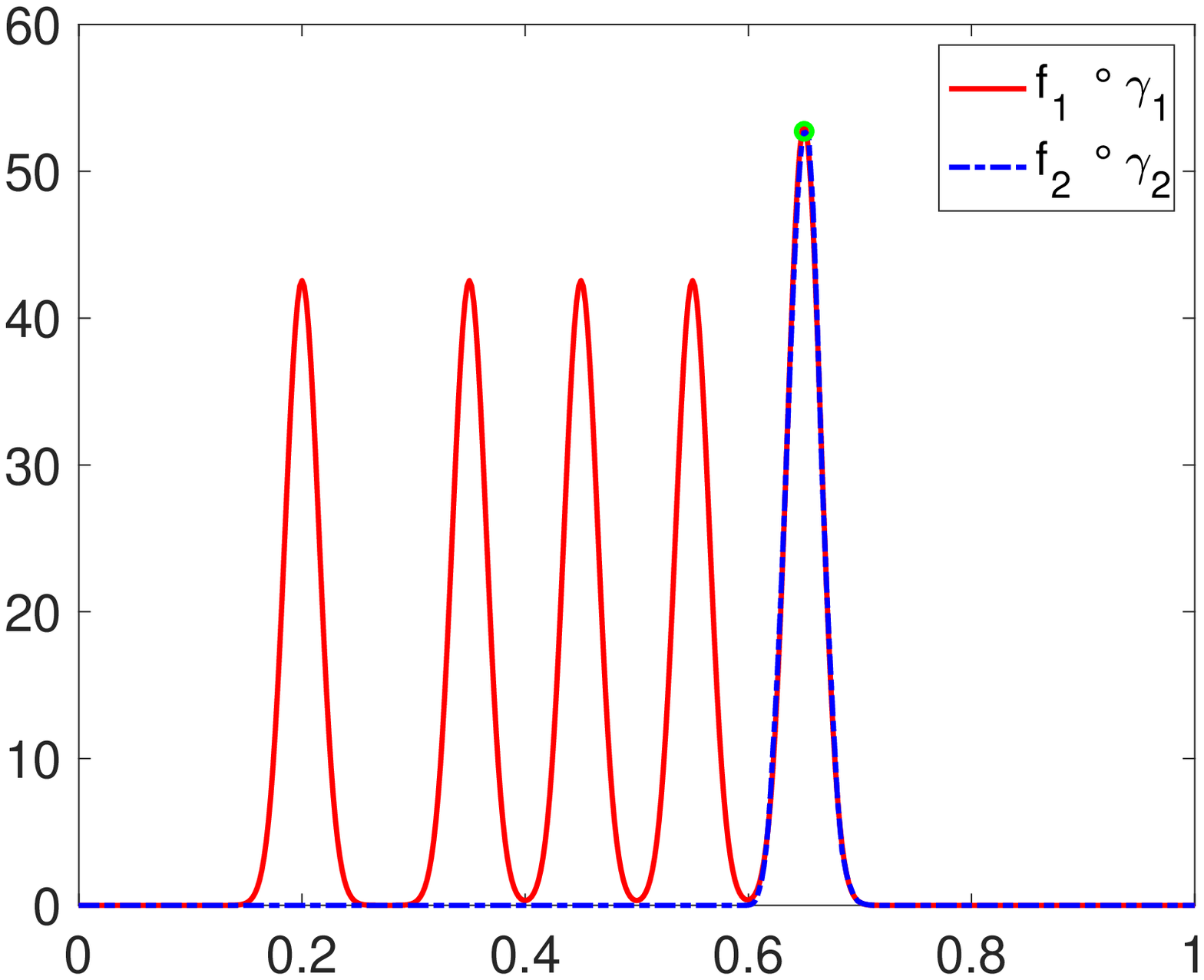}&
   \includegraphics[width=0.33\linewidth]{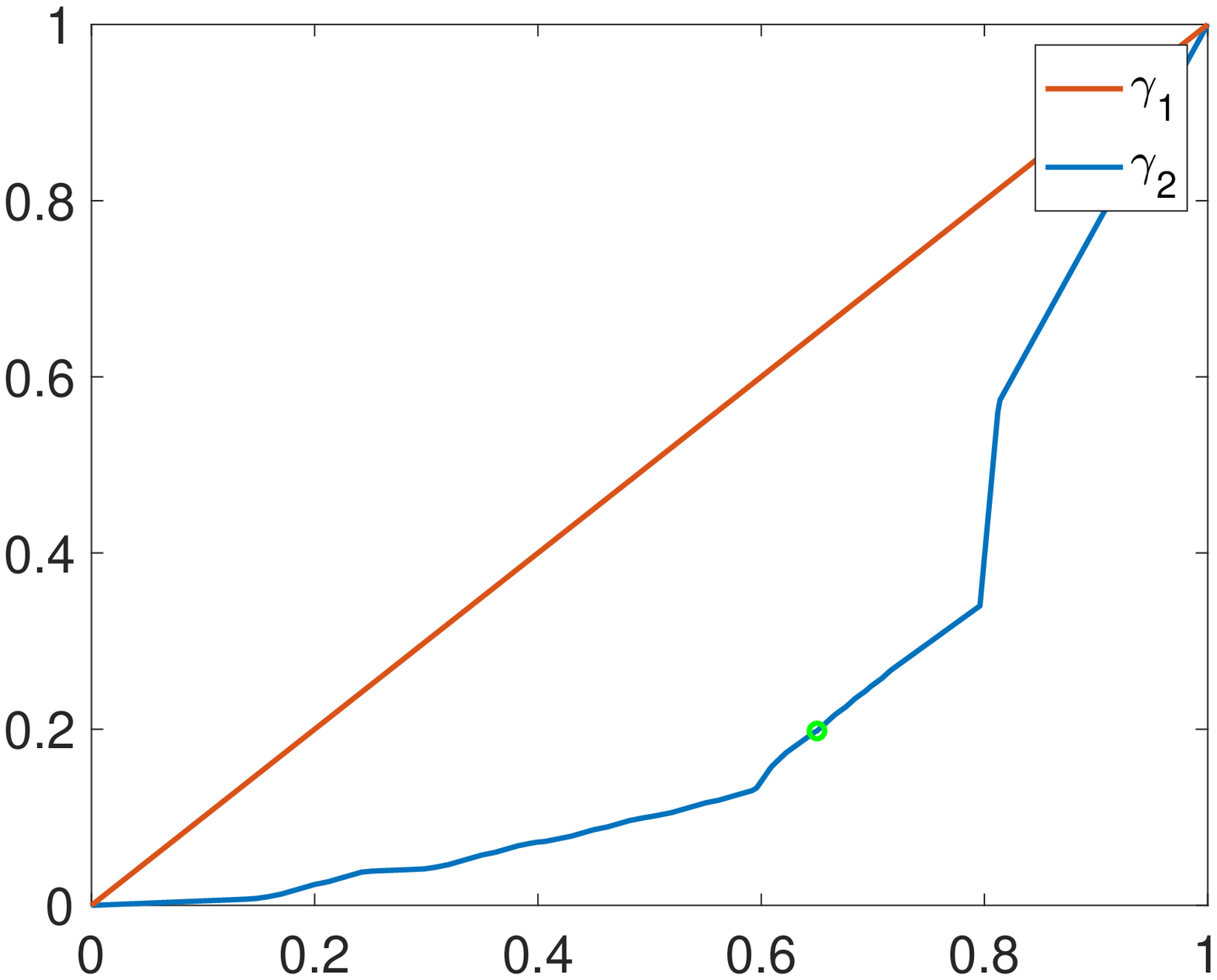}
\end{tabular}
\end{center}
   \caption{Example of hard registration. Left: original functions; Middle:
   result of a hard registration; Right: the warping functions.
   Green circle represents the function values at landmarks.}
\label{fig:simupre}
\end{figure}

\begin{figure}[H]
\centering
\begin{tabular}{ccc}
   \multirow{2}{*}[30pt]{\includegraphics[width=0.4\linewidth]{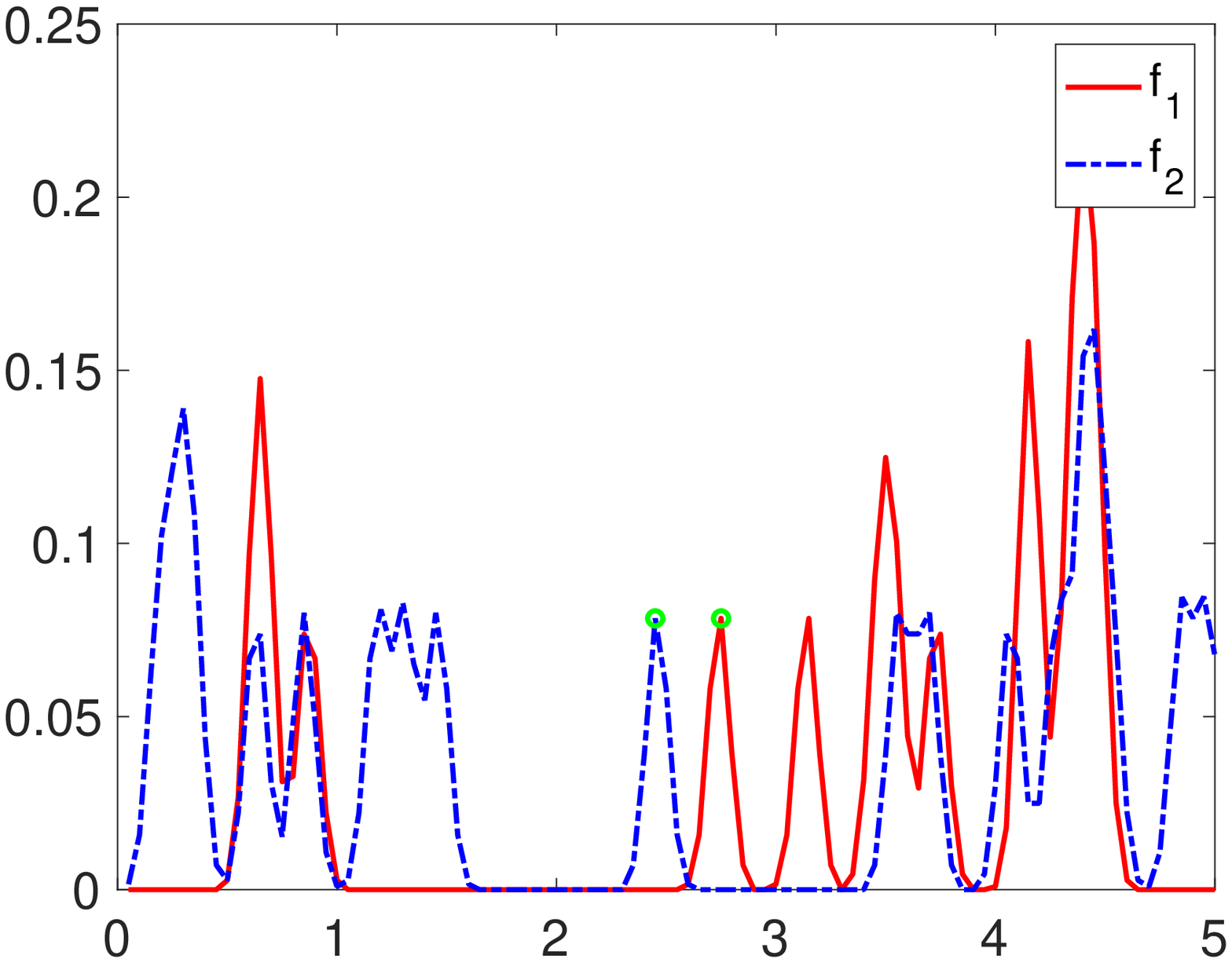}}
   &\includegraphics[width=0.3\linewidth]{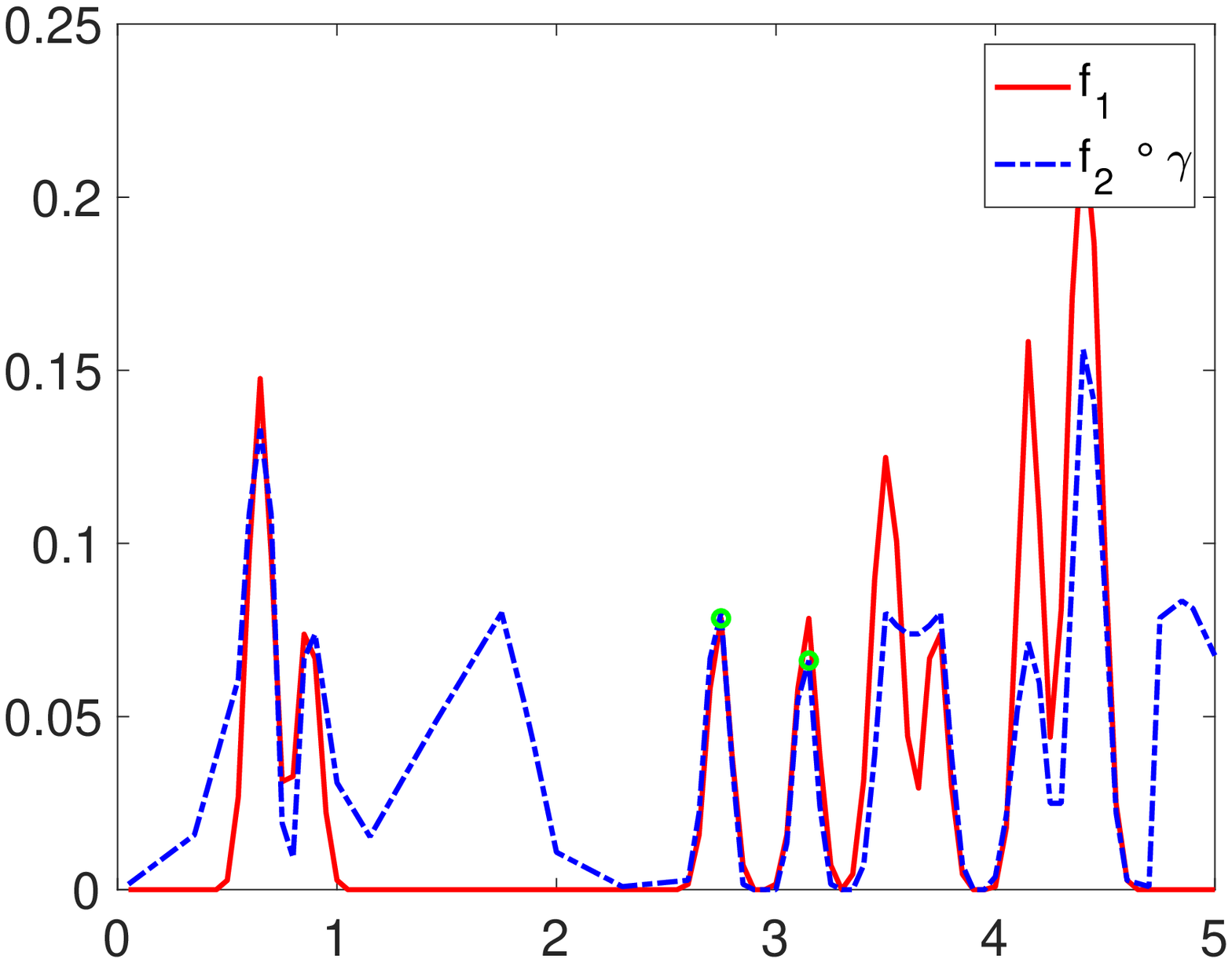}
   &\includegraphics[width=0.28\linewidth]{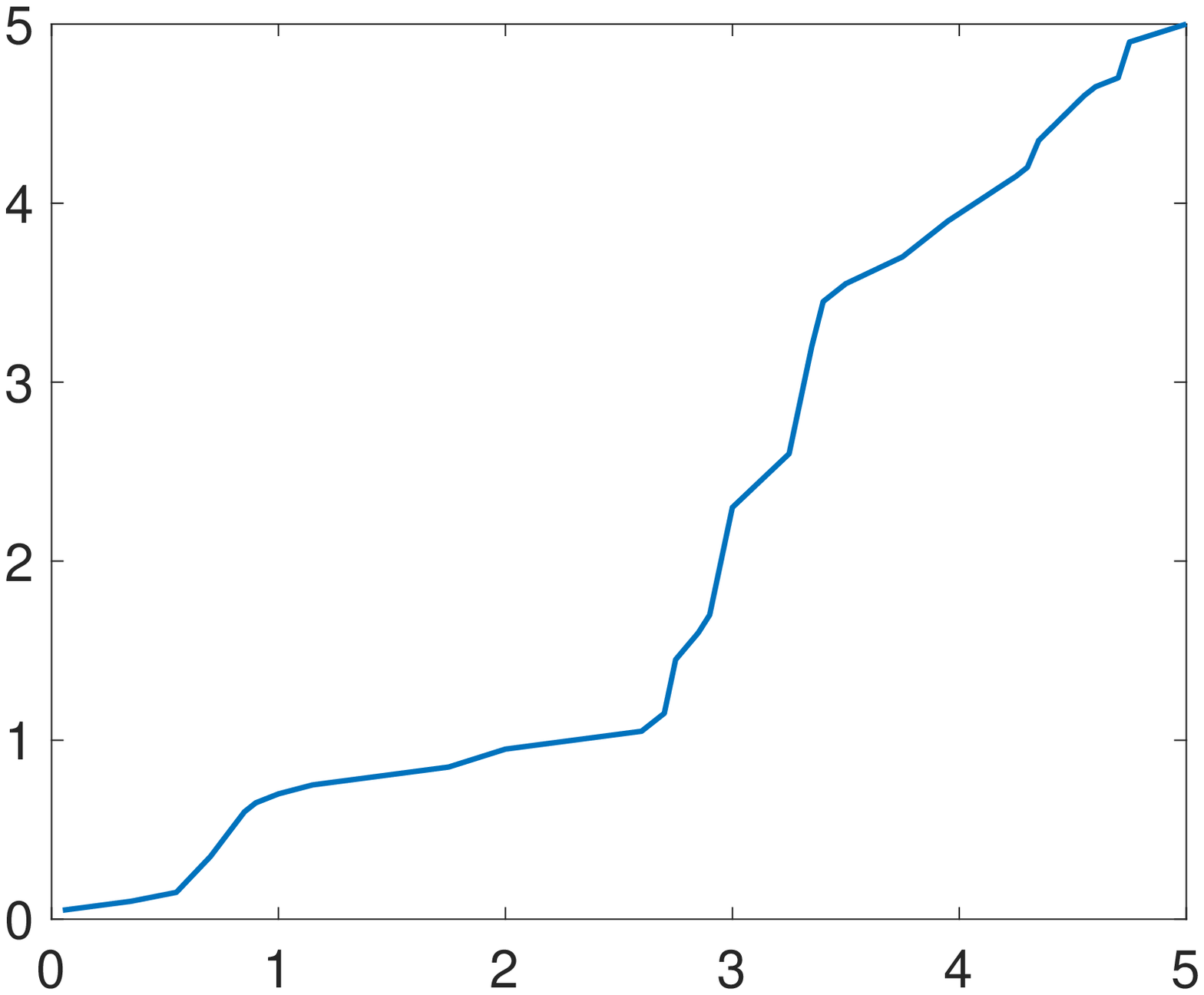} \\
   &\includegraphics[width=0.3\linewidth]{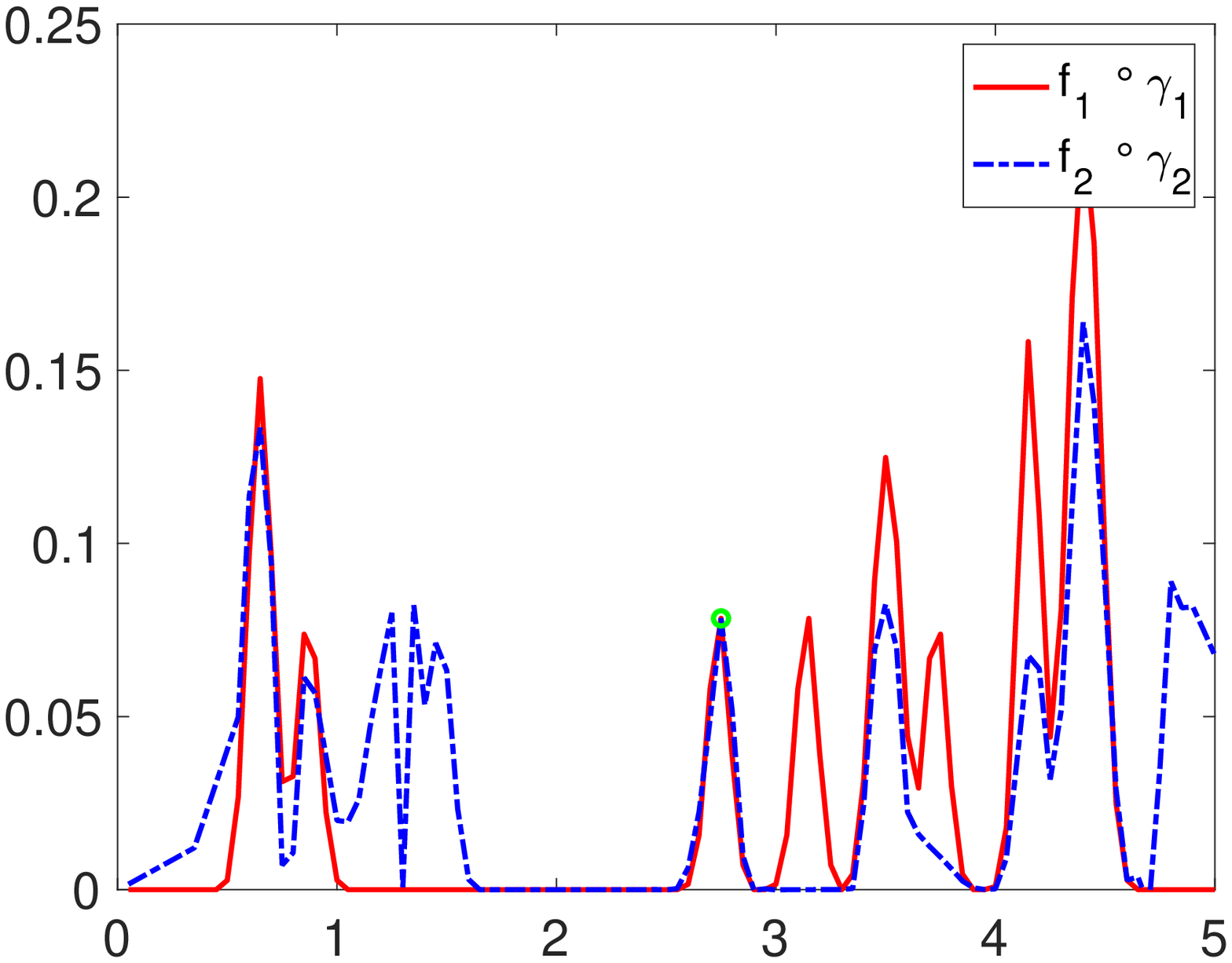}
   &\includegraphics[width=0.28\linewidth]{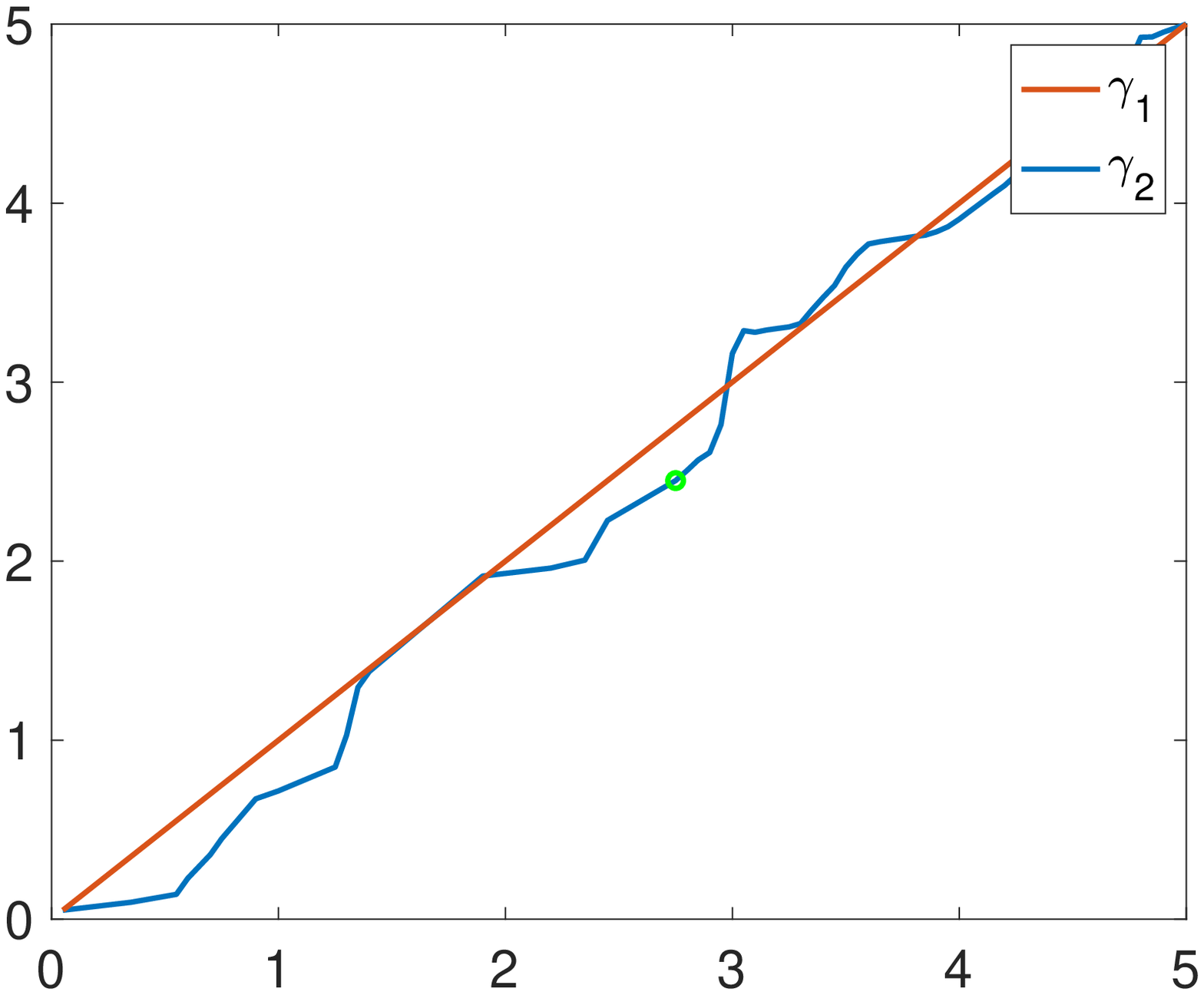}\\

\end{tabular}
   \caption{Example of hard registration of spike data. Left: 
   original functions $f_1, f_2$, each with a single landmark, marked by the green circles;
   Right: the top row shows the unconstrained registration while bottom shows the hard registration.}
\label{fig:spike_hard}
\end{figure}

To motivate the second item, note that
problem is ill defined if we only care about registering landmarks and ignore the shapes of functions. 
Given $f_1, f_2$, and their landmarks $\boldsymbol{\tau}^{(1)}$ and $\boldsymbol{\tau}^{(2)}$, 
the sets $\{\gamma_1 \in \Gamma|\gamma_1(\bar{\boldsymbol{\tau}}) =\boldsymbol{\tau}^{(1)} \}$ and 
$\{\gamma_2 \in \Gamma|\gamma_2(\bar{\boldsymbol{\tau}}) =\boldsymbol{\tau}^{(2)} \}$ have infinitely many elements.
To narrow down the solution space, one can restrict the $\gamma_1, \gamma_2$ to be piecewise linear.
Although this solution is simple, it completely ignores the shapes of functions and leads to badly registered functions.
Therefore, in addition to the landmark matching, one also requires the 
functions to be optimally registered to each other in the corresponding subintervals.
Towards that goal, one can simply apply the SRVF framework, discussed above for the full interval $[0,1]$,
to each of the matched subintervals independently. The detailed steps are as follows: 
\begin{enumerate}
\item Select the reference landmarks $\bar{\boldsymbol{\tau}}$, either from the problem 
context or from the data. 
Partition the time domain $[0,1]$ into $n+1$ subintervals with  
landmarks as boundaries, for each function. This defines the boundary points of 
the warping functions $\gamma_i^{-1}(\tau_j^{(i)}) = \bar{\tau}_j$, $i=1,2$. 

\item Align the respective functions on these subintervals separately using the SRVF framework. 
Concatenate the warping functions over these intervals to form full time warping functions 
over $[0,1]$. Call them $\gamma_1, \gamma_2$. 

\item Optionally, center the warping function according to $\gamma_i \mapsto \gamma_i \circ \bar{\gamma}^{-1}$, 
where $\bar{\gamma} = (\gamma_1 + \gamma_2)/2$. 

\end{enumerate}
We present an example of this approach in Figure \ref{fig:simupre} using a single landmark. 
The two functions: $f_1$ (multimodal) 
and $f_2$ (unimodal), are as shown on the left.
The landmark for $f_1$ is the location of its last peak while the landmark for $f_2$ is the location of its only peak. 
The reference landmark is chosen to be the same as the landmark of $f_1$.
The result of hard registration is shown in the middle panel, where landmarks as well as 
the corresponding functions in the two subintervals are matched perfectly.
An example with real data is presented in Figure \ref{fig:spike_hard}, which shows 
two spiking activities of a movement-encoded neuron in the primary motor cortex, taken from \citep{wu2011information}.
The two spike activities $f_1$ and $f_2$ are equipped with landmarks that are marked as green and black circles. 
We would like to align both the functional shapes as well as the landmarks. 
As the top row of Figure \ref{fig:spike_hard} shows, the unconstrained alignment by SRVF does not register landmarks correctly. However, by using hard registration, one can see in the bottom row that the landmarks are perfectly aligned. 

\subsection{Pairwise Soft Landmark Alignment}\label{subsec:pair_sfa}
While hard alignment provides us with a way of incorporating landmarks information to alignment, it relies too heavily on
the locations of the landmarks. 
The result can be misleading in situations where the landmarks are not precise.
Thus, we want to pursue a soft approach that can balance the contributions of 
landmarks and function shapes.
Unlike hard registration, a soft alignment does not require matched landmarks to be 
aligned exactly but pushes them towards 
each other.
In summary, soft alignment seeks intermediate solutions between unconstrained alignment 
(Section 2.1) and hard alignment (Section 2.2). 

We remind the reader that it is desirable to 
perform matching using an objective function that is either a proper metric or a pseudometric. 
That is, the desired cost function should satisfy non-negativity, identity of indiscernibles (for a metric), symmetry and triangle inequality.
This, in turn, requires that we have invariance of the objective function under the group action of time warping ({\tt refer to the proof in appendix}).

Now we are ready to present our approach for soft alignment. The basic idea is to perform pairwise alignment in two 
steps: (1) First, we perform hard registration to match the shapes
and landmarks of the given functions. 
(2) Then, using the resulting functions, we perform a further penalized SRVF alignment. That is, we minimize the 
SRVF objective function along with a term that forces the additional warping function to stay close to the identity. 
This second part 
is essentially a search in $\Gamma$ without any additional involvement of landmarks but constrained to stay closed to
$\gamma_{id}$. 
Depending on the choice of a tuning parameter $\lambda$, one can sweep the entire spectrum of solutions ranging 
from hard registration ($\lambda$ very high)
to fully unconstrained SRVF registration ($\lambda = 0$). 
We will describe how the propose soft alignment is superior in term of both theory and computation to the previous ideas. 

To define the objective function mathematically, start with any 
two functions $f_1, f_2 \in {\cal F}$, with SRVFs $q_1$ and $q_2$, and respective landmarks ${\boldsymbol{\tau}}^{(1)} \in D_n$ 
and $\boldsymbol{\tau}^{(2)} \in D_n$,
Let $\gamma_1$ and $\gamma_2$ be the optimal warping functions resulting from the hard registration
discussed previously (Section 2.2).
The individual landmarks are thus perfectly aligned to the 
reference landmarks $\bar{\bf \tau}$.

\begin{defn}(Pairwise Soft Landmark Registration)
In the setting described above, define the optimization problem for soft landmark, pairwise function registration to be: 
\begin{eqnarray} \label{eq:dis}
 && \inf_\gamma \left(\| (q_1*\gamma_1)-((q_2*\gamma_2)*\gamma) \|^2 +\lambda \| ({\bf 1}*\gamma)-{\bf 1} \|^2 \right) \\
 = && \inf_\gamma  \left( \| (q_1*\gamma_1)-((q_2*\gamma_2)*\gamma) \|^2+\lambda  \|\sqrt{\dot{\gamma}}-\bf{1} \|^2 \right),
 \nonumber
\end{eqnarray}
where $\bf{1}$ is a constant function with value one. 
\end{defn}
The positive constant $\lambda$ is a tuning parameter 
that can be used to control the influence of the landmarks on the overall solution. 
{\it What is the motivation for using the objective function suggested here}? Here are some properties that
makes this choice meaningful.

\begin{enumerate}
\item {\bf Difference from Eqn.~\ref{eqn:pen-L2}}: 
Even though this formulation takes the form of a penalized alignment, it is fundamentally different from 
the formulation in Eqn.~\ref{eqn:pen-L2}. Here, the default solution (without any alignment by $\gamma$) is hard SRVF alignment, 
while in Eqn.~\ref{eqn:pen-L2} the default solution is no alignment. {Here, $\gamma$ is being used to reduce local alignment (without constraint) 
and move towards a global (with constraint) alignment. In Eqn.~\ref{eqn:pen-L2}, $\gamma$ is being used to perform all alignment -- local and global. }

\item {\bf Pseudometric nature}: The optimization leads to a pesudo-metric for ensuing statistical analysis of functions. 
\begin{lemma} \label{lem:pseudom}
The resulting infimum of the objective function in Eqn. \ref{eq:dis}, call it $d_{\lambda}$, is a pseudometric on the space $\ltwo \times \real^n$. 
\end{lemma}
{\bf Proof}: See appendix.
Consequently, this objective function is symmetric and it satisfies triangle inequality.

\item {\bf Invariance condition}: 
An important property of this pseudometric is the following. 
\begin{lemma} \label{lem:isometry}
The action of $\Gamma$ on joint space $\ltwo \times \real^n$, given by 
$((q*\gamma), \gamma^{-1}({\boldsymbol{\tau}}))$ preserves $d_{\lambda}$ for a fixed reference landmark $\bar{\boldsymbol{\tau}}$.
That is, 
$$
d_{\lambda}((q_1*\gamma_0),(q_2*\gamma_0)) = d_{\lambda}(q_1,q_2)\ ,
$$
for all $q_1, q_2 \in \ltwo$ and $\gamma_0 \in \Gamma$. 
\end{lemma}
{\bf Proof}: Suppose $\gamma_1$ and $\gamma_2$ are the time warping functions from hard registration for $q_1$ and $q_2$. For an arbitrary $\gamma_0$, the new time warping functions for hard registration of $(q_1*\gamma_0)$ and $(q_2*\gamma_0)$ are $(\gamma_0^{-1}*\gamma_1)$ and $(\gamma_0^{-1}*\gamma_2)$, respectively. Therefore,
\begin{eqnarray*}
&& d_{\lambda}((q_1*\gamma_0),(q_2*\gamma_0)) \\
= && \inf_\gamma \left(\| (q_1*\gamma_0)*(\gamma_0^{-1}*\gamma_1)-((q_2*\gamma_0)*(\gamma_0^{-1}*\gamma_2)*\gamma) \|^2 +\lambda \|\sqrt{\dot{\gamma}}-{\bf 1} \|^2 \right) \\
= && \inf_\gamma  \left( \| (q_1*\gamma_1)-((q_2*\gamma_2)*\gamma) \|^2+\lambda  \|\sqrt{\dot{\gamma}}-\bf{1} \|^2 \right) =  d_{\lambda}(q_1,q_2) \ .
\end{eqnarray*}
\item {\bf Balance between hard and unconstrained registration}: 
When $\lambda=0$, the resulting distance $d_0$ is exactly 
the same as cost function used in Eqn.~\ref{eq:SRVF-align} for unconstrained alignment. 
As $\lambda$ increases, the impact of landmarks on the alignment increases, and eventually
when $\lambda$ becomes very large then $\gamma$ tends to $\gamma_{id}$, resulting in a purely 
hard registration. 
For intermediate $\lambda$s, we get solutions that represent combinatiosn of these two solutions.
\end{enumerate}

\noindent {\bf Selecting $\lambda$ Using Cross Validation}: The choice of the 
relative contributions of two terms, controlled by $\lambda$, is 
important in reaching a good solution. This choice depends on the application, the 
data and the end-user.
In addition, one can automate the process using cross-validation. 
We will use some examples in Section \ref{subsec:cv} to discuss this issue further.

\section{Soft Landmark Alignment of Multiple Functions}\label{sec:multiple_sfa}
Now we consider the problem of alignment of multiple functions, 
each equipped with their own landmarks.  
This multiple alignments is based on the repeated use of the pairwise alignment
framework derived previously. 
The basic idea is to use $d_{\lambda}$, a pseudometric on $\ltwo$ space, 
to define a 'consensus' function, 
and use this consensus as a template to align individual functions to this template.  

Let $\mu_q \in \ltwo$, with fixed landmark $\bar{\boldsymbol{\tau}} \in D_n$, 
denote the consensus, defined as follows. We are given $m$ individual SRVFs $\{q_i\}$ with 
respective landmarks $\{\boldsymbol{\tau}^{(i)}\},\  i=1, 2, ... , m$. {Define the 
landmarks of the consensus in some pre-determined fashion, for instance, using landmarks from one of given function;  call them $\bar{\boldsymbol{\tau}}$}. Then, 
the consensus function (in the SRVF space) is defined to be: 
\begin{equation}\label{eq:mean}
\mu_q = \arg \inf_q \left( \sum_{i=1}^m d_\lambda (q, q_i)^2 \right) \ ,
\end{equation}
while keeping $\bar{\boldsymbol{\tau}}$ fixed. 
(We call this function consensus instead of a mean because we don't have a full metric but only 
a pseudometric on the representation space.)
We solve this optimization problem iteratively as follows. First we perform a hard registration of 
each $(q_i, \boldsymbol{\tau}_i)$ to the current estimate $\mu_q$ (note that $\bar{\boldsymbol{\tau}}$ is 
kept fixed in these iterations). 
After the hard registration, Eqn. \ref{eq:mean} is reduced to a penalized SRVF multiple alignment problem:
\begin{equation*}
\mu_q =\arg \inf_{{\gamma}_i,\mu_q}  \left( \sum_{i=1}^m \| \mu_q -((q_i*\tilde{\gamma_i})*{\gamma}_i) \|^2+\lambda  \|\sqrt{\dot{\gamma}_i}-\bf{1} \|^2 \right) \ ,
\end{equation*}
where $\tilde{\gamma_i}$ is already pre-computed from the hard registration.
Then, for fixed $\{\gamma_i\}$, $\mu_q = \frac{1}{m}\sum_{i=1}^m(({q}_i*\tilde{\gamma_i})*{\gamma}_i)$. 
The overall algorithm for soft registration of multiple functions is given as in Algorithm \ref{algo:msa}. 

\begin{algorithm}
\caption{Multiple Soft Alignment}\label{algo:msa}
Given functions $f_1, f_2, \dots, f_m$ and the associated landmark vectors $\boldsymbol{\tau}^{(1)}, \boldsymbol{\tau}^{(2)}, \dots, \boldsymbol{\tau}^{(m)}$. 
\begin{algorithmic}[1]
\State {\bf Initialization}: Compute SRVFs $q_1, q_2, \dots, q_m$ of the given functions. Determine 
the reference landmarks $\bar{\boldsymbol{\tau}}$ and 
set $\mu_q = q_i$ where $i = \arg \inf_{1 \leq i\leq m} \sum_{j \neq i} d_{\lambda}(q_i,q_j)^2$
\State {\bf Hard Registration}: For each $(q_i,\boldsymbol{\tau}^{(i)})$, find the initial warping function $\tilde{\gamma_i}$ by hard registration to $(\mu_q, \bar{\boldsymbol{\tau}})$.

\item {\bf Consensus}: Update $\mu_q = \frac{1}{m}\sum_{i=1}^m({q}_i*\tilde{\gamma_i})$.
\State {\bf Further Alignment}: For each $i=1,2,\dots,n$, solve for 
$$
{\gamma}_i = \arg\inf_\gamma  \left( \| \mu_q -((q_i*\gamma_i)*\gamma) \|^2+\lambda  \|\sqrt{\dot{\gamma}}-\bf{1} \|^2 \right)\ , 
$$ 
and 
set $\tilde{q}_i = ((q_i*\tilde{\gamma}_i)*{\gamma}_i)$. 
\State  
Return to step 3 until convergence ($\mu_q$ is stable).
\end{algorithmic}
\end{algorithm}

This is an iterative algorithm and one needs to consider its
convergence properties. Since the algorithm is based on
alternative optimization over $\{\gamma_i\}$
and $\mu_q$, the convergence to a global solution is not guaranteed.
At best, one expects a local solution solution from this optimization. 
Algorithm 1 returns $m$ soft-aligned functions $(f_i \circ \tilde{\gamma_i} \circ {\gamma}_i)$, $i = 1, 2, \dots, m$ 
in which landmarks are pushed together but not necessarily registered.

\subsection{Selection of $\lambda$ by Cross Validation}\label{subsec:cv}
The relative contribution of landmarks and functions' shapes 
is controlled by $\lambda$, so the choice of $\lambda$ becomes an important factor.
In case $\lambda$ is to be chosen automatically from the data, 
we propose to use leave-one-out cross validation (LOOCV) as follows:
\begin{equation*}
\lambda = \arg\min_{\lambda \in [0,\Lambda]}
\left(  \sum_{j=1}^{m} \inf_{\gamma} \|(q_j *\gamma) - \frac{1}{m-1}\sum_{i \neq j} ((q_i*\tilde{\gamma}_i)*\gamma_{i,\lambda}) \|^2  \right)\ ,
\end{equation*}
where $\tilde{\gamma}_i$ is the time warping function from hard registration (Step 2 of Algorithm \ref{algo:msa}) and 
$\gamma_{i,\lambda}$ is the additional warping (Step 3 of Algorithm \ref{algo:msa}).
{The intuition is to find the $\lambda$ such that the resulting consensus is the center of the shape distribution of functions. }
Given a fixed $\lambda$, we first leave a function 
$f_j$ out and calculate the consensus of the remaining $m-1$ functions using Algorithm \ref{algo:msa}. 
This consensus acts as the estimate for $f_j$, and we compute 
the difference in the shape of $f_j$ and its estimate. Total error over all functions makes up the objective function 
for selecting $\lambda$. 
We use grid search over an interval $[0,\Lambda]$ to find the optimal $\lambda$.

We demonstrate this idea using two simulated but extreme examples. 
\begin{enumerate}

\item {\bf Completely Arbitrary Landmarks}: 
We create 5 functions using $f_i = f_0 \circ \gamma_i$ for different 
$\gamma_i$s. 
We select the midpoint $(0.5)$ as the landmark for all functions, irrespective of their shapes.
Thus, the landmarks do not contain any useful information.
We use the LOOCV procedure to find the best $\lambda$ and results are shown in Figure \ref{fig:cv_wrong_lm}.
As expected, the best $\lambda$ is found to be $0$, implying that landmarks actually do not help and are discarded. 

\begin{figure}[H]
\centering
\subfloat[Raw Simulated Functions]
{
\includegraphics[width=0.33\linewidth]{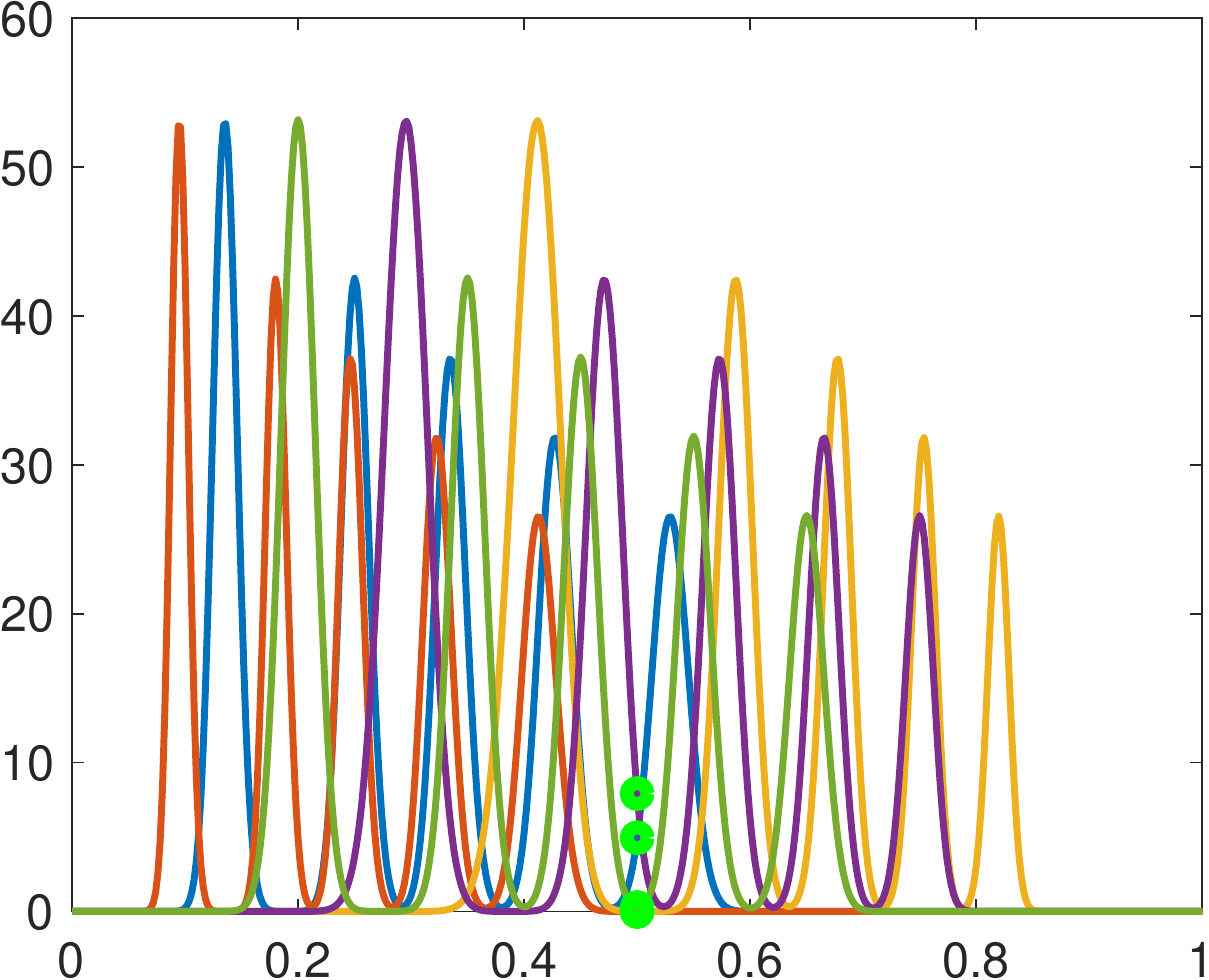}    
}
\subfloat[Hard Registration]
{
\includegraphics[width=0.33\linewidth]{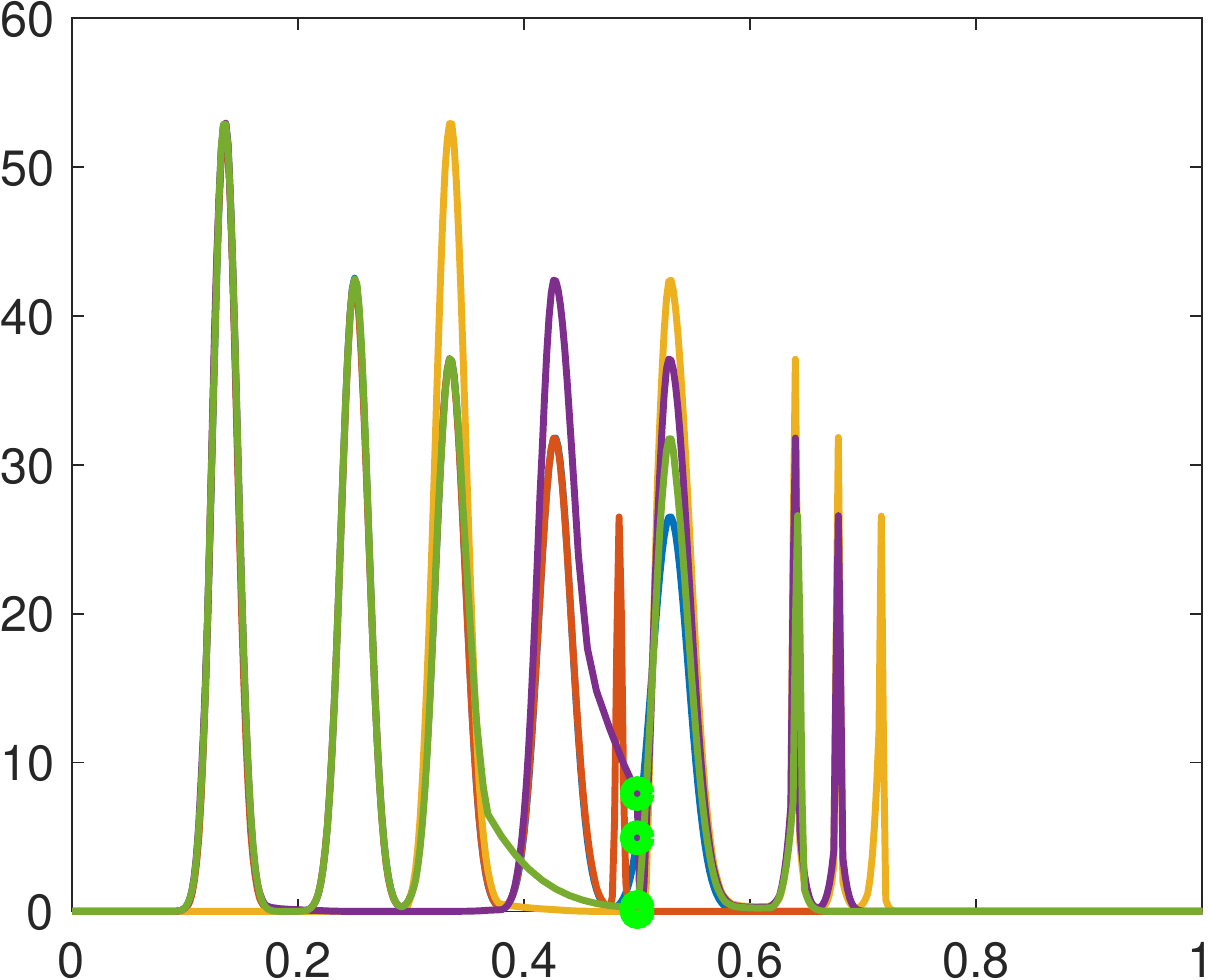}    
}
\subfloat[Unconstrained Registration]
{
\includegraphics[width=0.33\linewidth]{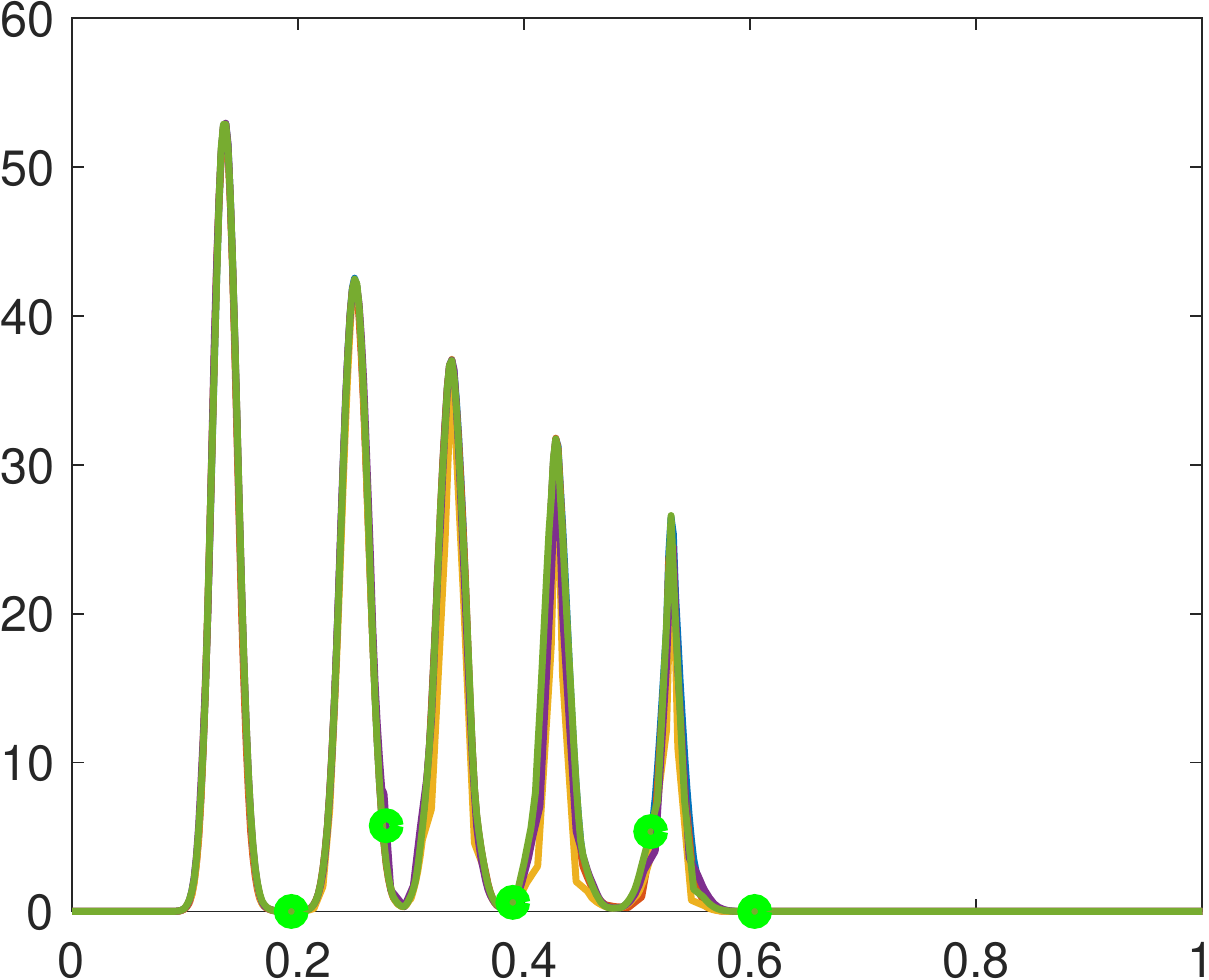}    
}
\\
\subfloat[LOOCV Shape Error]
{
\includegraphics[width=0.33\linewidth]{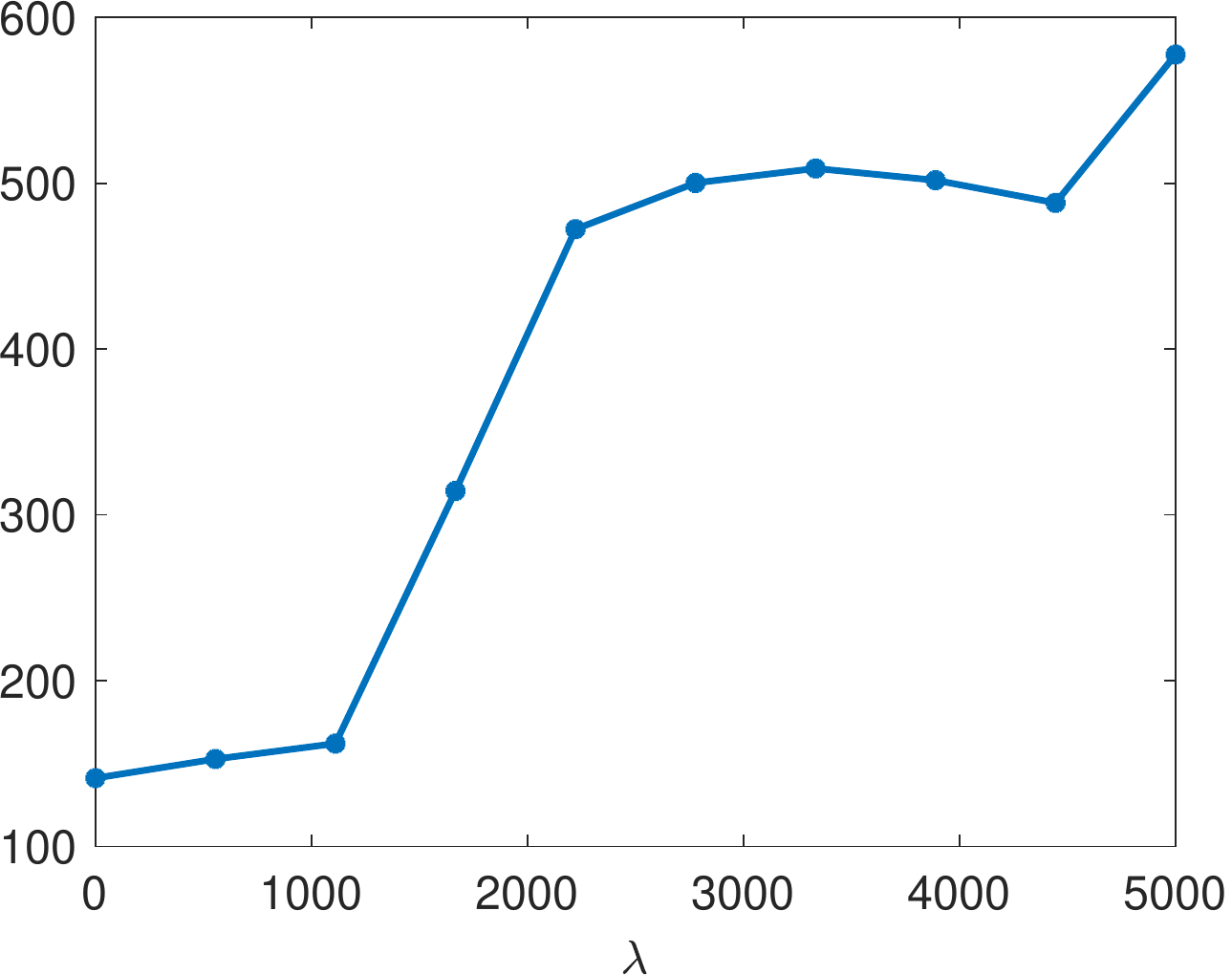}    
}
\subfloat[\tiny Time Warping Functions in Hard Registration]
{
\includegraphics[width=0.33\linewidth]{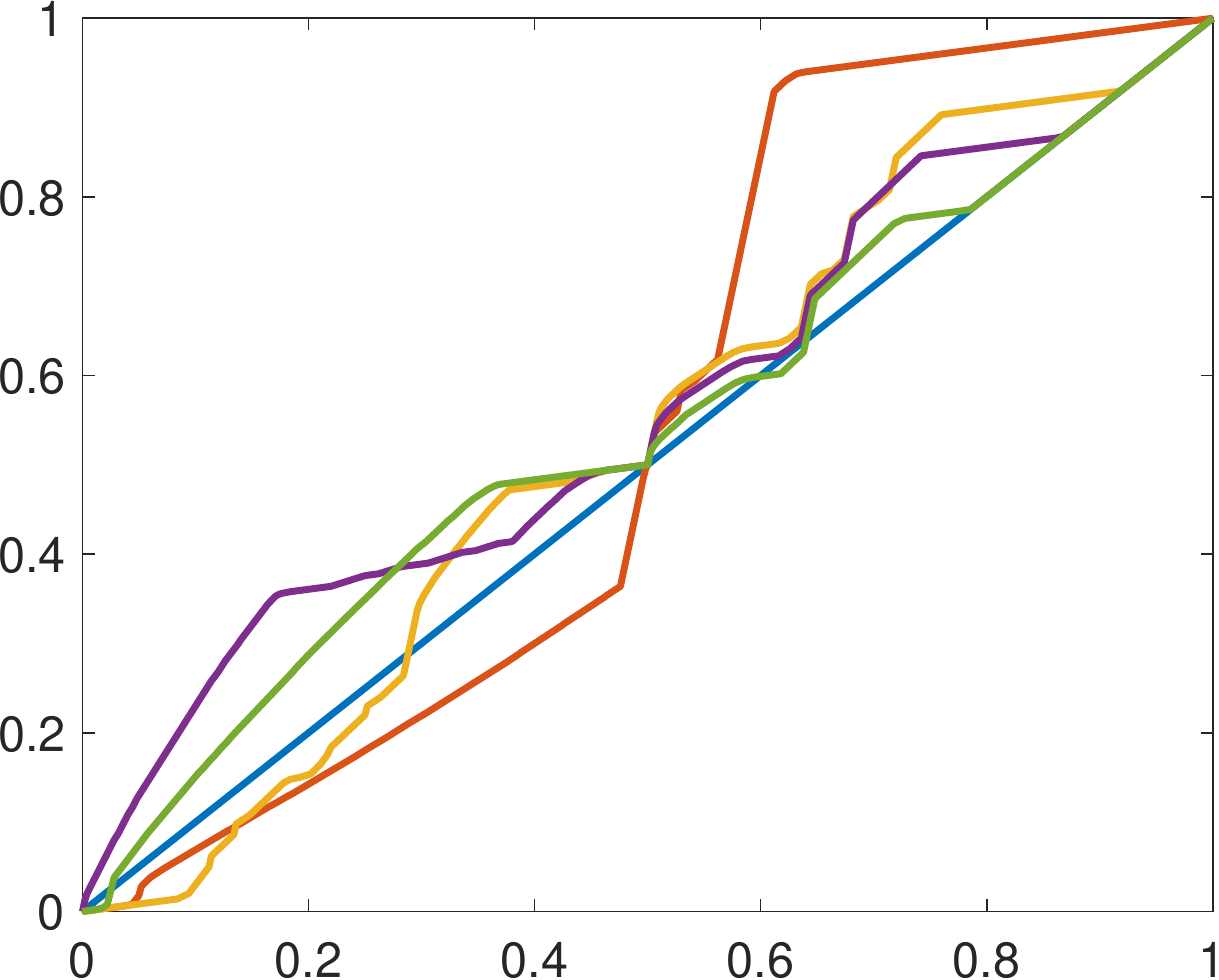}    
}
\subfloat[\tiny Additional Time Warping Functions, $\lambda = 0$]
{
\includegraphics[width=0.33\linewidth]{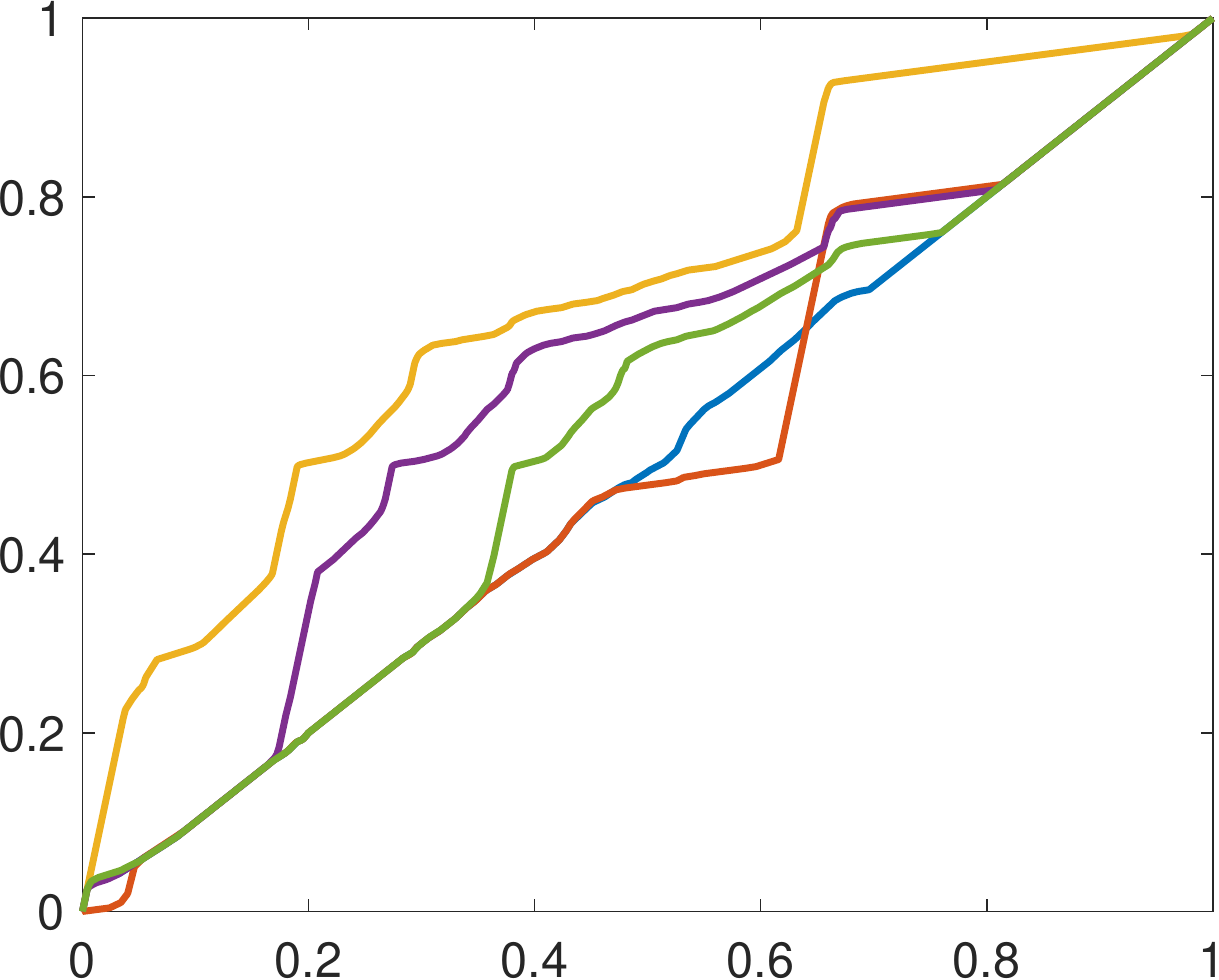}    
}
\caption{Selecting best $\lambda$ based on LOOCV shape error when landmarks are meaningless}
\label{fig:cv_wrong_lm}
\end{figure}

\item {\bf Precise Landmarks}: This time we create some multimodal 
functions and pick the locations of third peak as landmarks in each functions.
In addition, these functions are corrupted by additive white Gaussian noise.
We compute soft alignment experiments, and show the results in Figure~\ref{fig:cv_good_lm}.
In this case, the unconstrained alignment fails to register the correct peaks, 
and LOOCV shape error points to the large $\lambda$ that approximates hard registration.
Incidentally, there are two near-optimal values of $\lambda$ and they both lead to similar alignment results. 

\begin{figure}[H]
\centering
\subfloat[Raw Simulated Functions]
{
\includegraphics[width=0.33\linewidth]{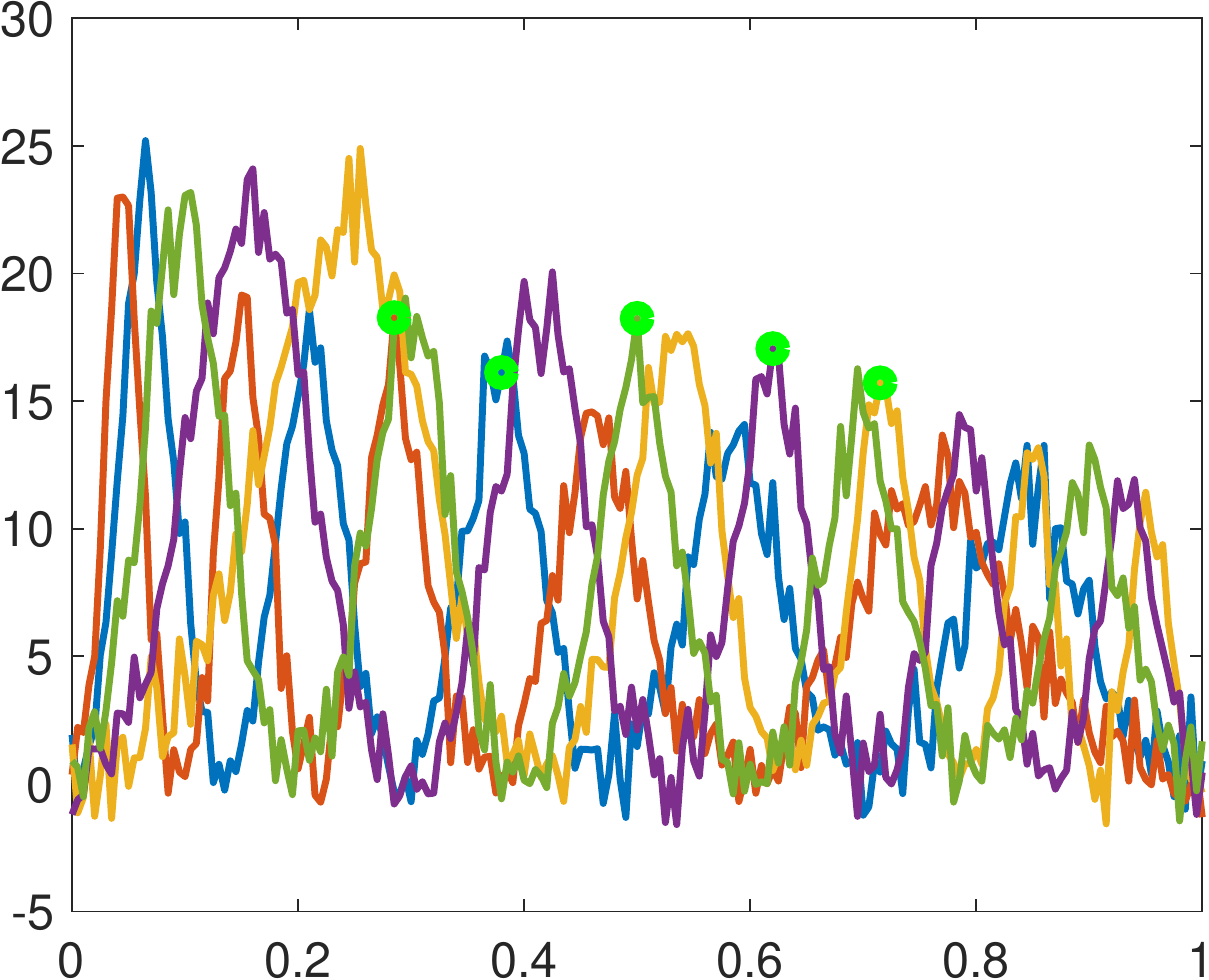}    
}
\subfloat[Unconstrained Alignment]
{
\includegraphics[width=0.33\linewidth]{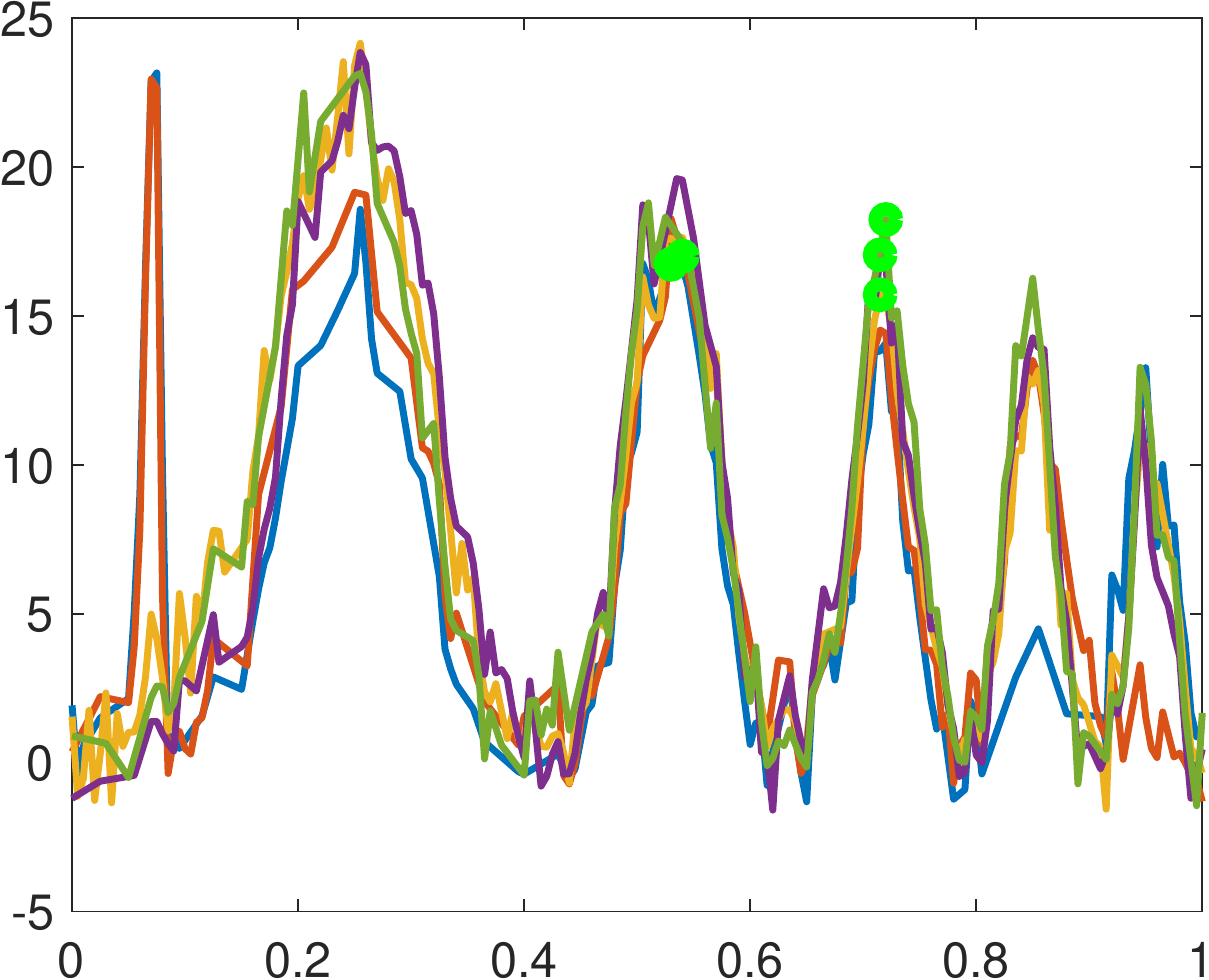}    
}
\subfloat[Soft Alignment, $\lambda = 88.8889$]
{
\includegraphics[width=0.33\linewidth]{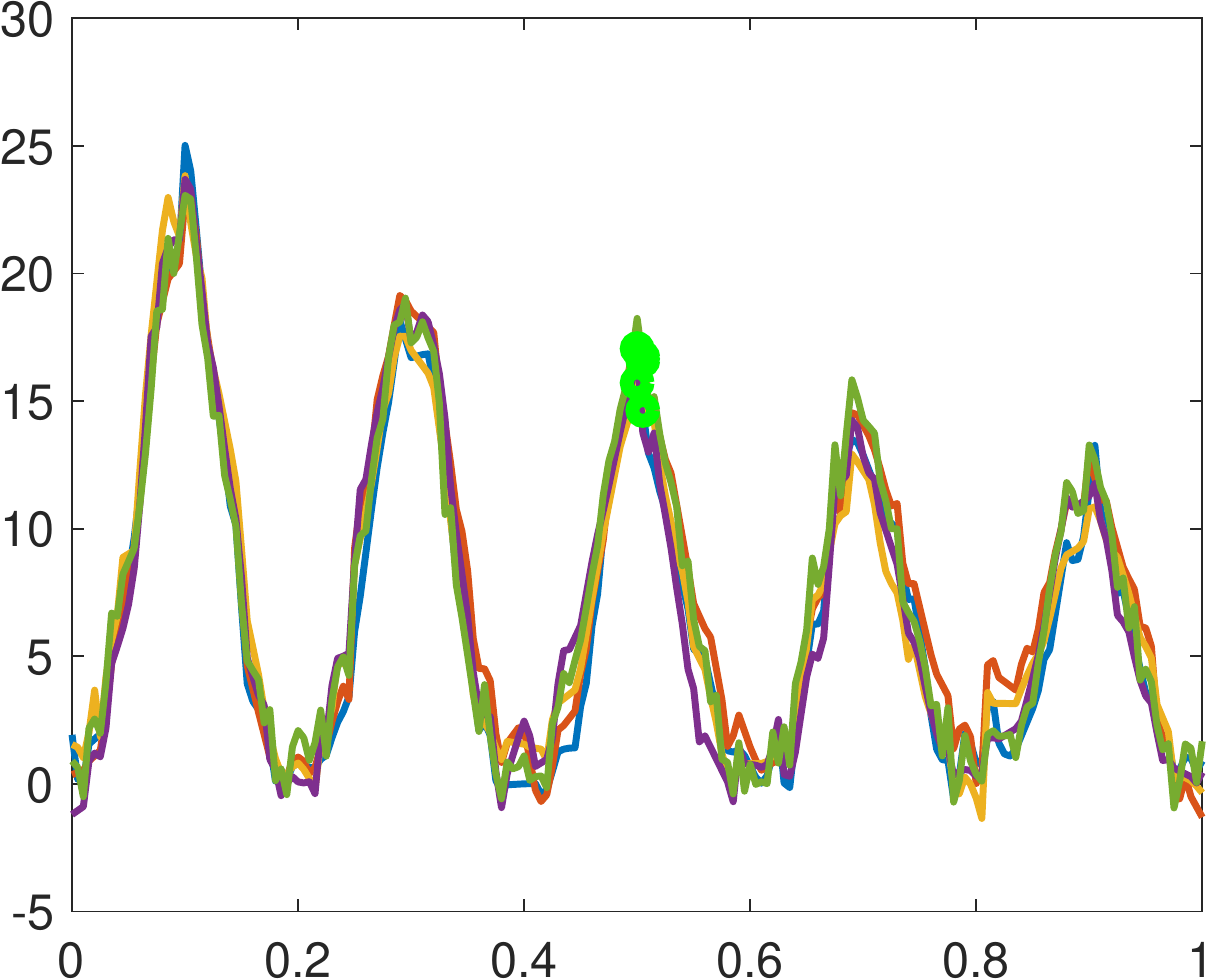}    
}
\\
\subfloat[LOOCV Shape Error]
{
\includegraphics[width=0.33\linewidth]{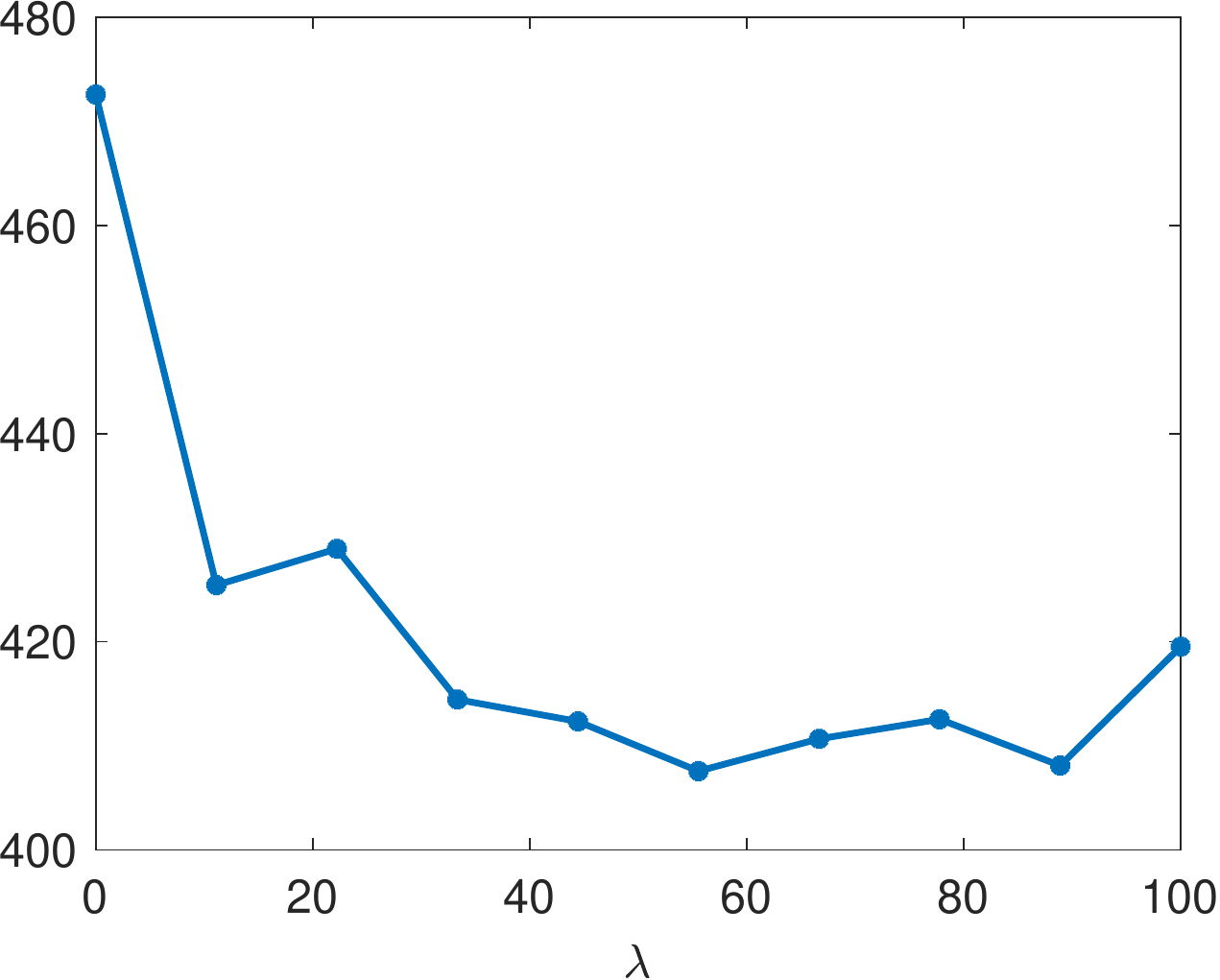}    
}
\subfloat[Hard Registration Warpings]
{
\includegraphics[width=0.33\linewidth]{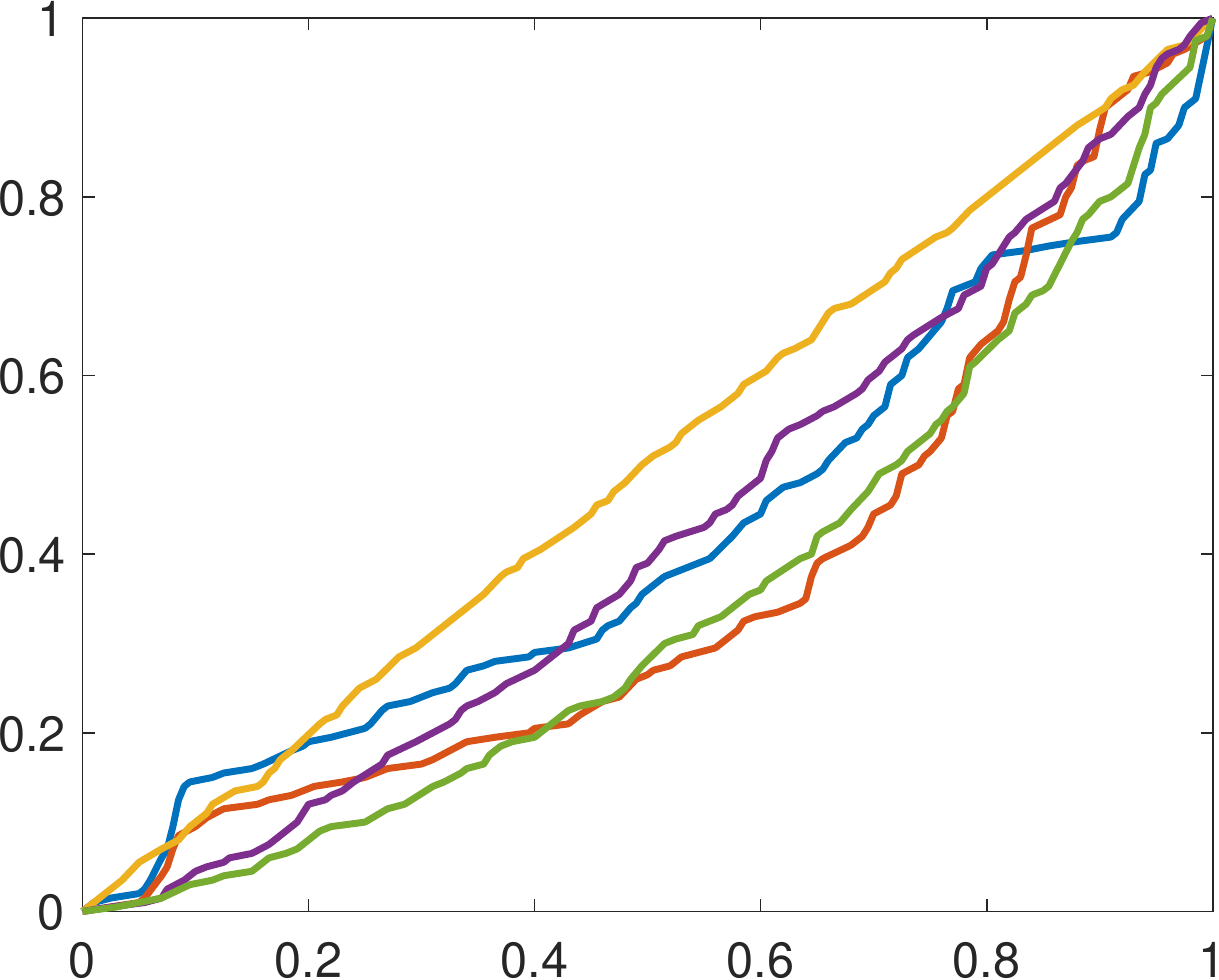}    
}
\subfloat[Additional Warping, $\lambda = 88.8889$]
{
\includegraphics[width=0.33\linewidth]{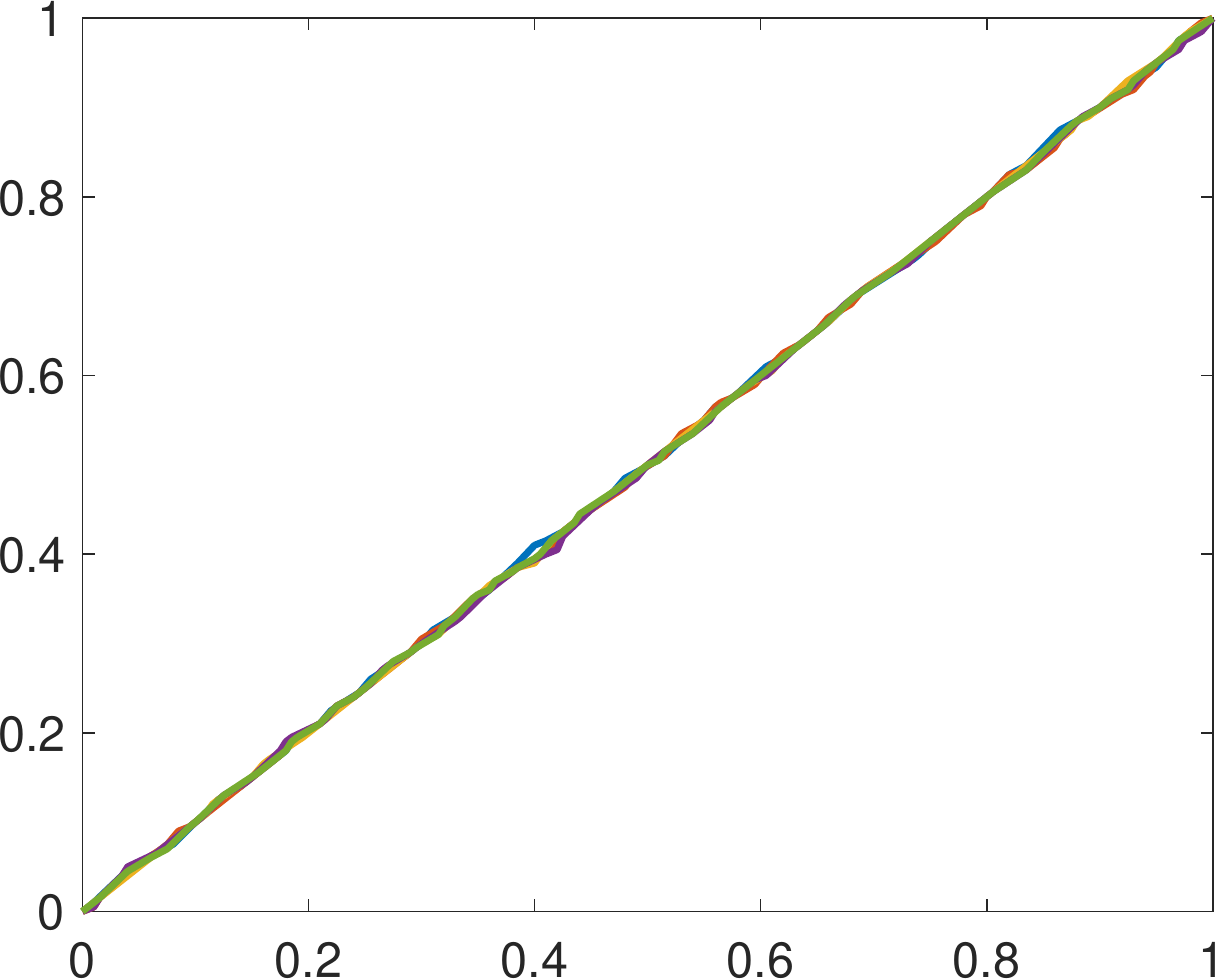}    
}
\caption{Selecting best $\lambda$ based on LOOCV shape error when landmarks are meaningful}
\label{fig:cv_good_lm}
\end{figure}

\end{enumerate}

\section{Experiment and Applications}\label{sec:exp_sfa}
Next, we will present some experimental results involving both simulated as well as the real data to demonstrate the proposed method. 

\subsection{Synthetic Data}
\begin{figure}[H]
\begin{center}
\begin{tabular}{ccc}
\includegraphics[width=0.3\linewidth]{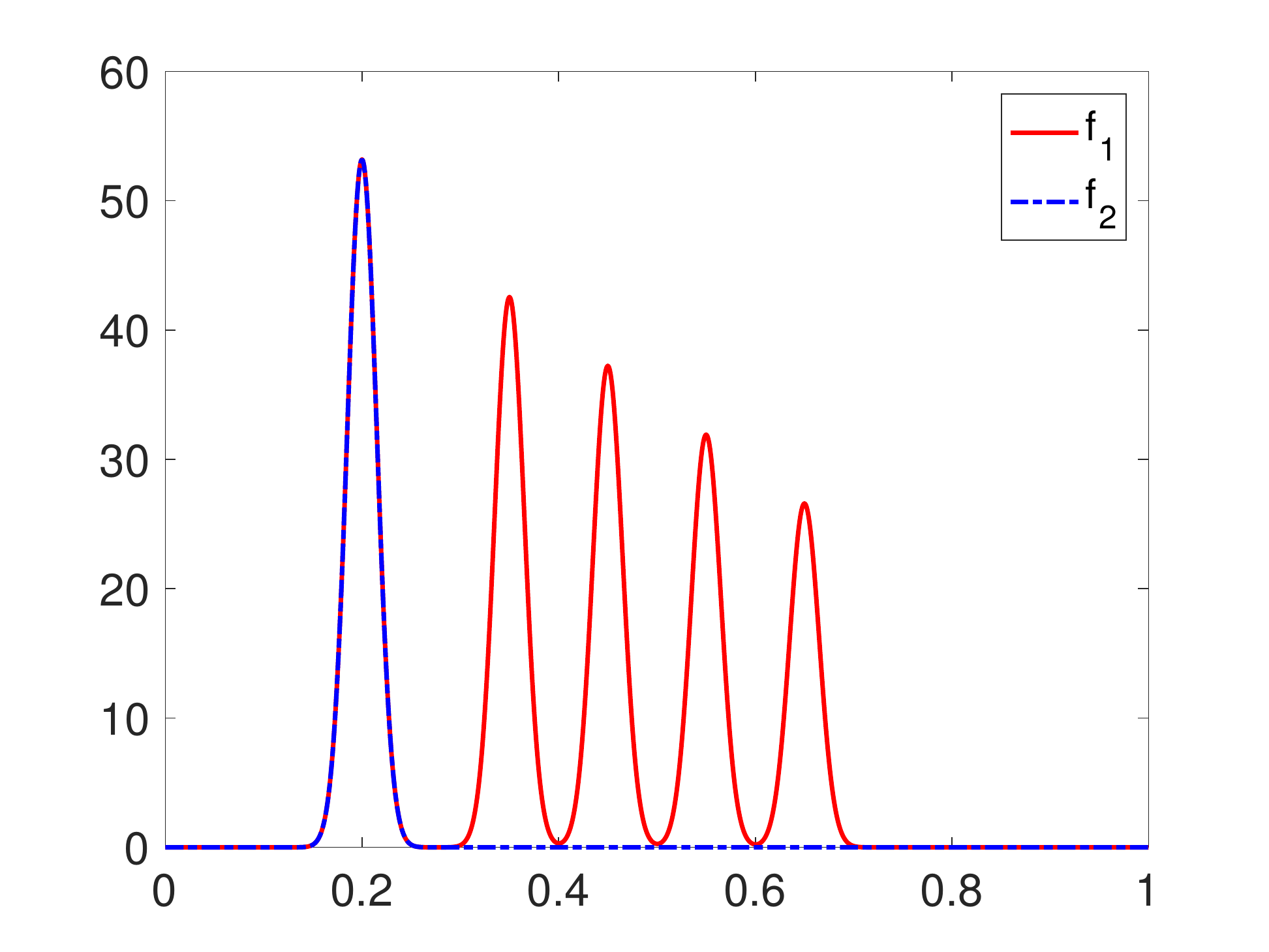}
\includegraphics[width=0.3\linewidth]{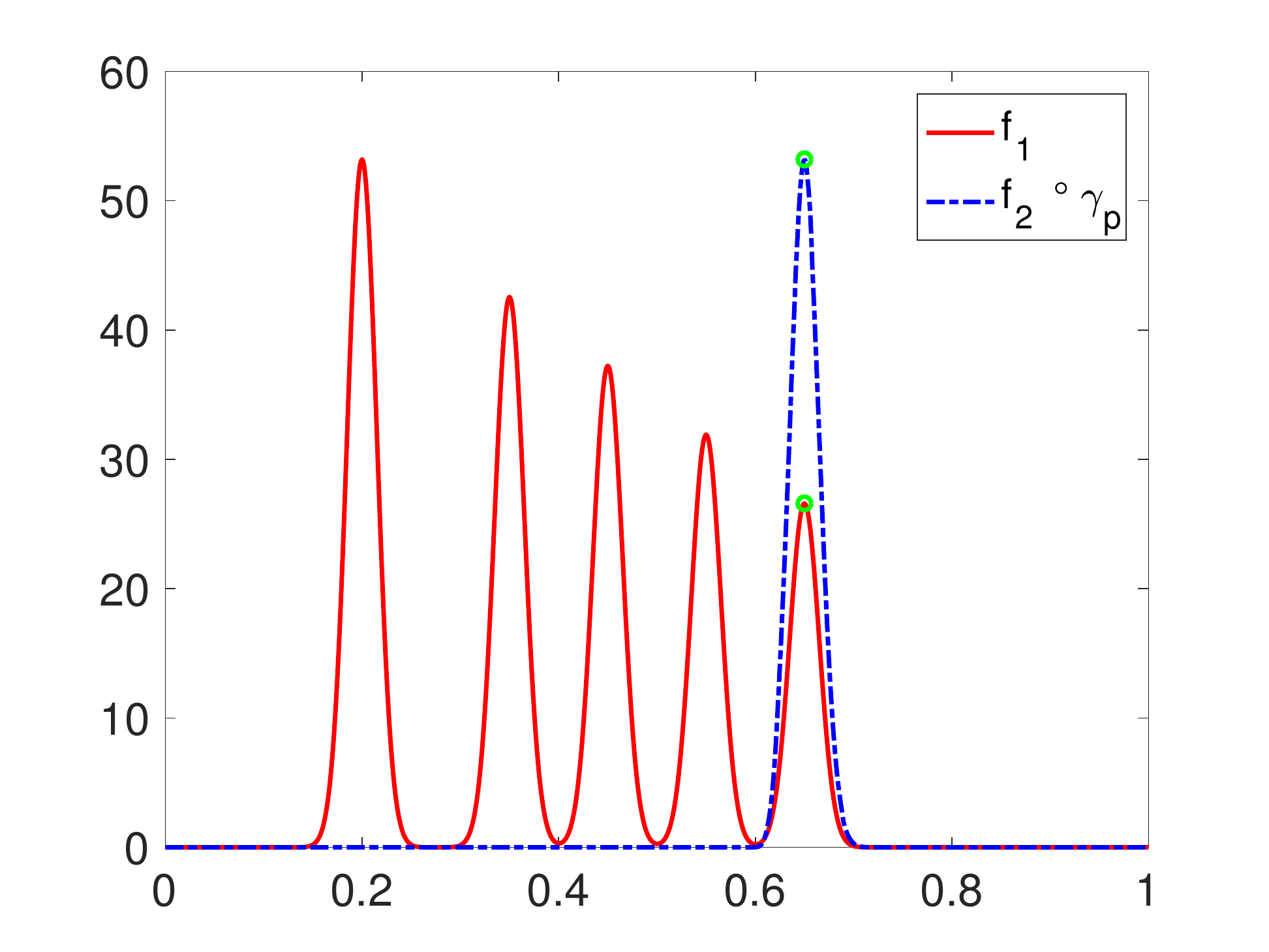}
\includegraphics[width=0.3\linewidth]{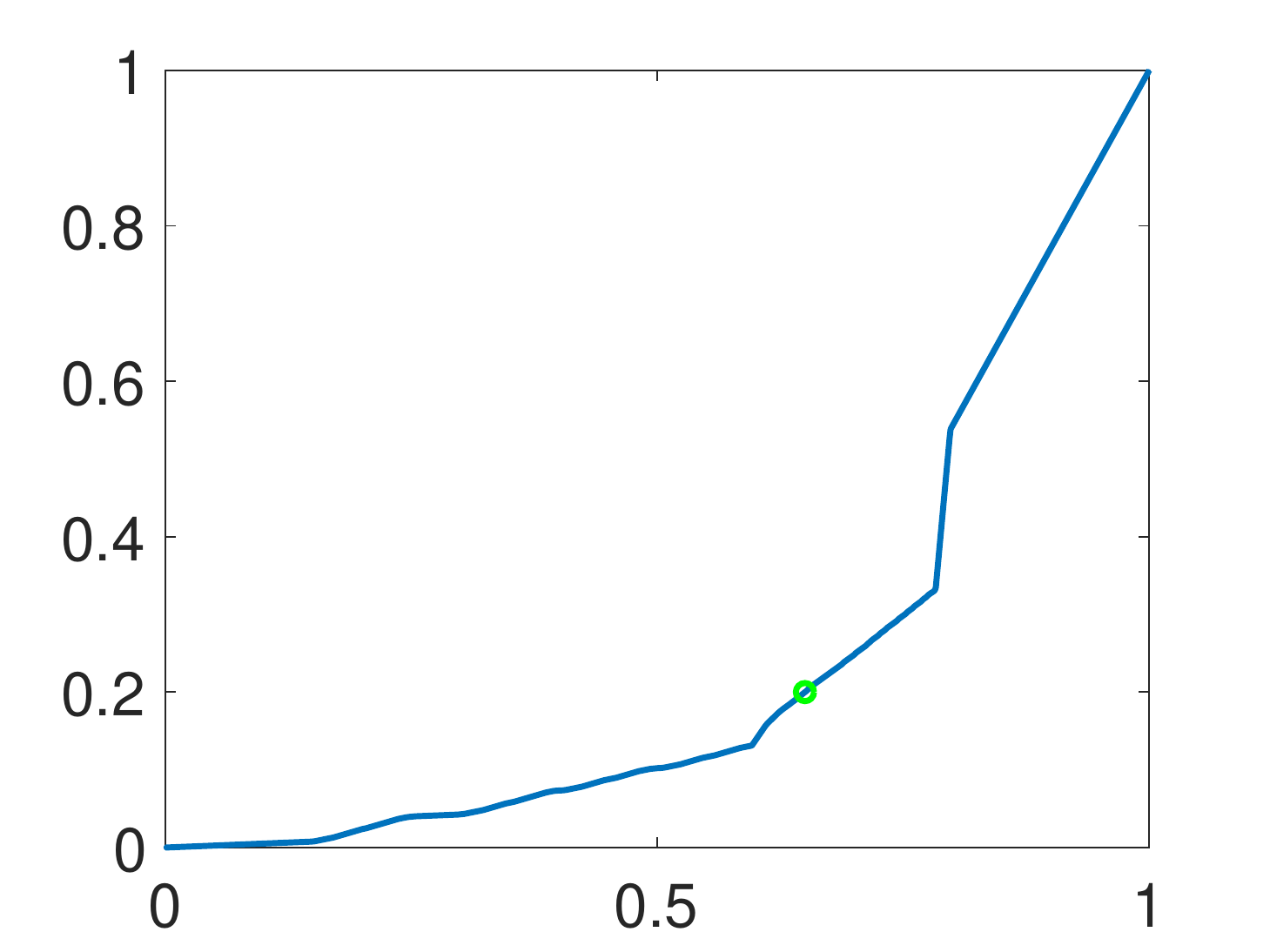}
\end{tabular}

\begin{tabular}{ccccc}
   \includegraphics[width=0.2\linewidth]{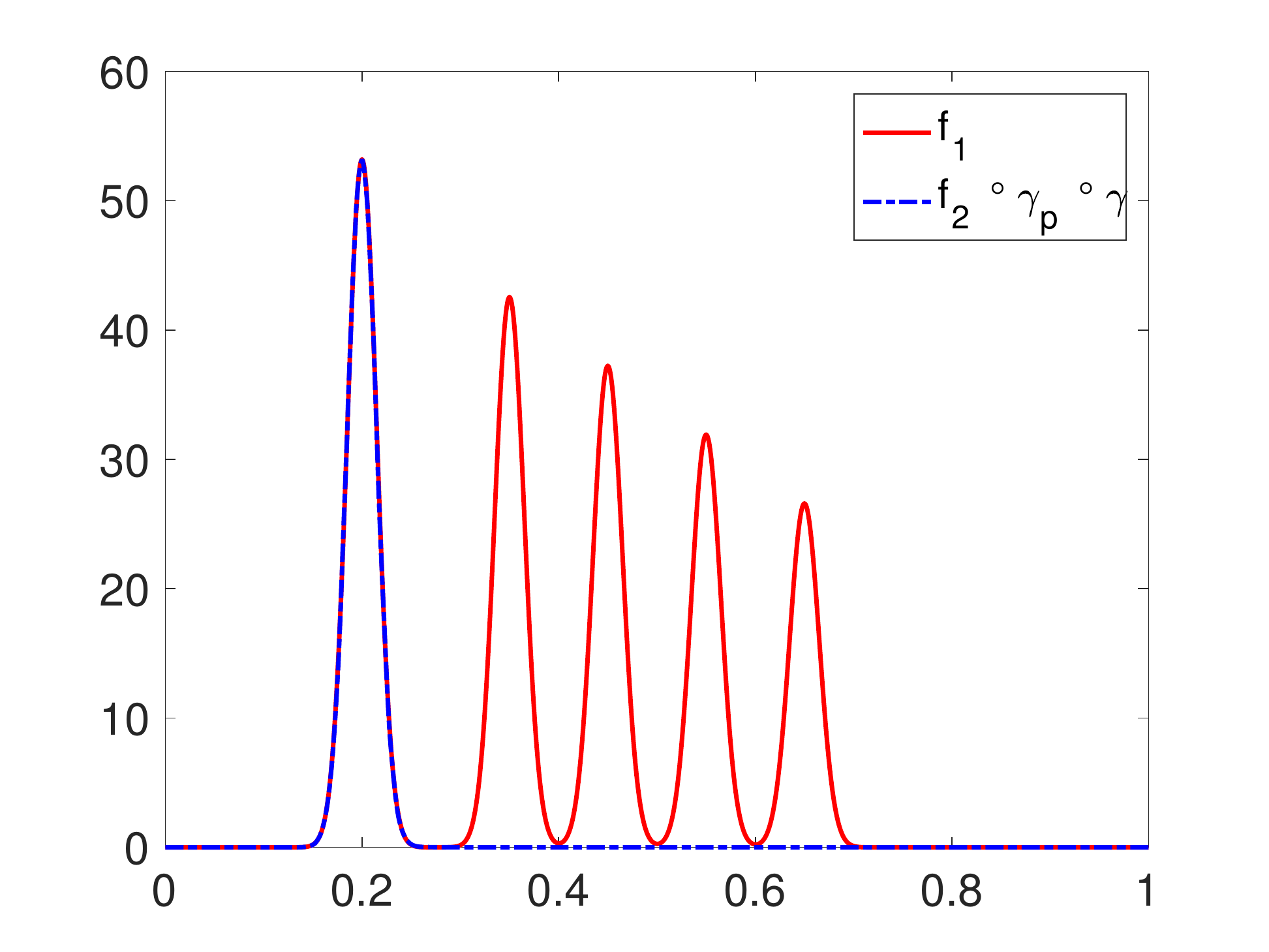}&
   \includegraphics[width=0.2\linewidth]{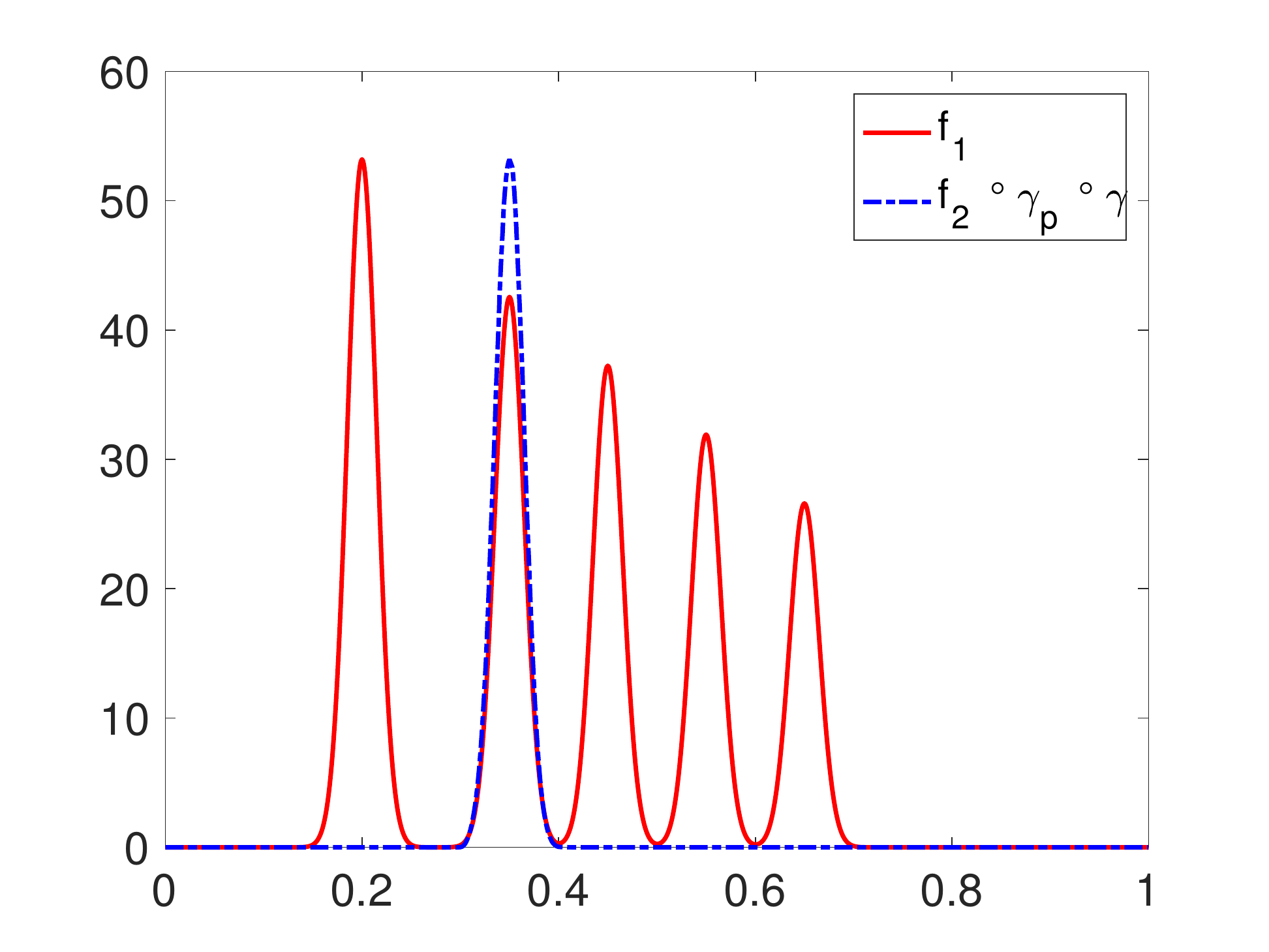}&
   \includegraphics[width=0.2\linewidth]{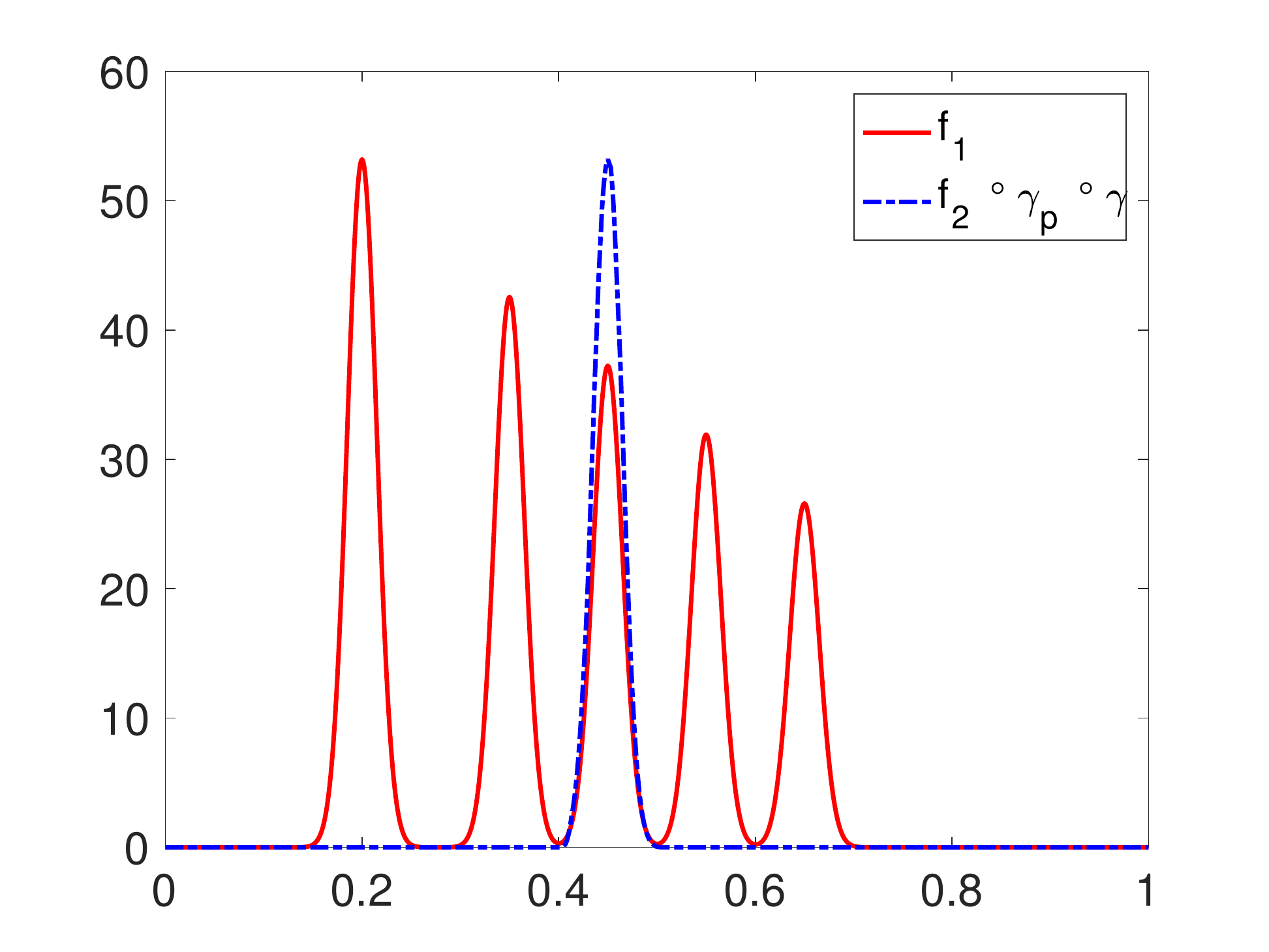}&
   \includegraphics[width=0.2\linewidth]{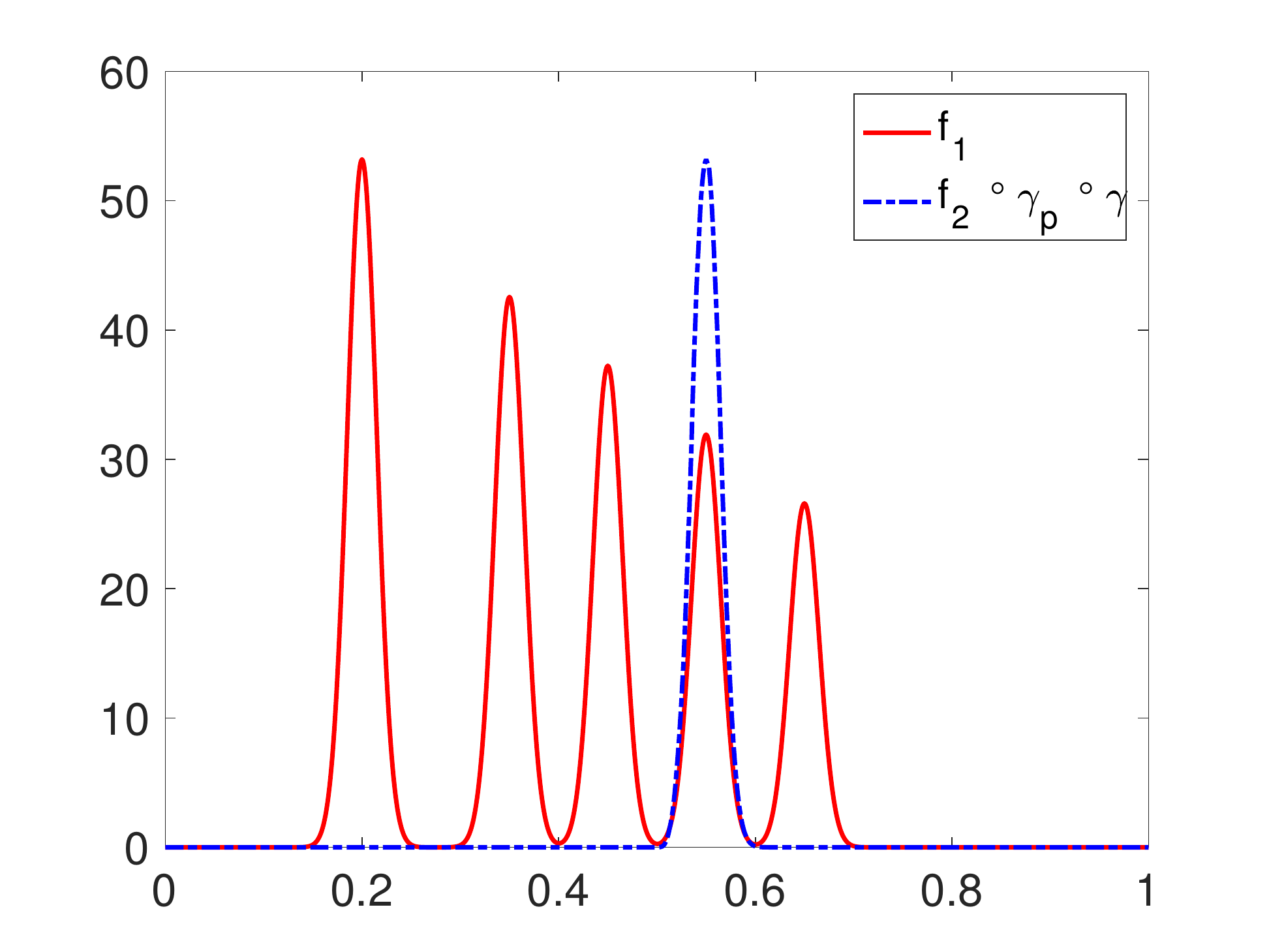}&
   \includegraphics[width=0.2\linewidth]{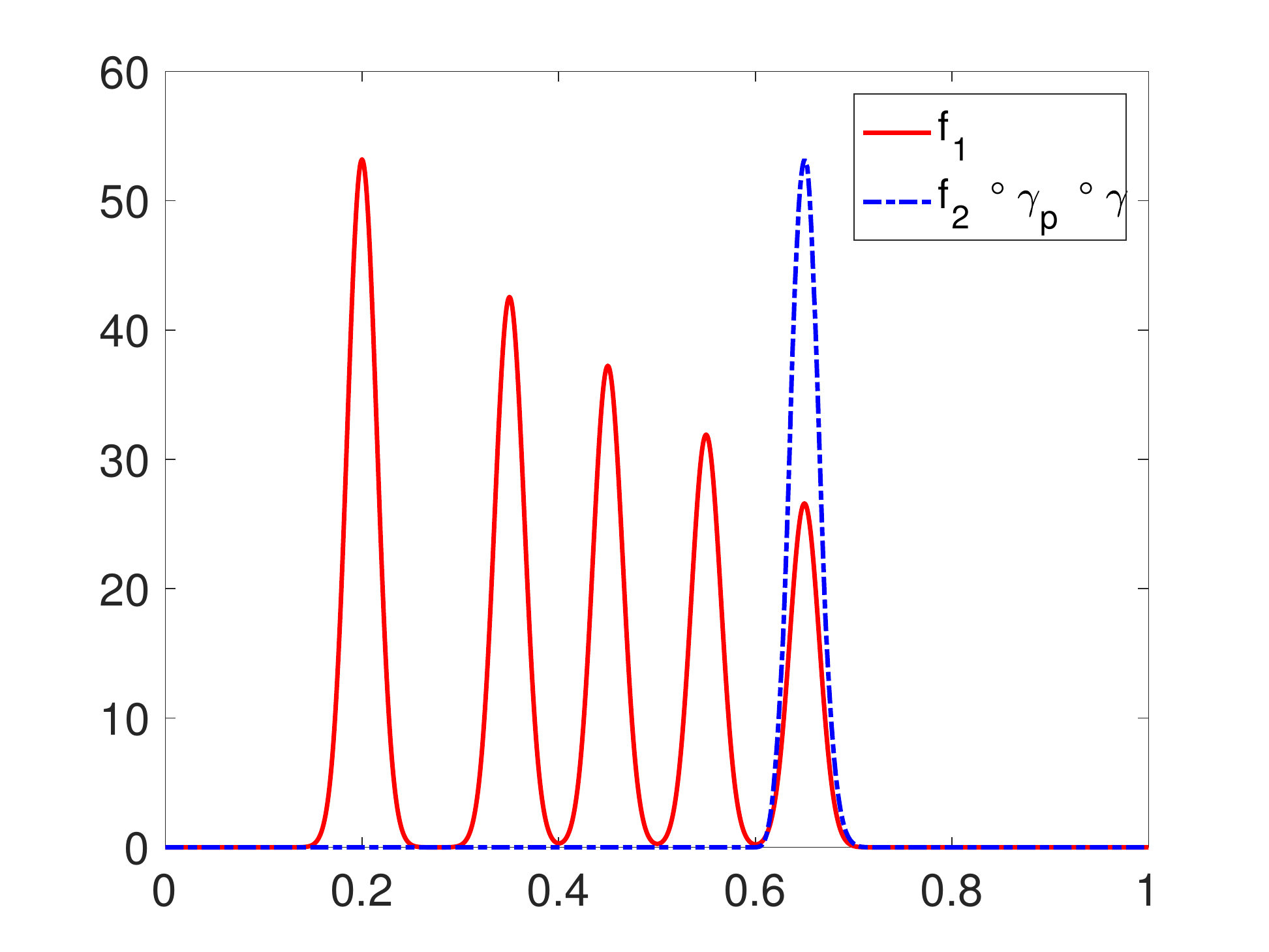}\\

   \includegraphics[width=0.2\linewidth]{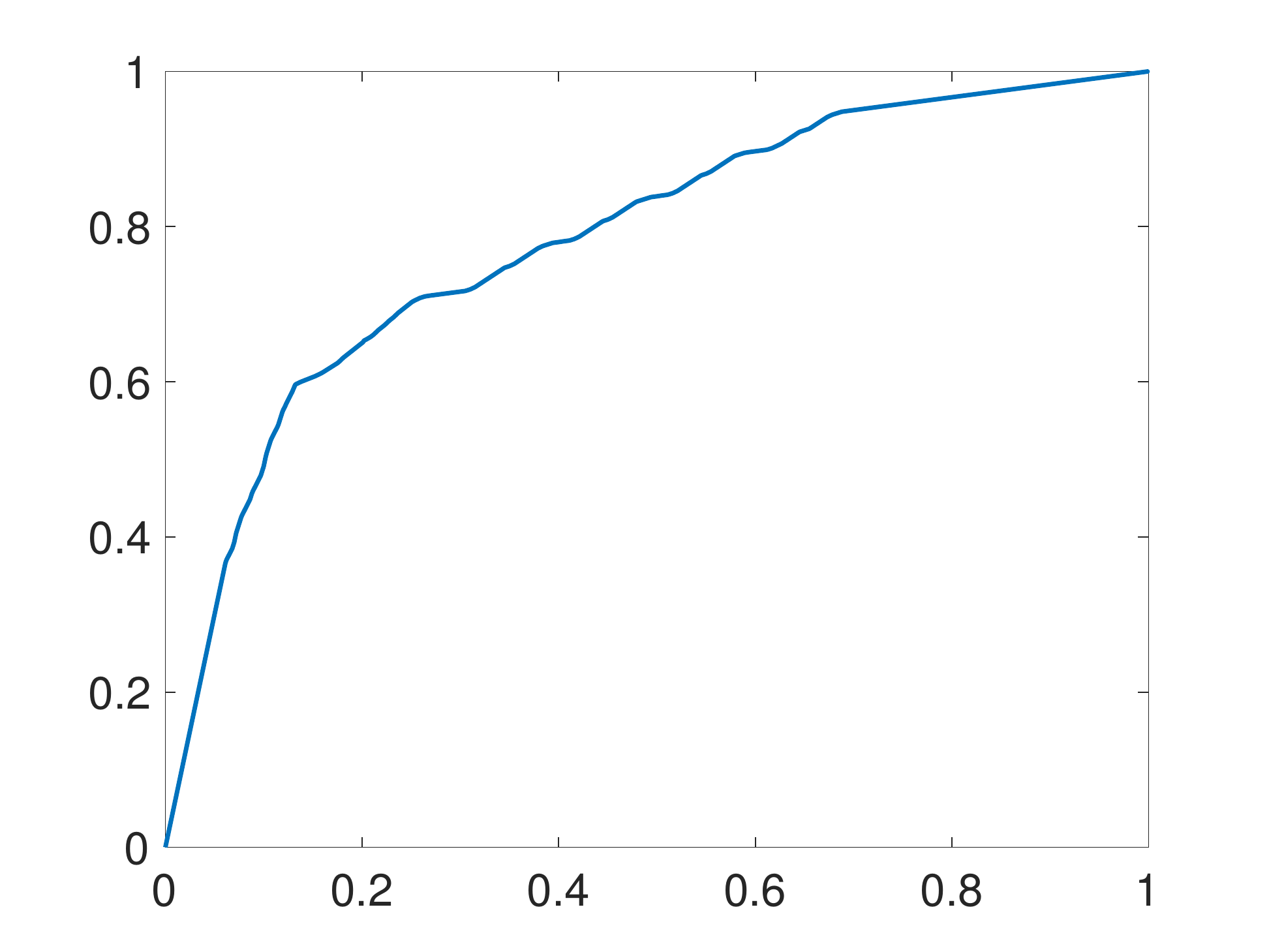}&
   \includegraphics[width=0.2\linewidth]{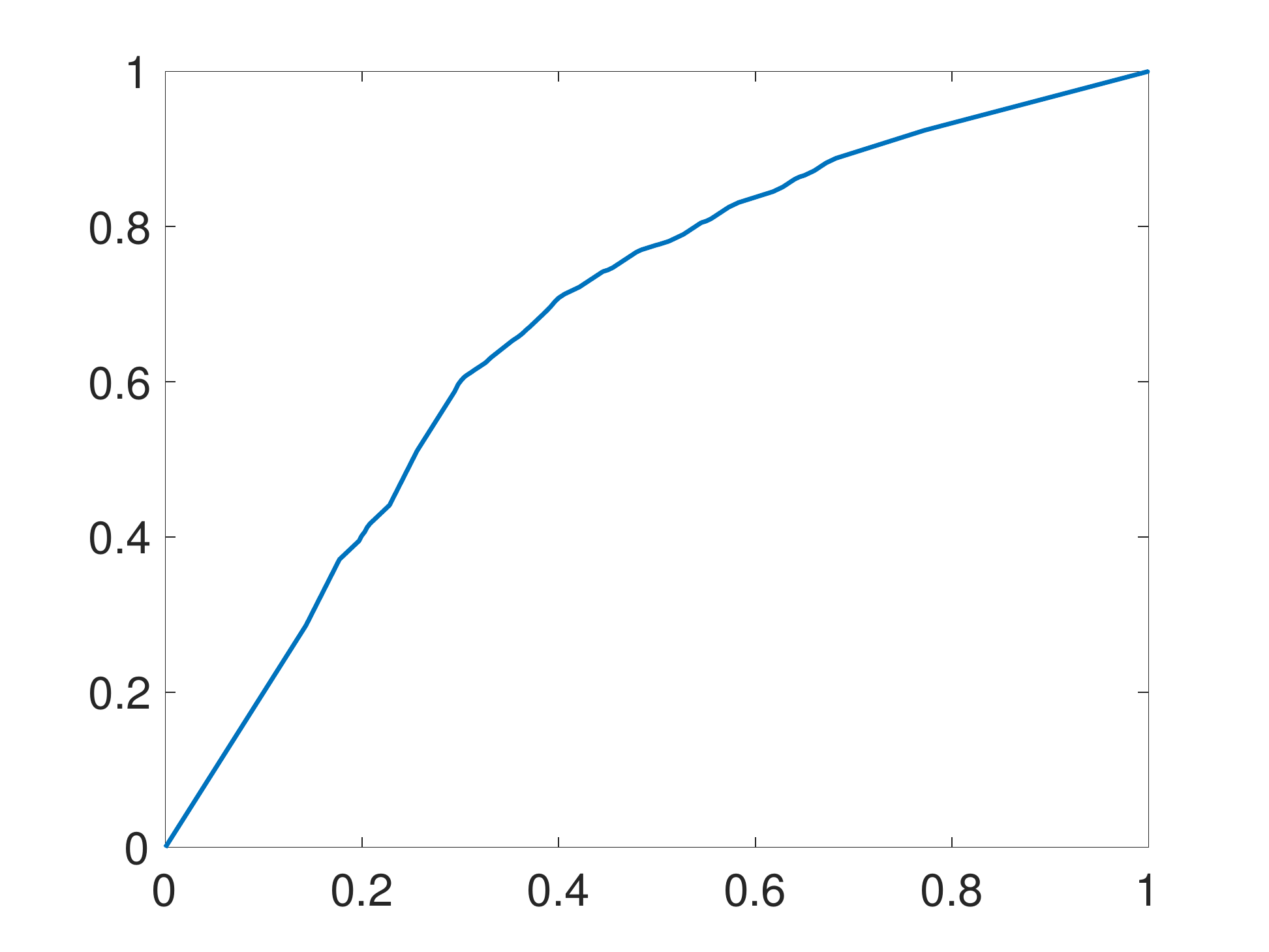}&
   \includegraphics[width=0.2\linewidth]{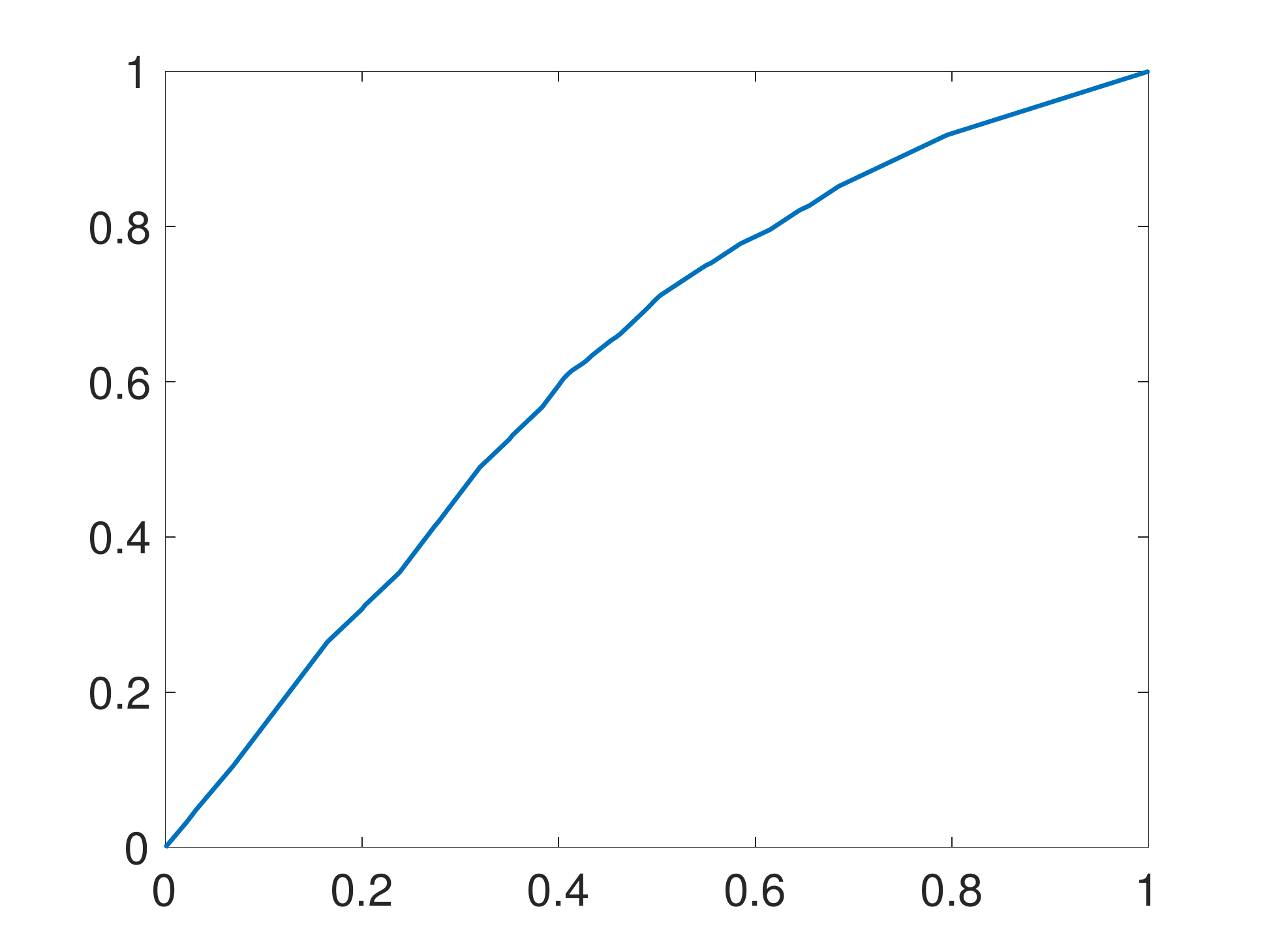}&
   \includegraphics[width=0.2\linewidth]{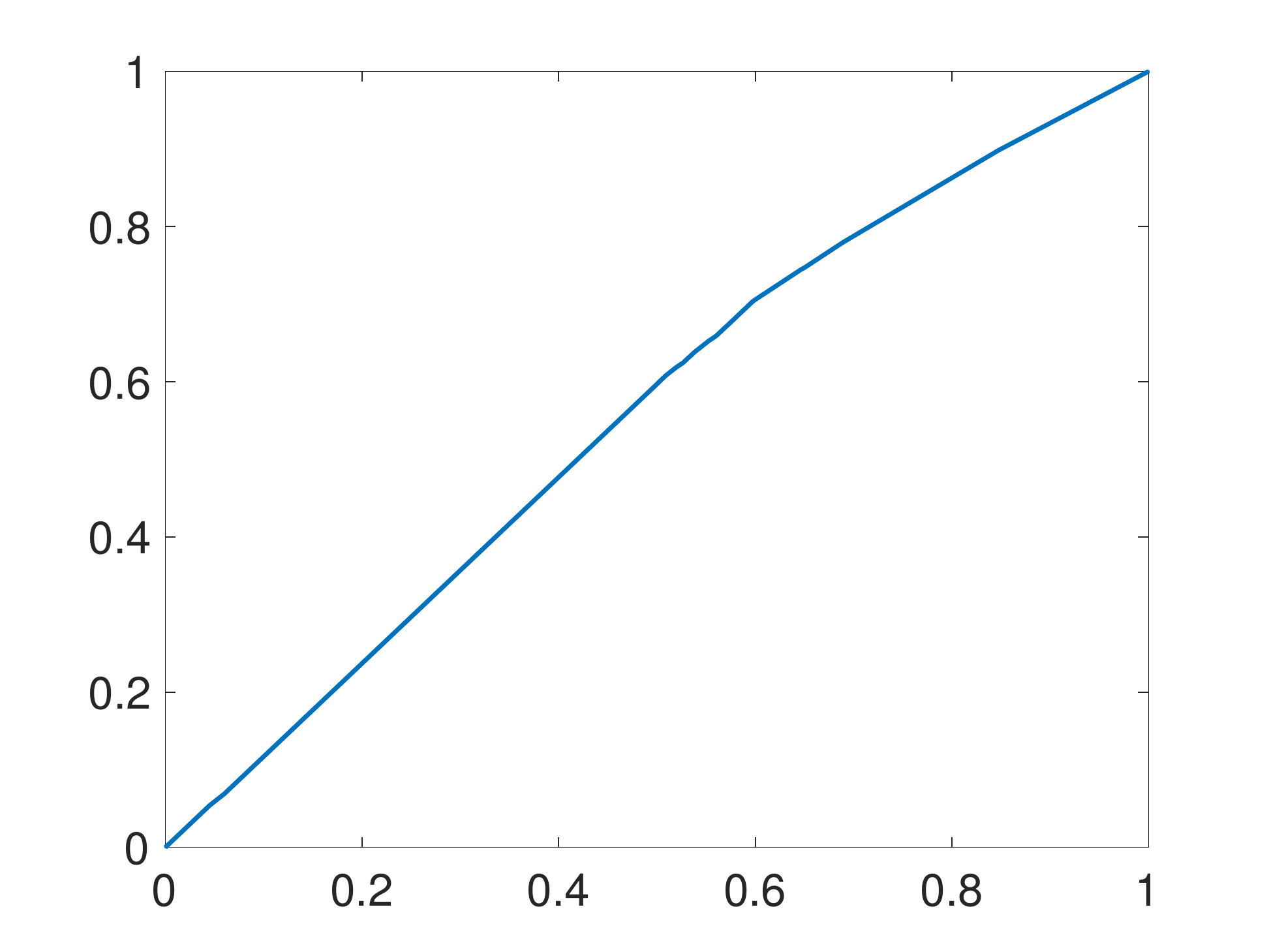}&
   \includegraphics[width=0.2\linewidth]{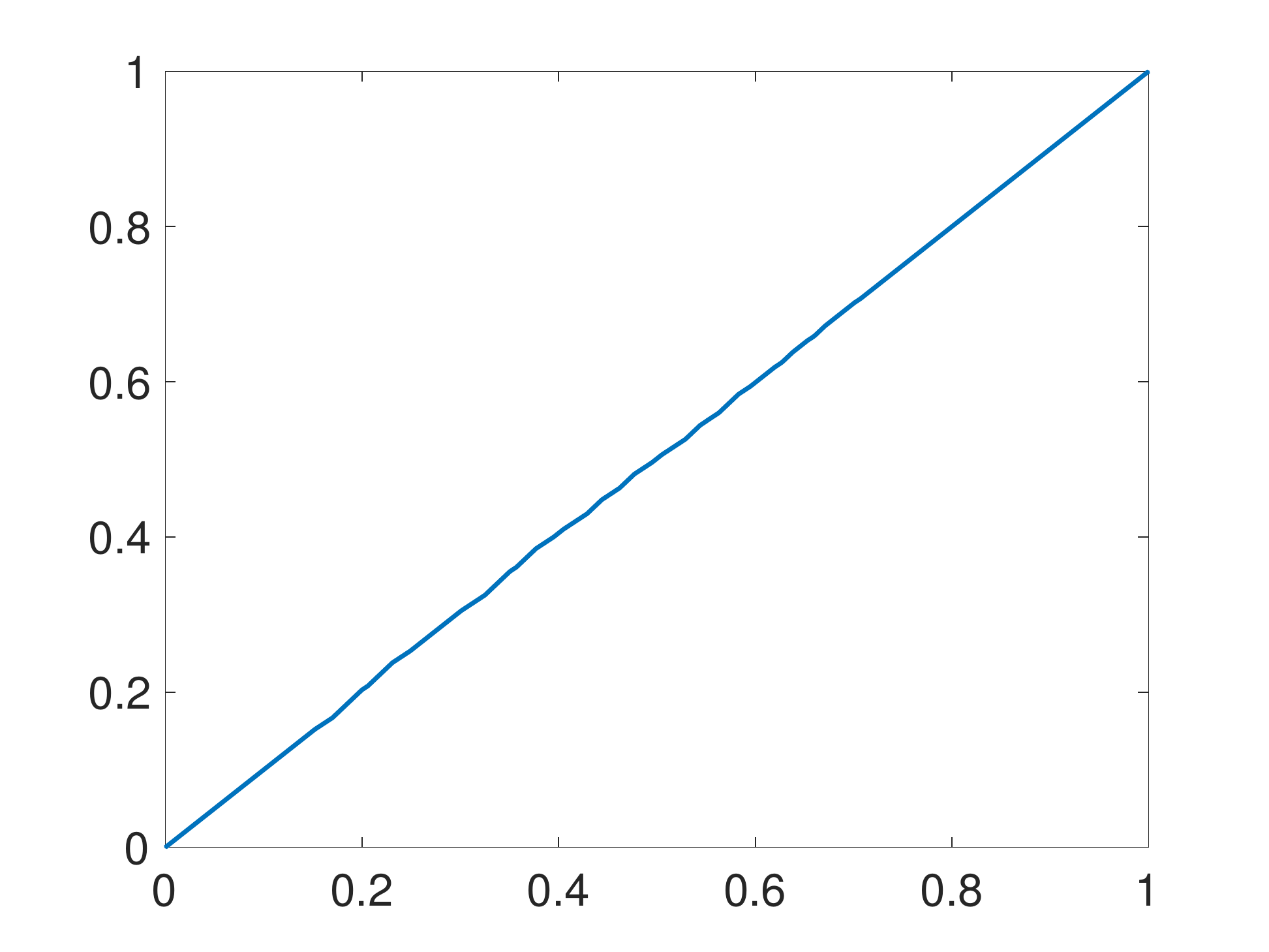}\\
\end{tabular}
\end{center}
   \caption{Soft alignment. Top row: left panel shows the original functions, middle shows hard registration and right
   shows corresponding time warping. After further alignment (step 2), middle row shows the aligned functions while bottom row shows warping functions. From left to right, $\lambda$=0, 120, 160, 500, 1500, respectively.}
\label{fig:simusoft}
\end{figure}

The top left panel of Figure \ref{fig:simusoft} shows two simulated functions, $f_1$ and $f_2$
formed by superimposing some Gaussian density functions. 
The function $f_1$ has five peaks while 
the $f_2$ has one peak. 
We select the last peak of $f_1$ and the only peak of $f_2$ as 
a single landmark for each function. We first pre-align two sets of landmarks using hard registration, as shown in 
the top row. 
Next, 
for different values of the tuning parameter $\lambda$, we obtain different soft alignments
as shown in the middle. 
From left to right, the value of $\lambda$ is increasing, causing a steady increase in the 
influence of the landmarks. 
As we can see, the landmark of $f_2$ goes through every peak of $f_1$ for different $\lambda$.

\begin{figure}[H]
\centering
    \subfloat[\scriptsize Original Functions]{
    \includegraphics[width=0.33\linewidth]{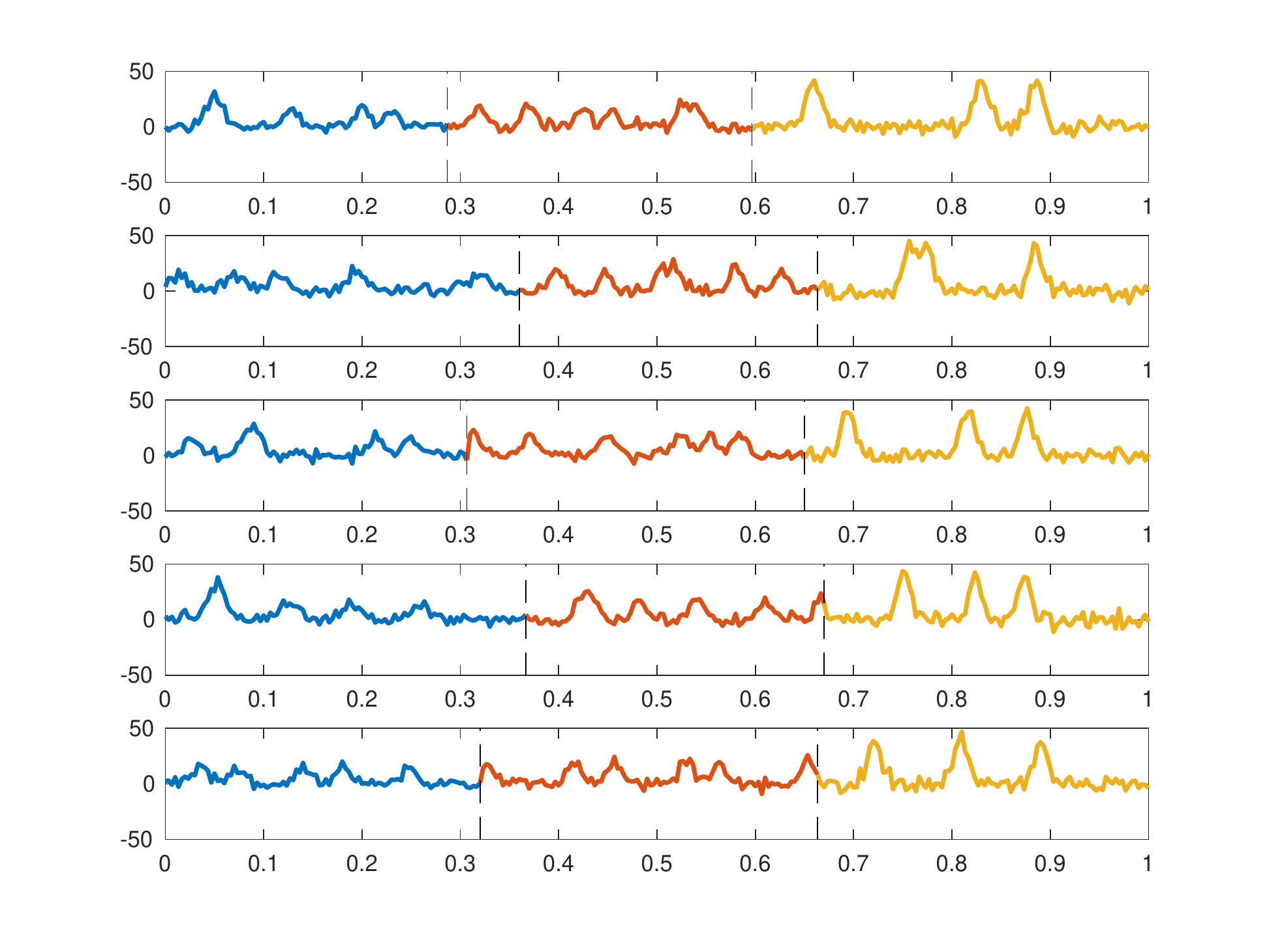}}
    \subfloat[\scriptsize Hard Registration of Functions]{
    \includegraphics[width=0.33\linewidth]{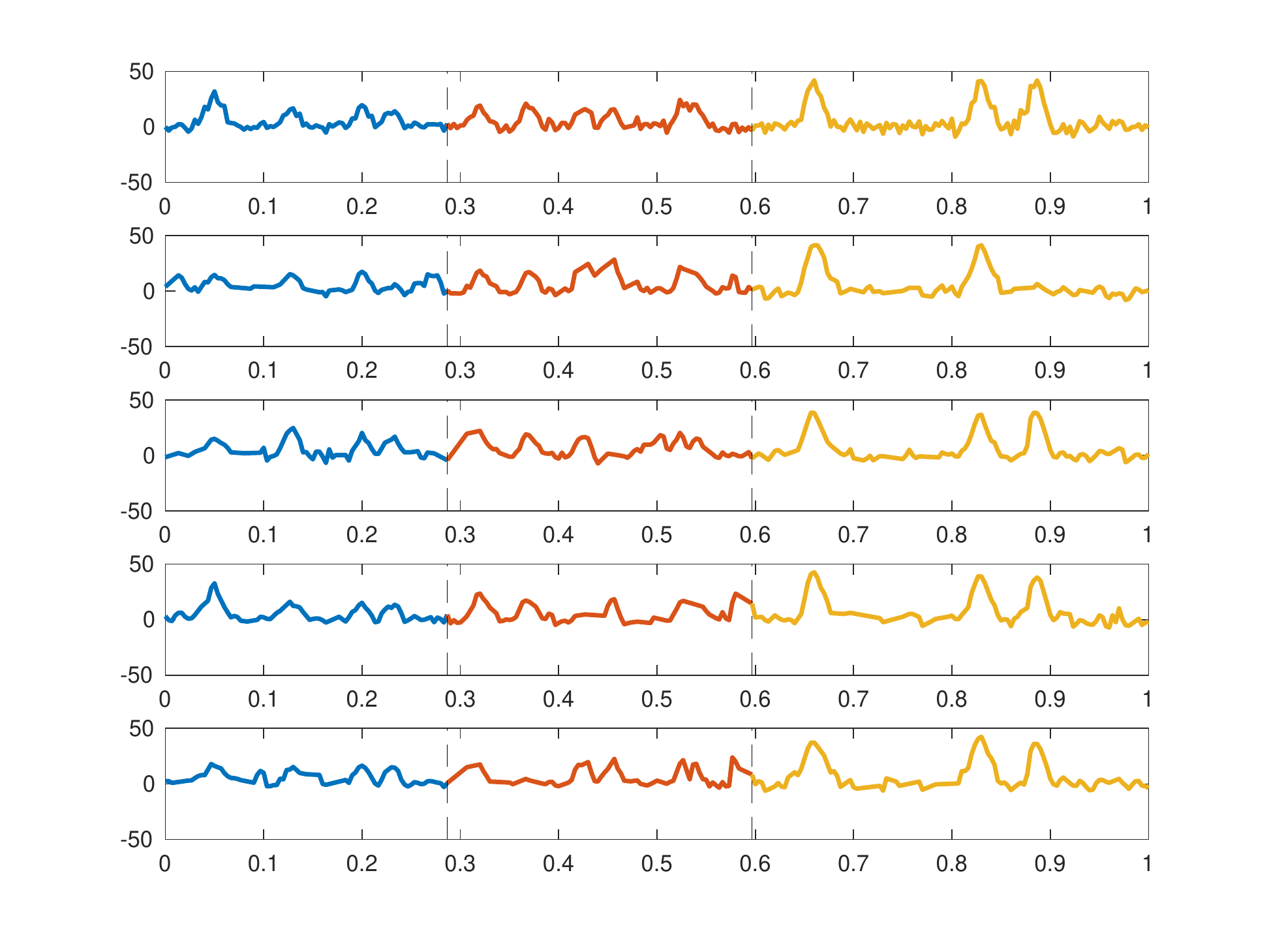}}
    \subfloat[\scriptsize Standard Deviation of Landmarks]{
    \includegraphics[width=0.33\linewidth]{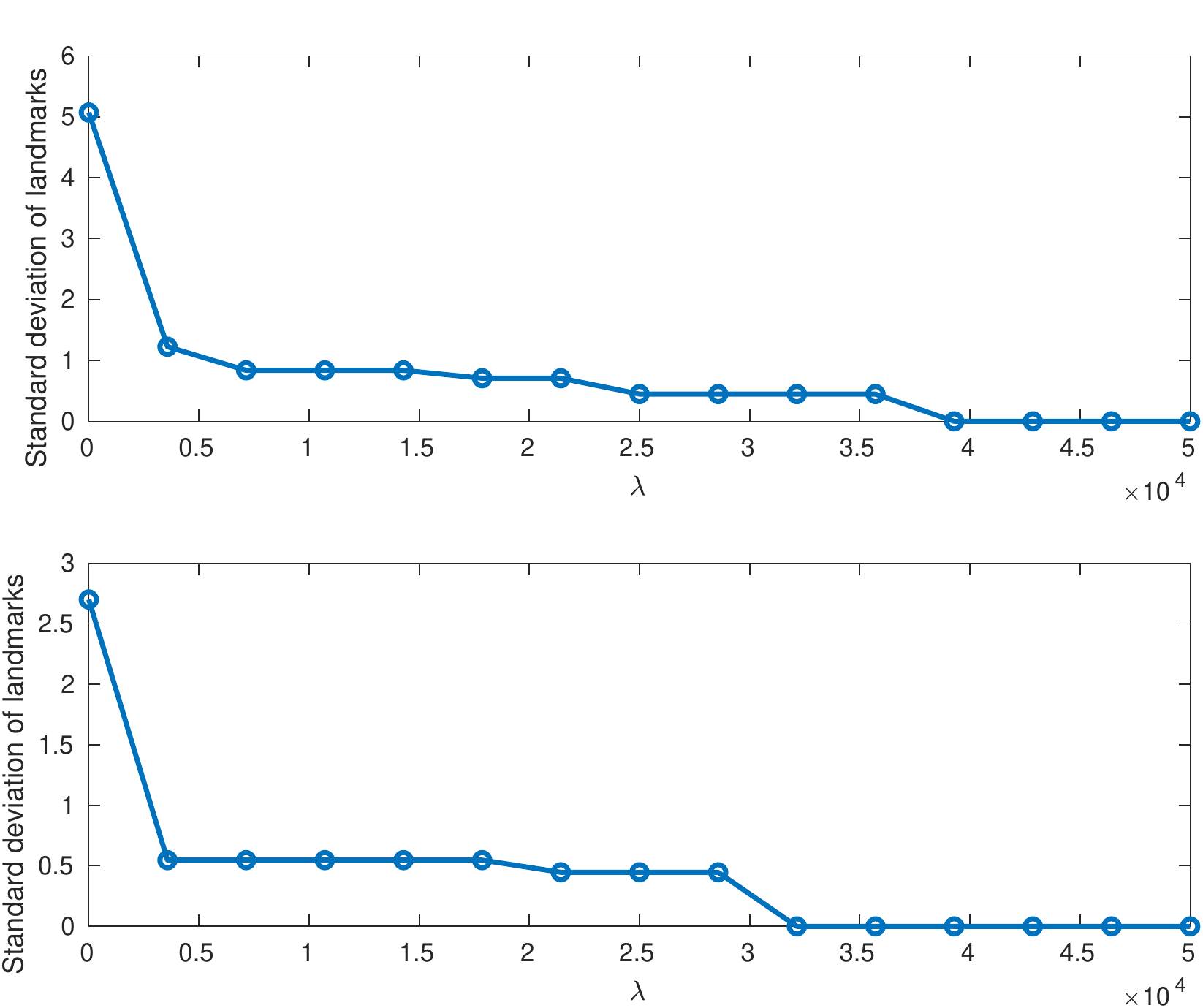}}
   \caption{Multiple soft alignment. Left shows the original functions with 2 landmarks each, 
shown using vertical dashed line. Middle shows the functions aligned using the hard registration. Right plots 
the standard deviation of two landmarks as $\lambda$ varies: top plot is for the first landmarks and the bottom for the second landmark.}
\label{fig:multiop}
\end{figure}

In the second simulated scenario, we focus on 
the soft nature of the solution for the multiple alignment. The left panel of Figure \ref{fig:multiop} shows five simulated functions, each 
with two landmarks
each marked by vertical dashed lines. The corresponding segments across functions are colored in the same way: blue, red and yellow. The pre-alignment is given by hard registration, which can be seen in the middle panel.  
As we increase $\lambda$, the resulting standard deviation of two sets of landmarks is plotted in the right panel. It goes from a large value to 0, implying the alignment is changed from unconstrained alignment to hard registration. 
The intermediate values represent soft alignment solutions.

\subsection{Real Data}
Now we show three examples of soft alignment using real data. {We also show results from a hard landmark alignment method~\citep{ramsay2021fda} for comparison.}\\

\begin{figure}[H]
\centering
\begin{tabular}{cccc}
\includegraphics[width=0.25\textwidth]{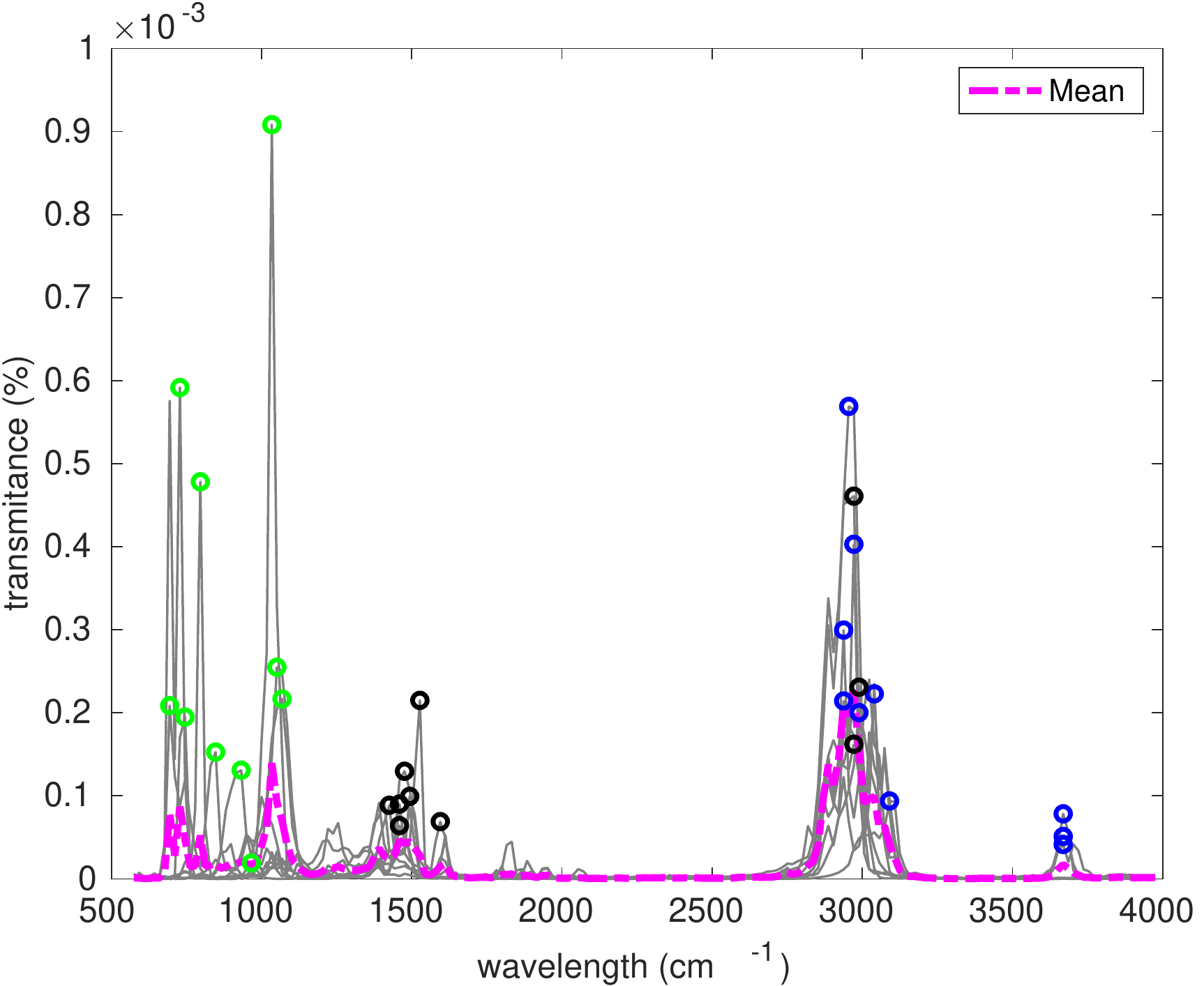}
& \includegraphics[width=0.25\textwidth]{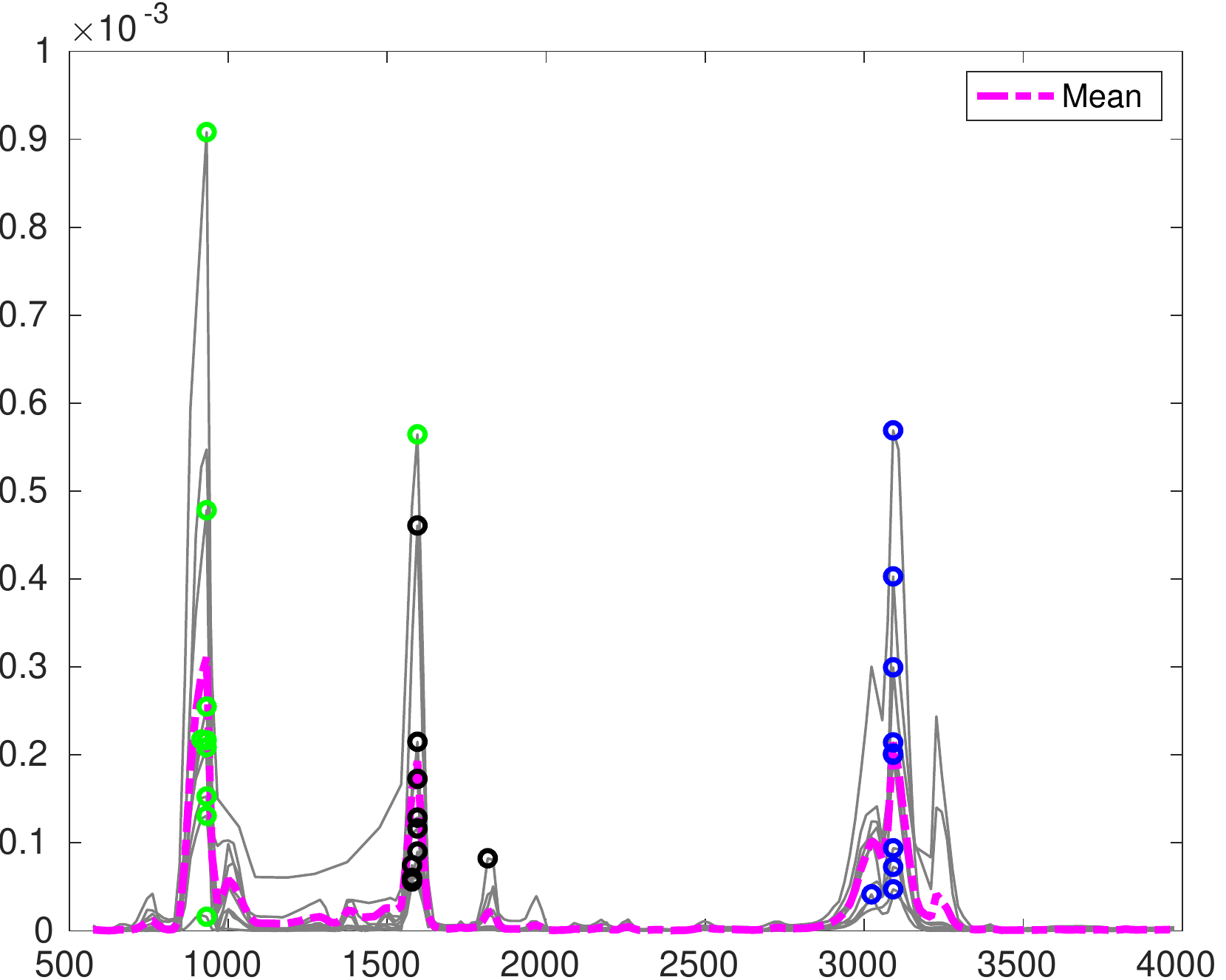}
& \includegraphics[width=0.25\textwidth]{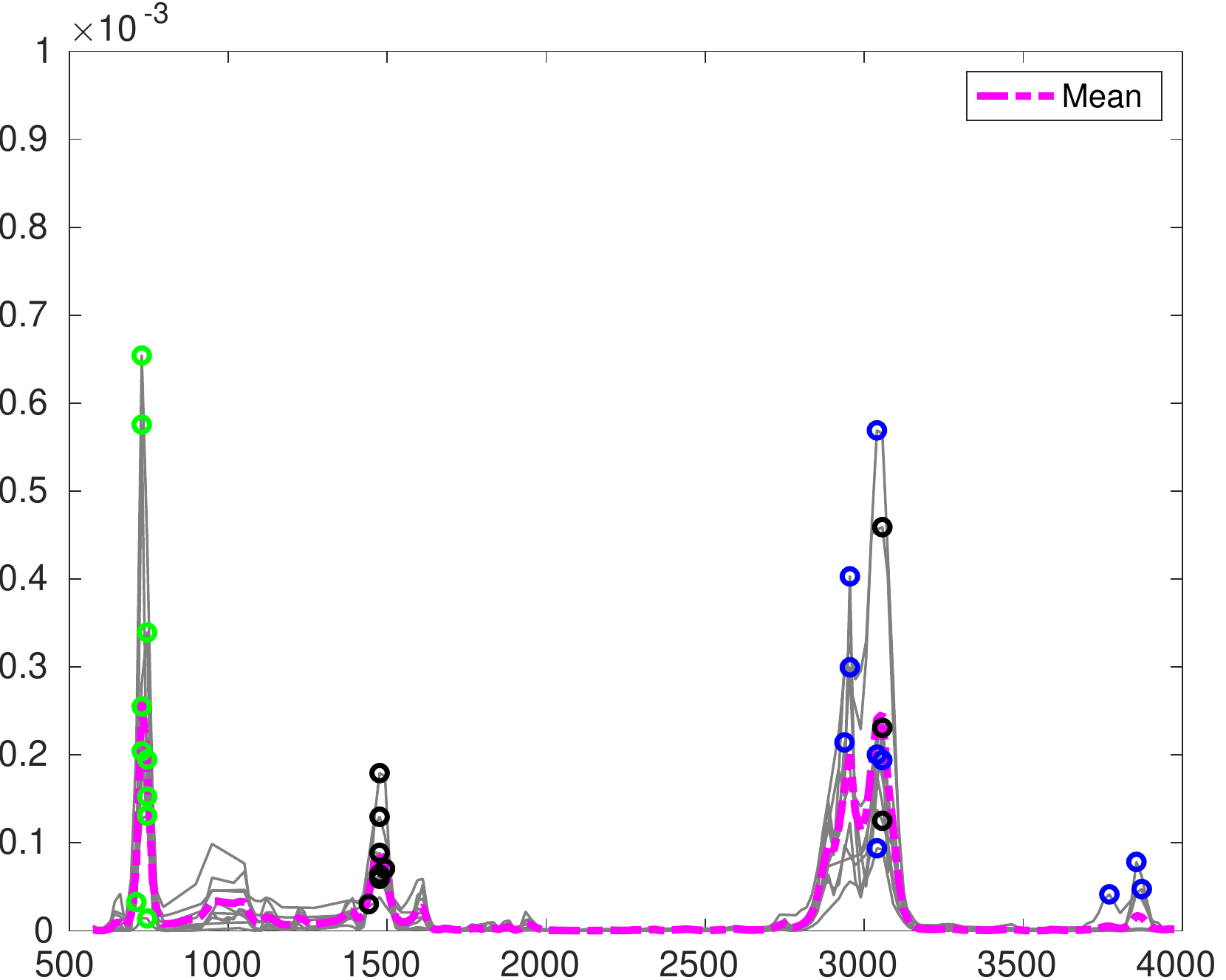}
& \includegraphics[width=0.25\textwidth]{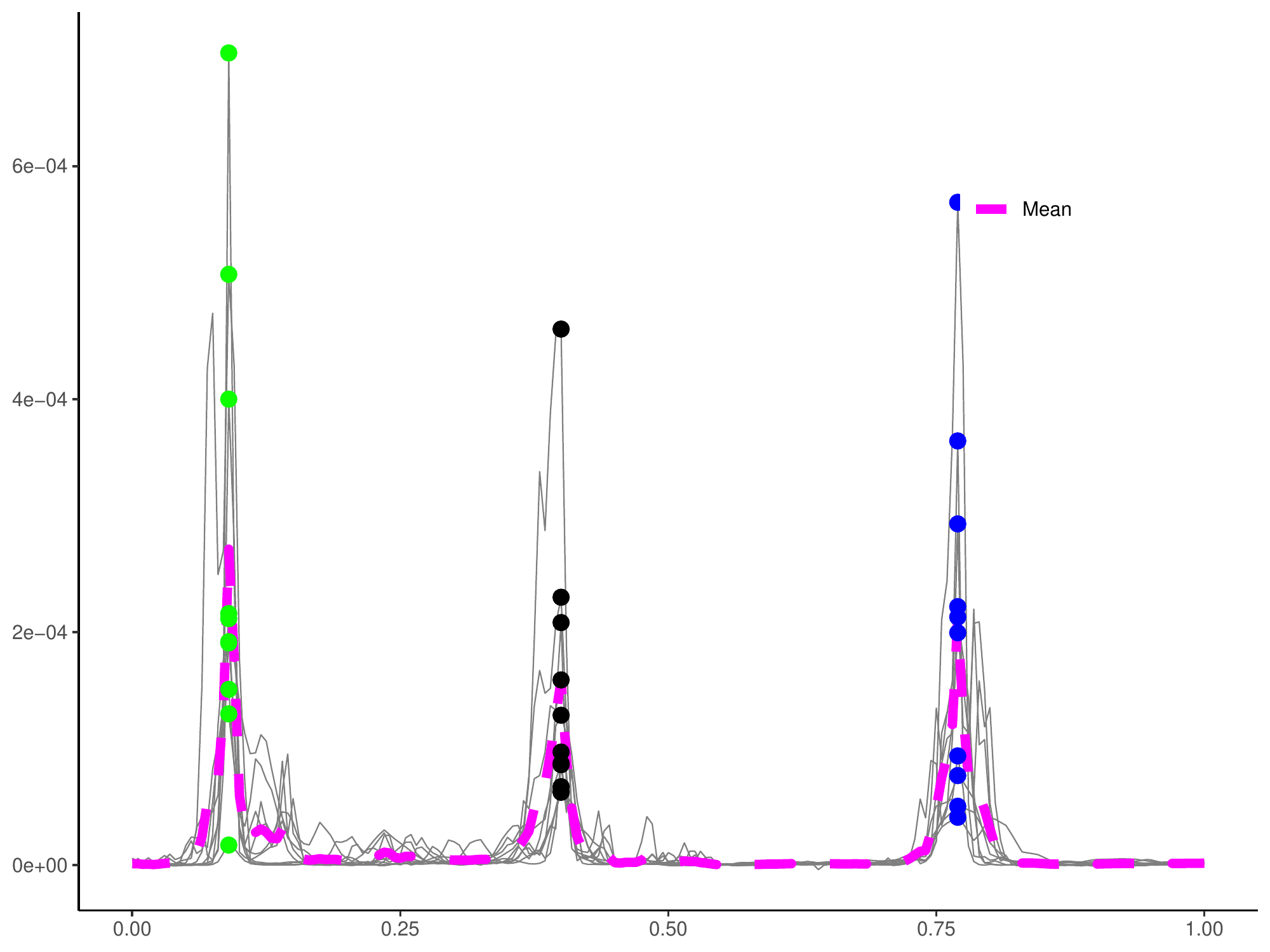}
\\
\tiny{(a) Raw Data}
&  \tiny{(b) Soft Alignment, $\lambda = 8 \times 10^{-4}$}
&  \tiny{(c) Unconstrained Alignment}
& \tiny{(d) \cite{ramsay2021fda}}
\\
\includegraphics[width=0.23\textwidth]{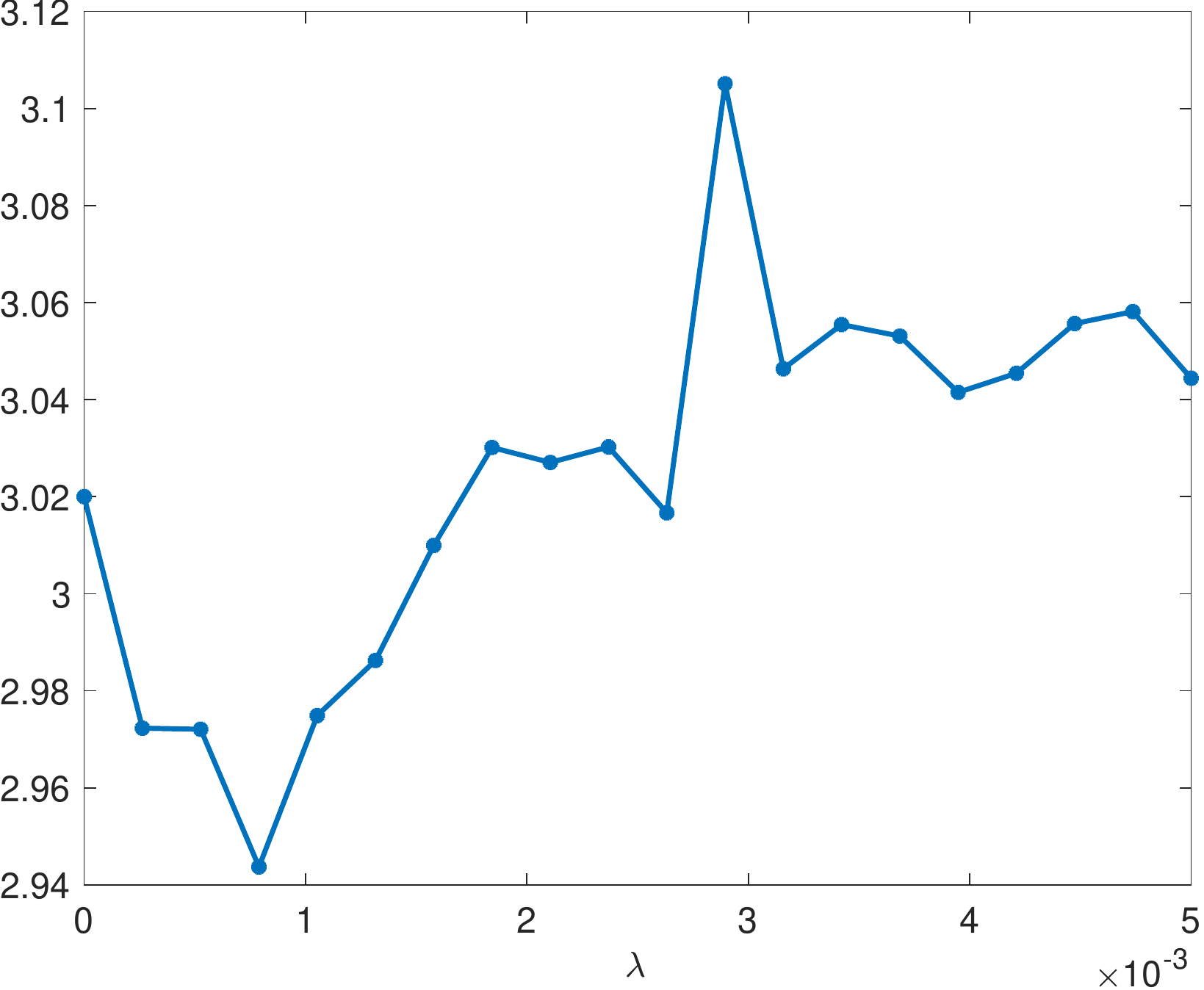}
& \includegraphics[width=0.25\textwidth]{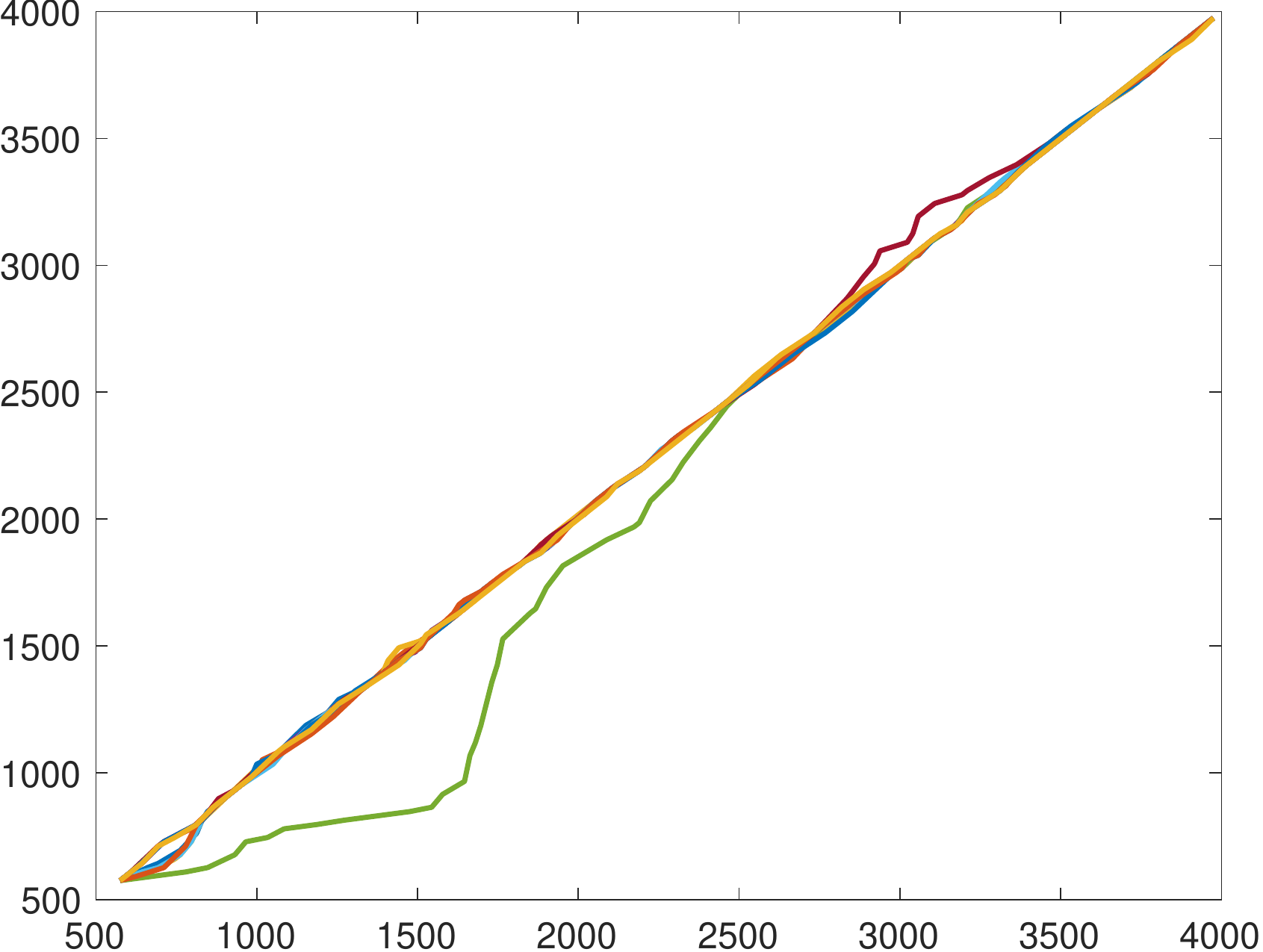}
& \includegraphics[width=0.25\textwidth]{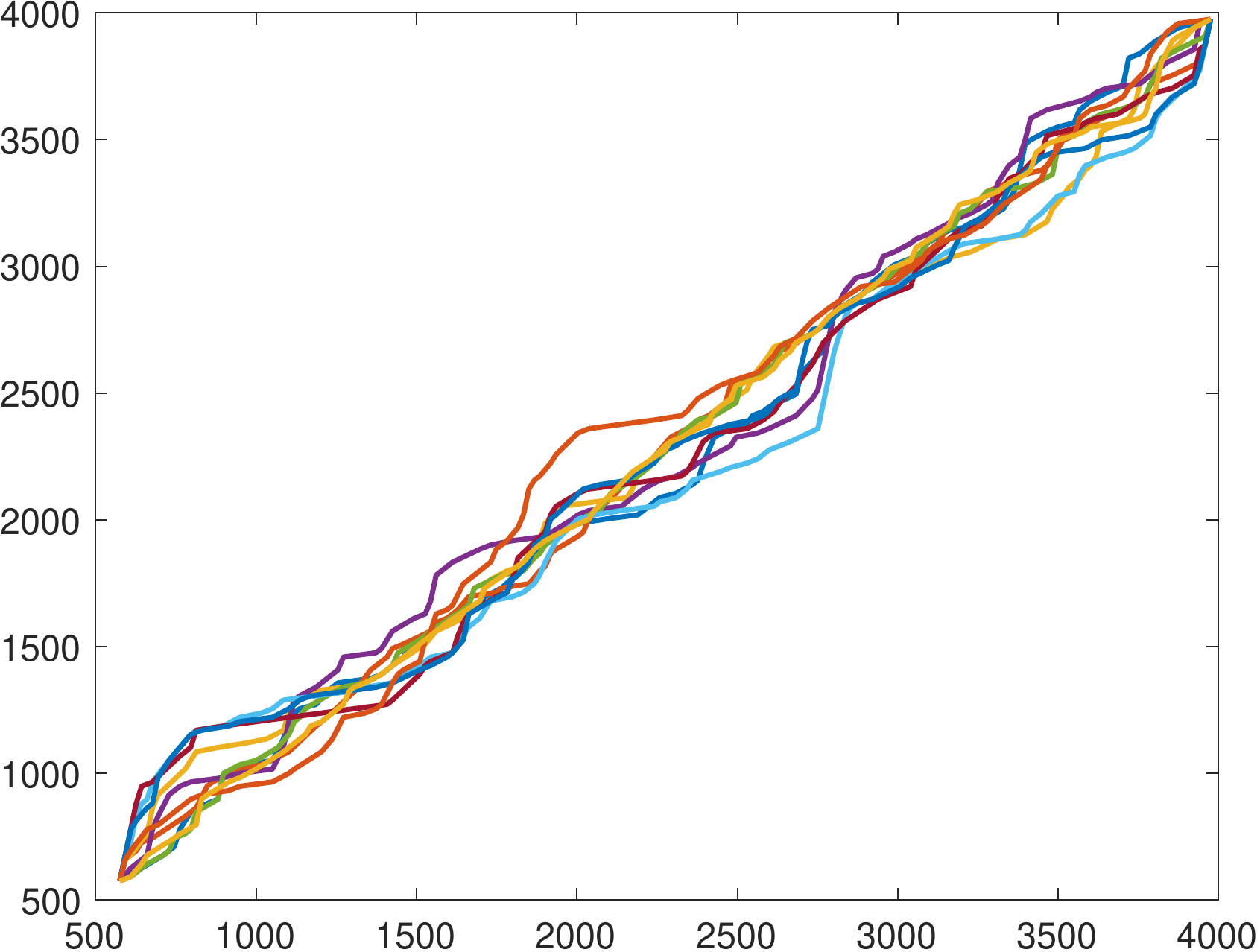}
& \includegraphics[width=0.27\textwidth]{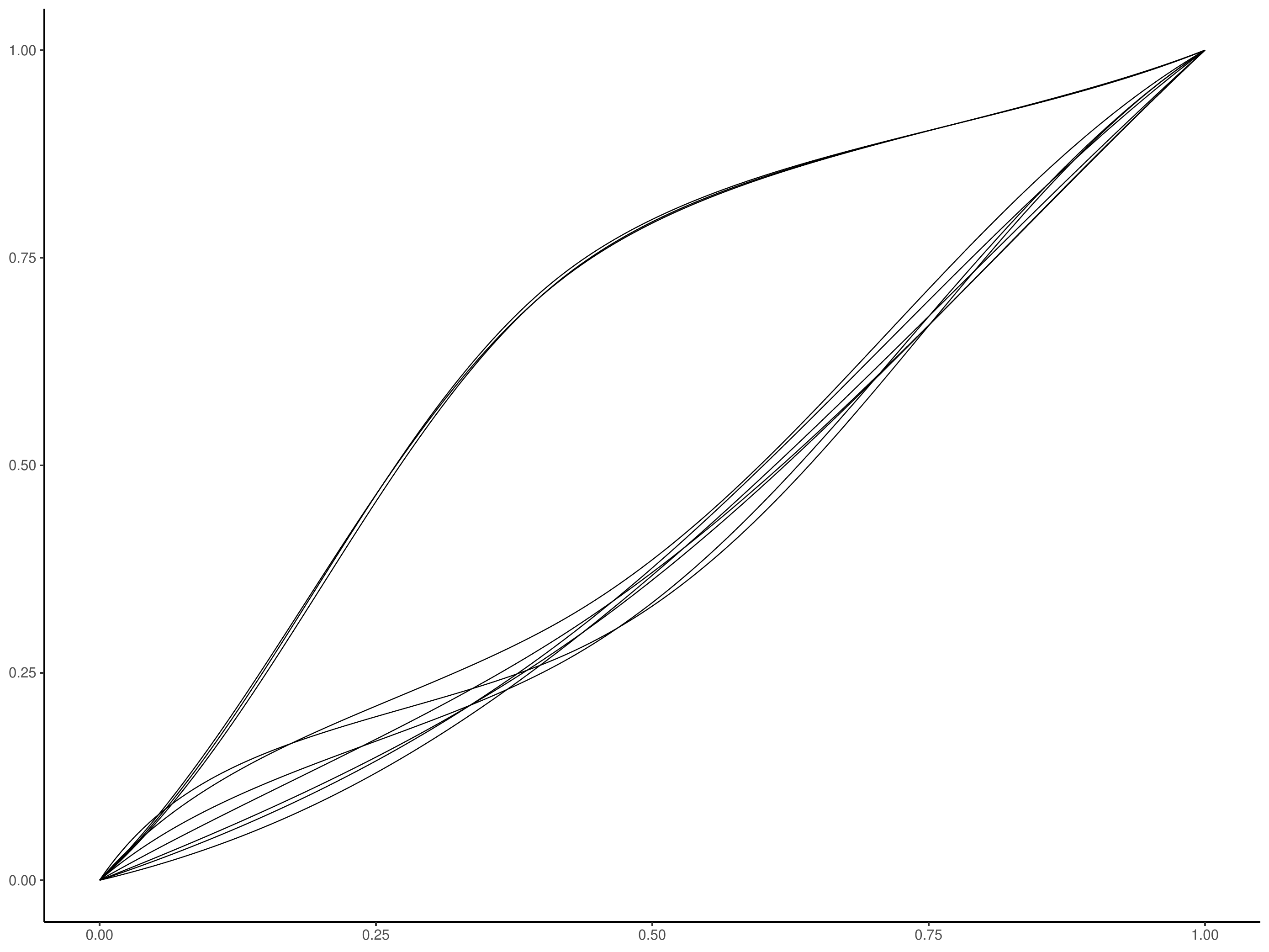}
\\
\tiny{(e) LOOCV Shape Error}
& \tiny{(f) Additional Warpings, $\lambda=8 \times 10^{-4}$}
& \tiny{(g) Unconstrained Time Warpings}
& \tiny{(h) \cite{ramsay2021fda}}
\end{tabular}
\caption{Soft alignment of FTIR data. Green, black and blue circles show the first, second and third landmarks.}
\label{fig:ftir}
\end{figure}

\noindent {\bf Fourier-transform infrared spectroscopy (FTIR) Data}: 
The first application is from Fourier-transform infrared spectroscopy (FTIR) data 
studied in \citep{srivastava2017automated}. 
FTIR is widely used in geology, chemistry, materials, and biology to 
to measure infrared absorption and emission spectra of a solid, liquid or gas, 
and to characterize material properties.
The dataset contains 10 observations and each observation contains the changes in transmittance as a function of wavelength. 
The common pattern in this data is that all functions have three major peaks that are expected to be matched. 
However, there are often some small noise peaks in the data because of experiment conditions 
such as instrument calibration, specimen contamination, and so on.  
Due to the existence of numerous small
peaks, the unconstrained alignment fails to match the three major peaks with each other, as 
shown in Figure~\ref{fig:ftir}(c).
In order to assist with the alignment, we select landmarks representing the three major peaks. 
{The landmarks are not guaranteed to be precise thus hard landmarks registration is a strong constraint, as shown in Figure \ref{fig:ftir}(d).} 
Figure~\ref{fig:ftir}(b) illustrates the result of soft alignment, which reflects the common pattern of these functions. 
As we can see, the landmarks of the second and the third peak are not matched exactly while for the first peak, they are precisely registered.
In this case, $\lambda = 8 \times 10^{-4}$ provides the smallest prediction of LOOCV Shape error. {The resulting mean functions from soft alignment and hard alignment show the three major patterns while the one from unconstraint alignment has more undesired small peaks.}
\\

\noindent {\bf Tallahassee Electricity Data}: The next 
example involves daily household electricity consumption profiles for homes in Tallahassee, FL, USA \citep{Dasgupta2019Clustering}. 
We select 75 households from a neighborhood and 
for each household, the data shows the daily electricity consumption (on weekdays).
The common pattern of these functions is bimodal, because of the typically high consumptions in the mornings and evenings.
However, the unconstrained alignment does not result in a bimodal pattern, as shown in (c), due to the presence of multiple small peaks.
To apply landmark registration, we automatically select the locations of largest peaks in the first and second half 
of domain $[0, 24]$ as landmarks. 
The resulting soft alignment result is presented in (b), where $\lambda=6.3$ are obtained in (d).
{Looking at mean functions, unconstrained alignment shows many smaller peaks and valleys while soft alignment exhibits a prominent bimodal pattern, reflecting the utility consumption pattern of the neighborhood. The hard landmark registration also shows a bimodal mean. However, the mean function seems to over flat for the smaller peaks. Most importantly, the LOOCV curve explains that soft alignment achieves lowers shape error.}

\begin{figure}[H]
\centering
\begin{tabular}{cccc}
\includegraphics[width=0.25\textwidth]{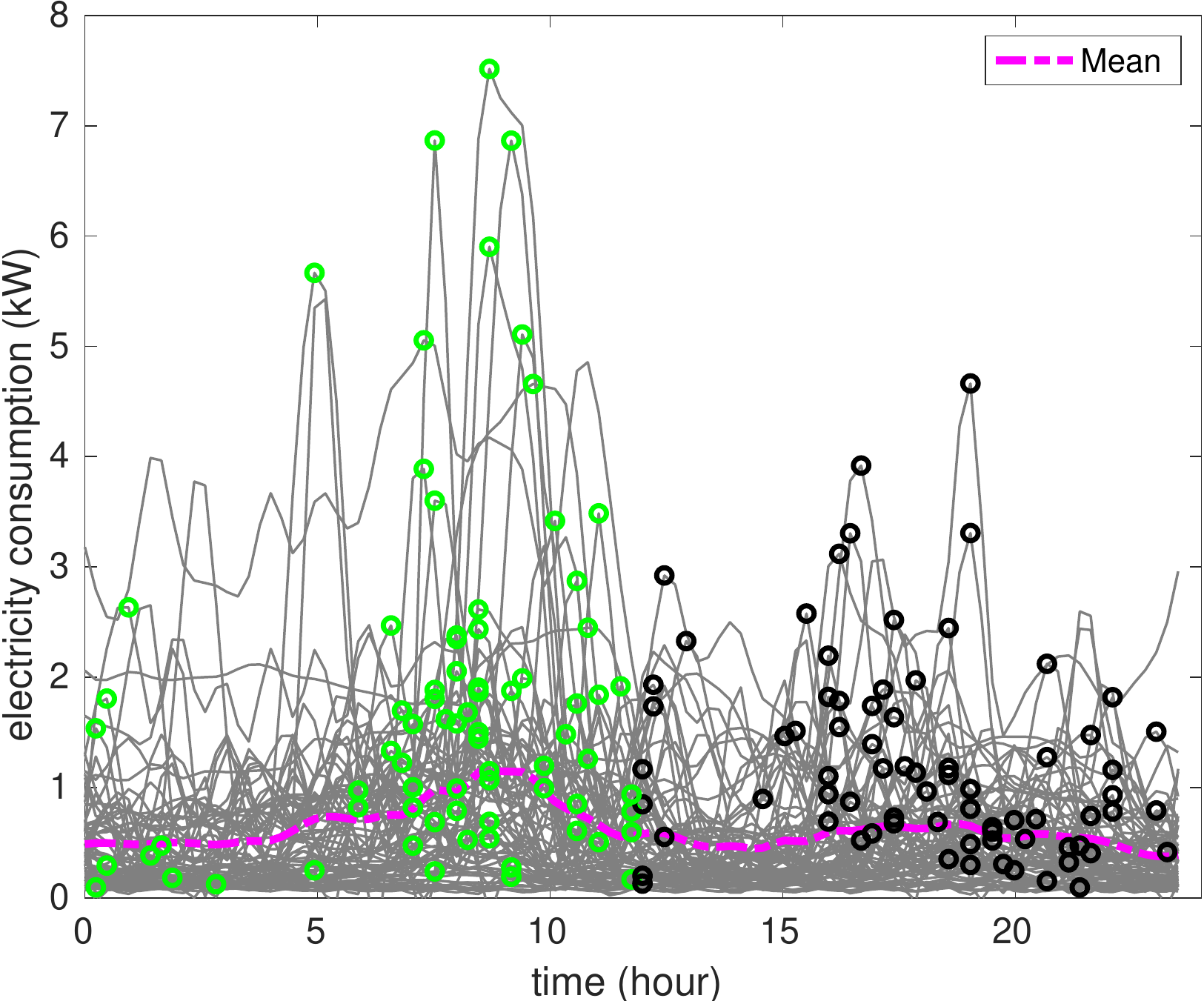}
& \includegraphics[width=0.25\textwidth]{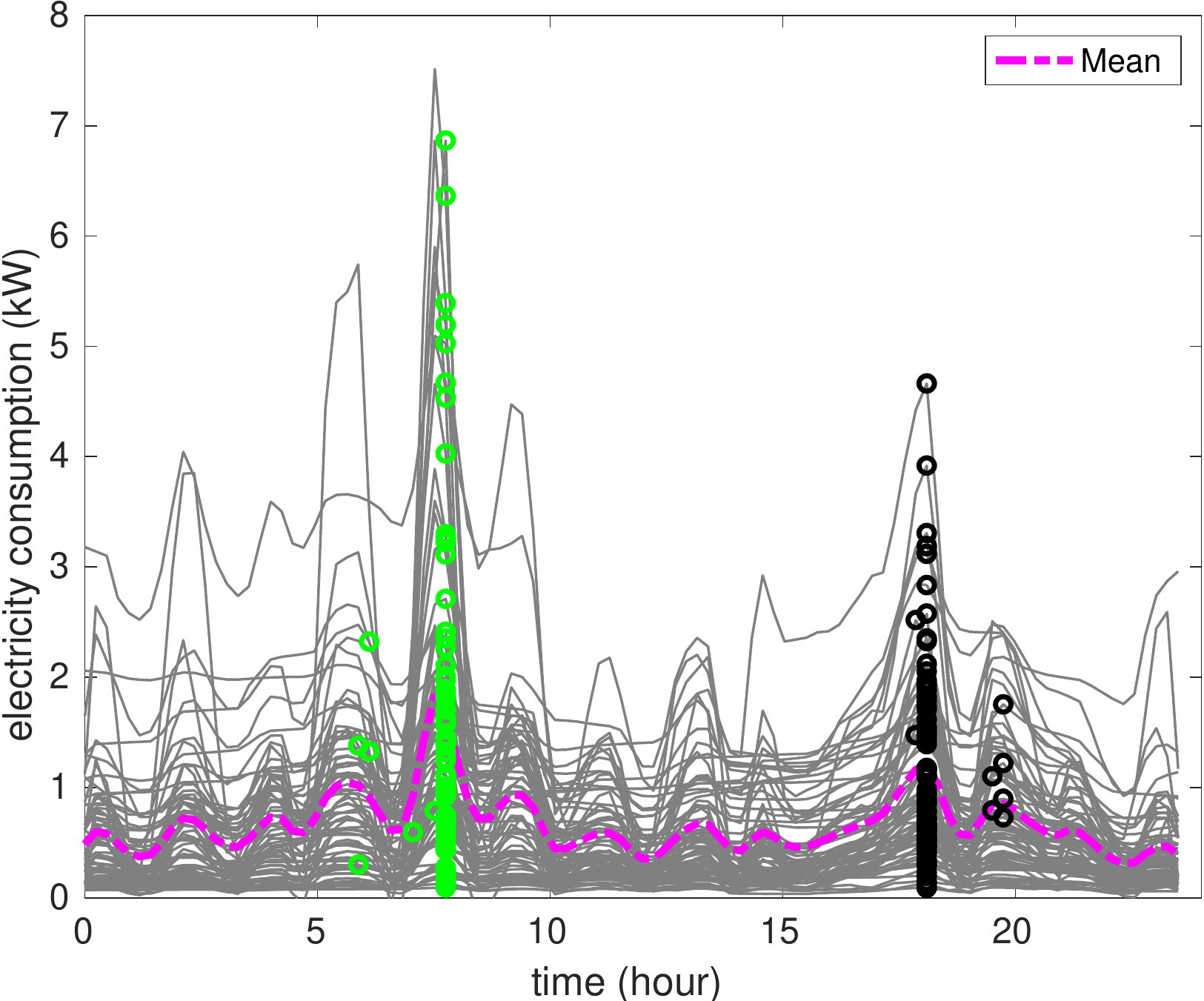}
& \includegraphics[width=0.25\textwidth]{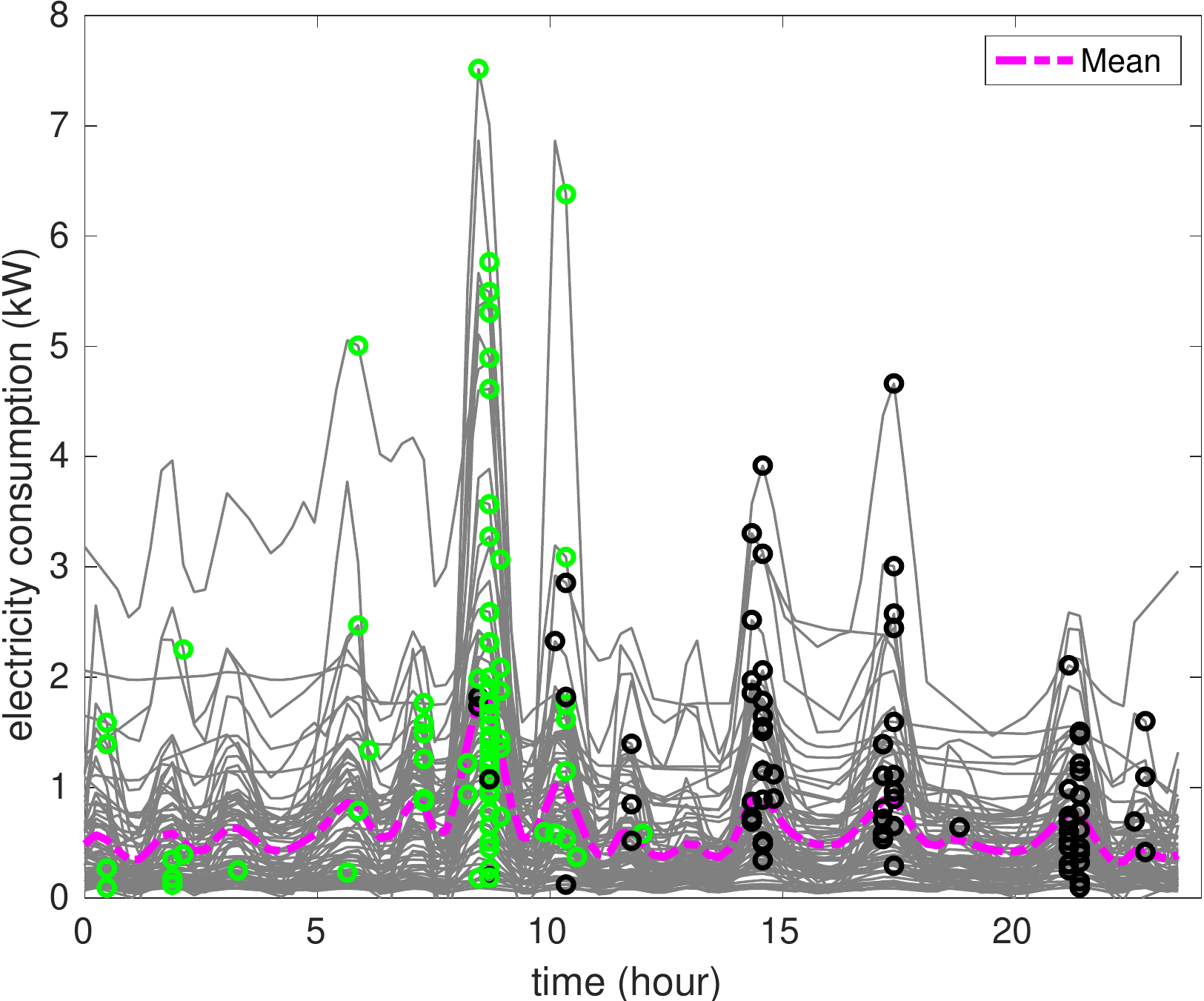}
& \includegraphics[width=0.25\textwidth]{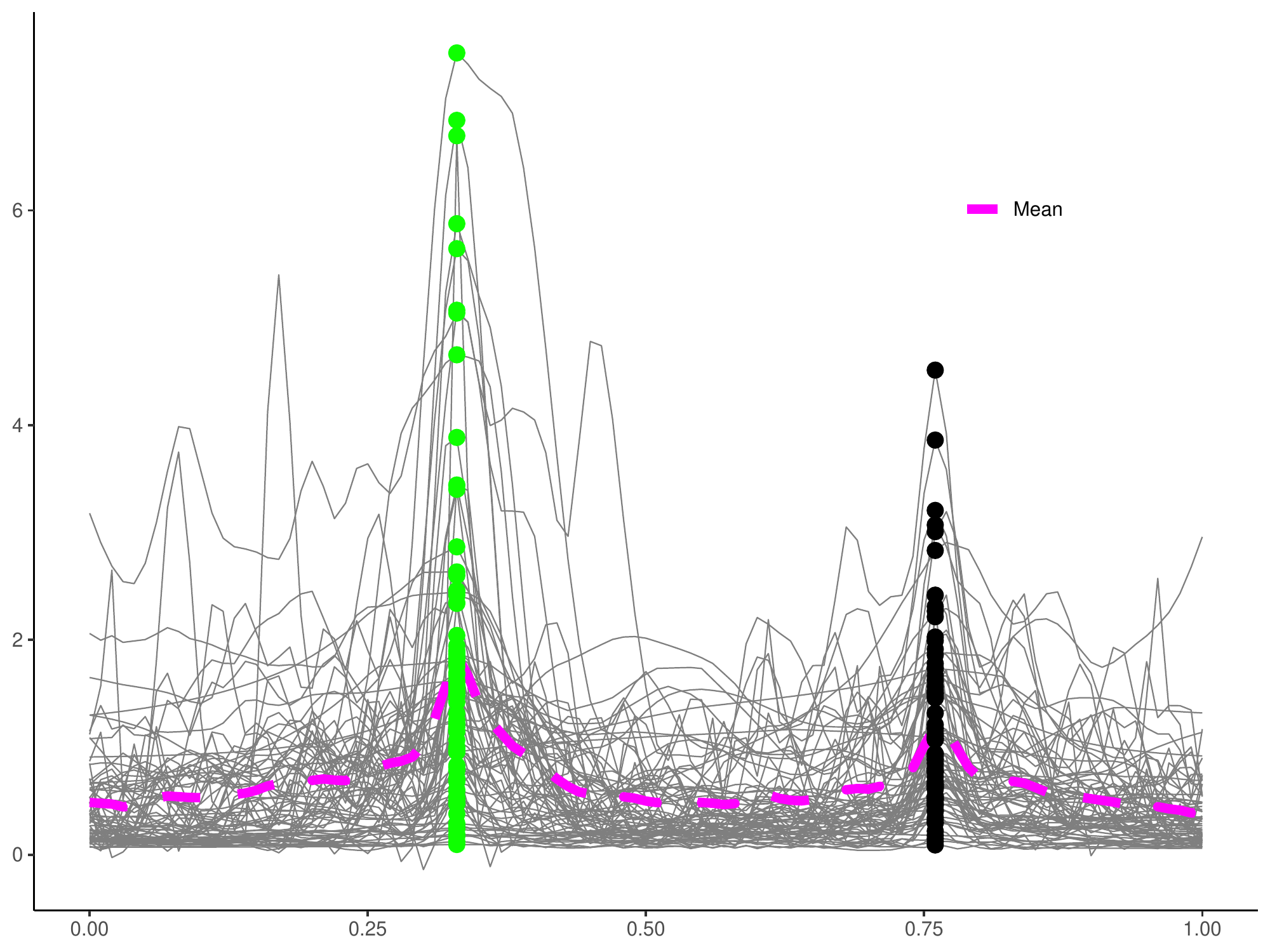}
\\
\tiny{(a) Raw Data}
& \tiny{(b) Soft Alignment, $\lambda = 6.3$}
& \tiny{(c) Unconstrained Alignment}
& \tiny{(d) \cite{ramsay2021fda}}
\\
\includegraphics[width=0.25\textwidth]{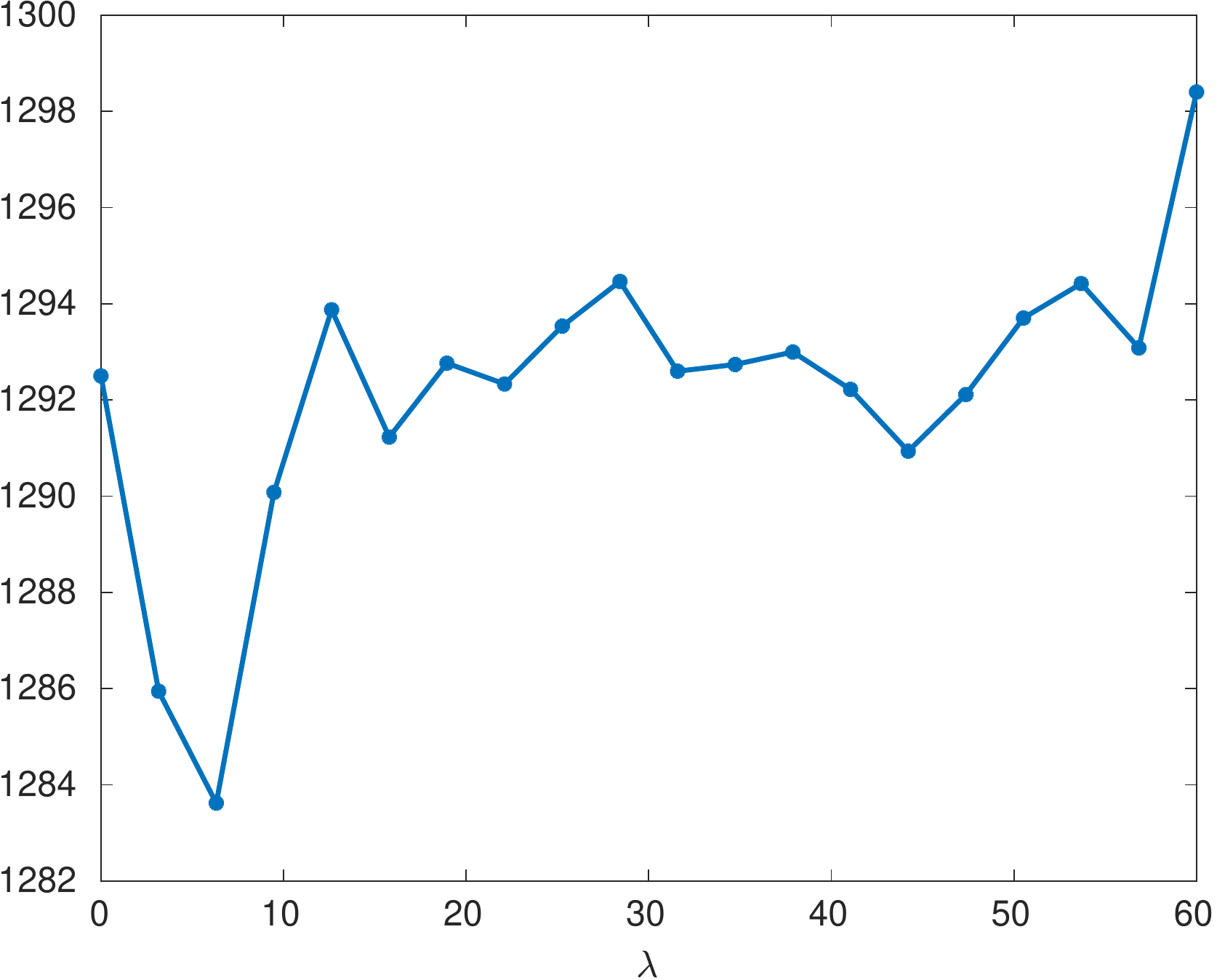}
& \includegraphics[width=0.25\textwidth]{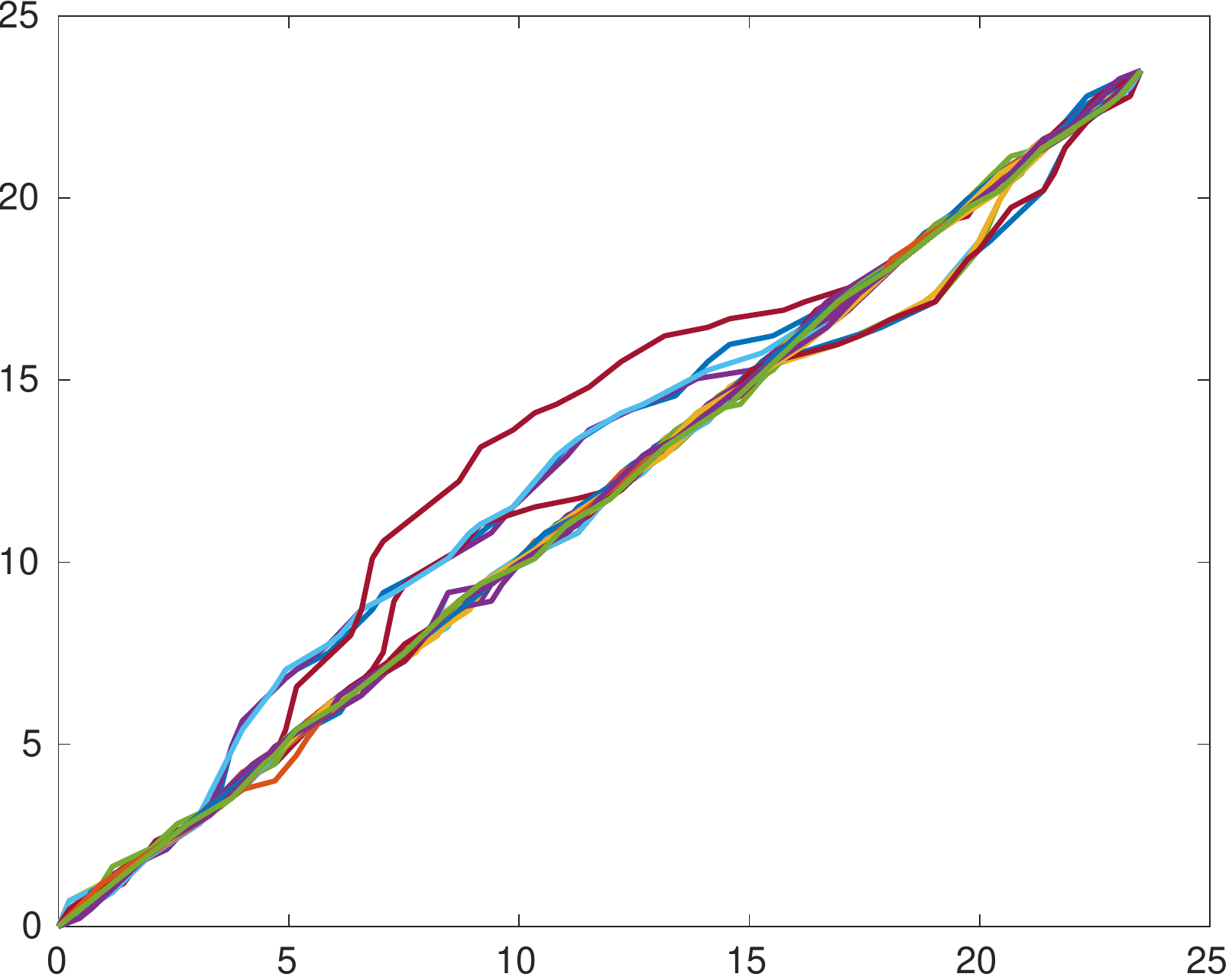}
& \includegraphics[width=0.25\textwidth]{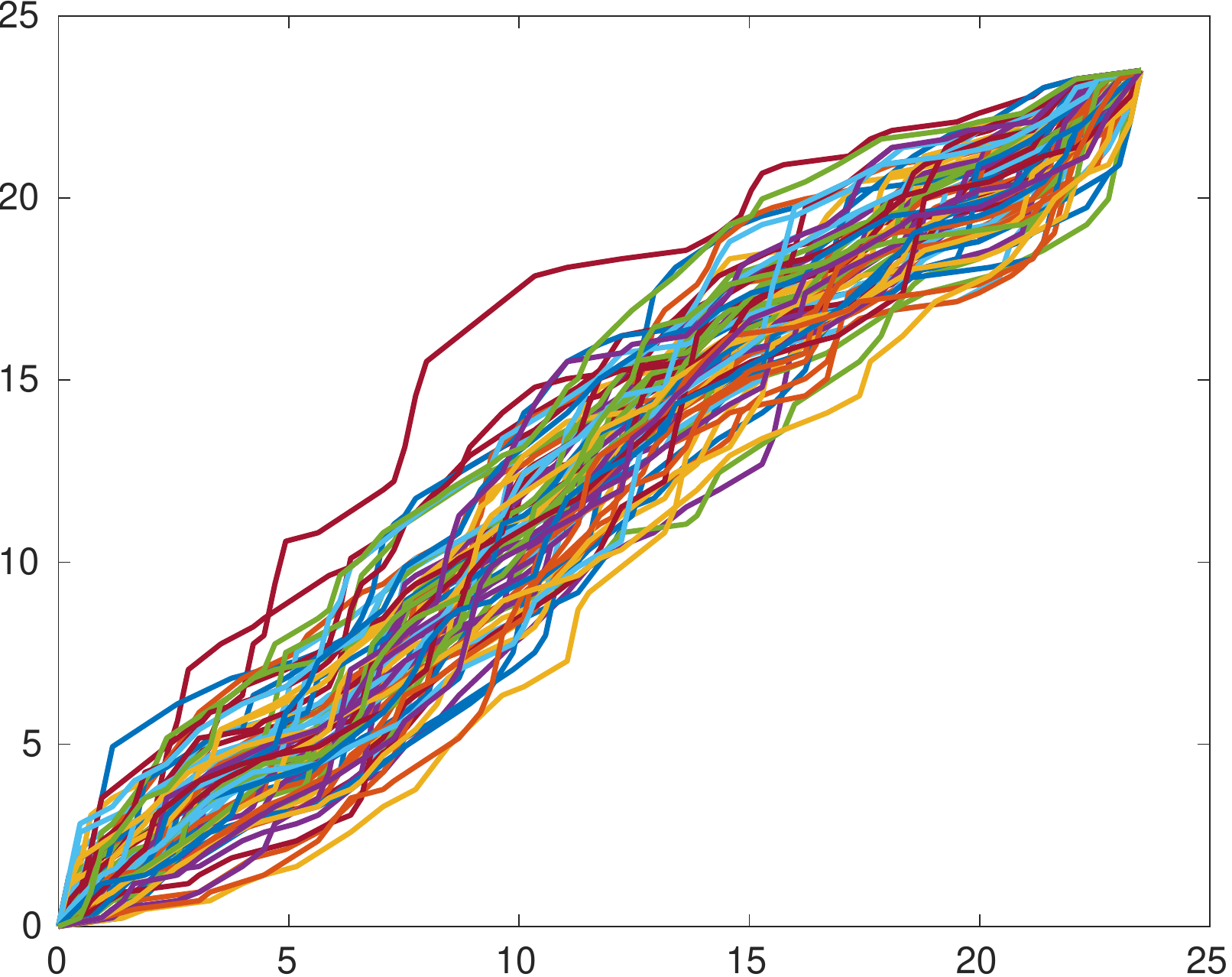}
& \includegraphics[width=0.27\textwidth]{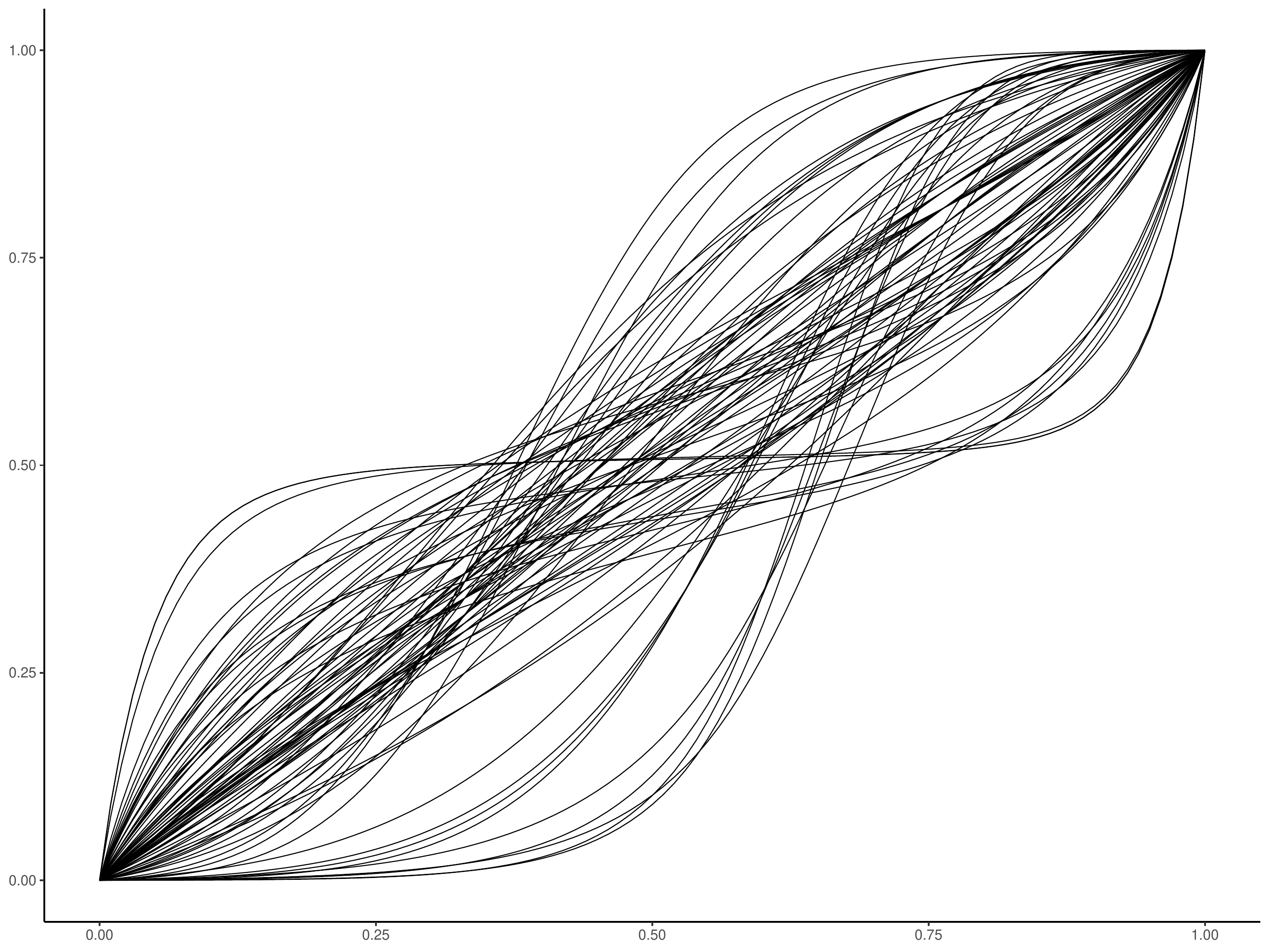}
\\
\tiny{(e) LOOCV Shape Error}
& \tiny{(f) Additional Warpings, $\lambda = 6.3$ }
& \tiny{(g) Unconstrained Warpings}
& \tiny{(h) \cite{ramsay2021fda}}
\end{tabular}  
\caption{Soft alignment for Tallahassee electricity data. Green and  black circles show the first and second landmarks.}
\label{fig:tally}
\end{figure}

\begin{figure}[H]
\centering
\begin{tabular}{cccc}
\includegraphics[width=0.25\textwidth]{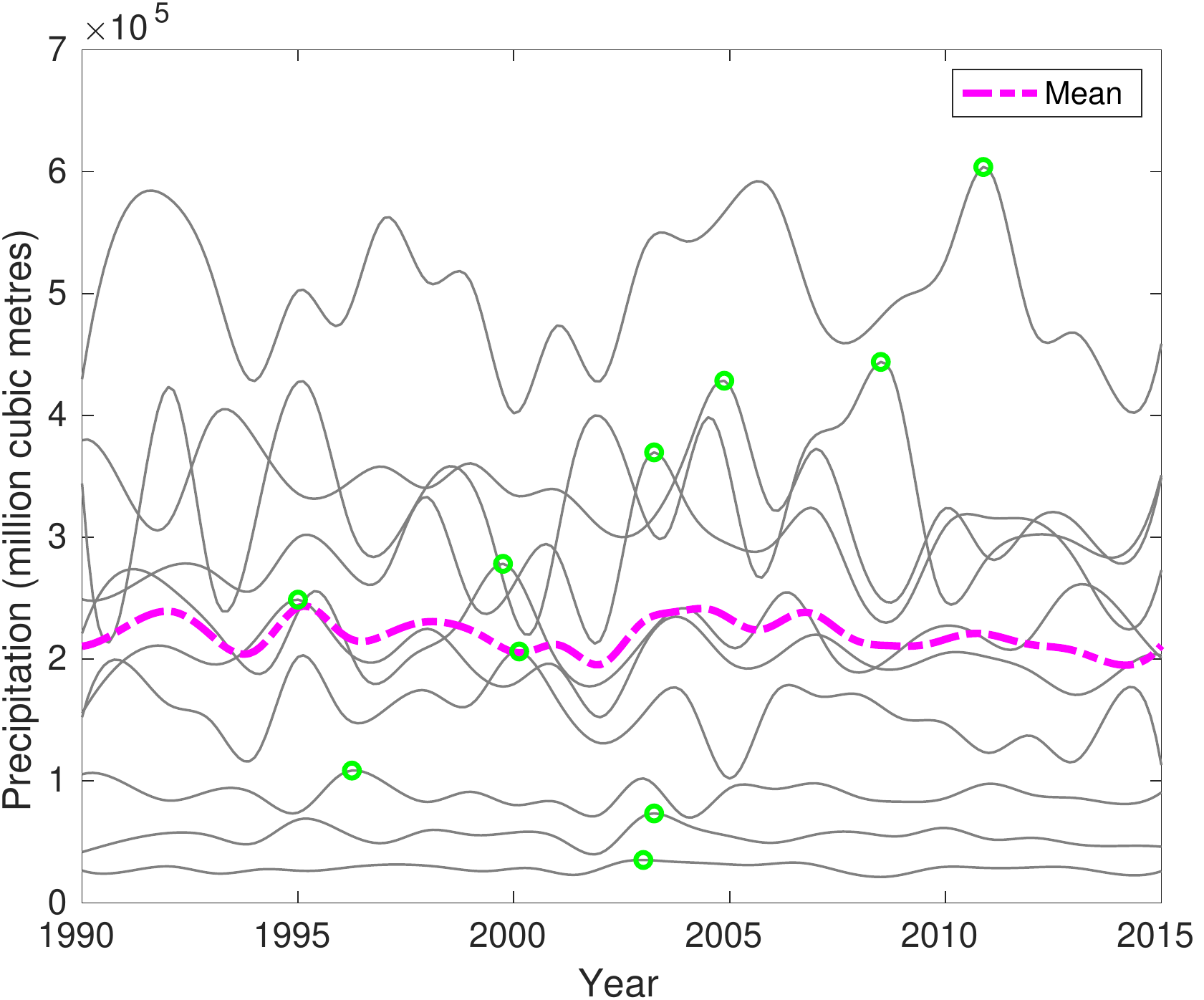}
& \includegraphics[width=0.25\textwidth]{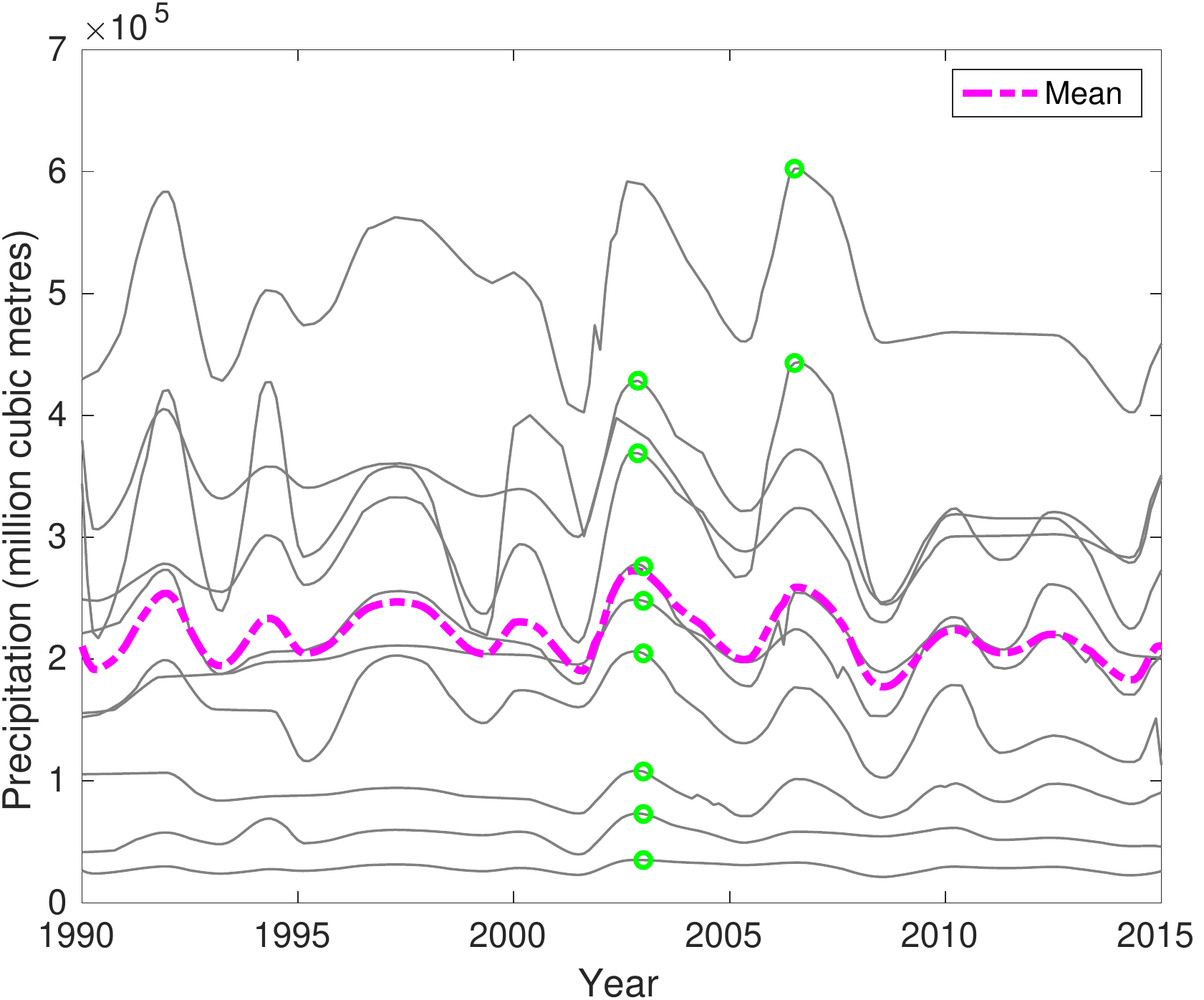}
& \includegraphics[width=0.25\textwidth]{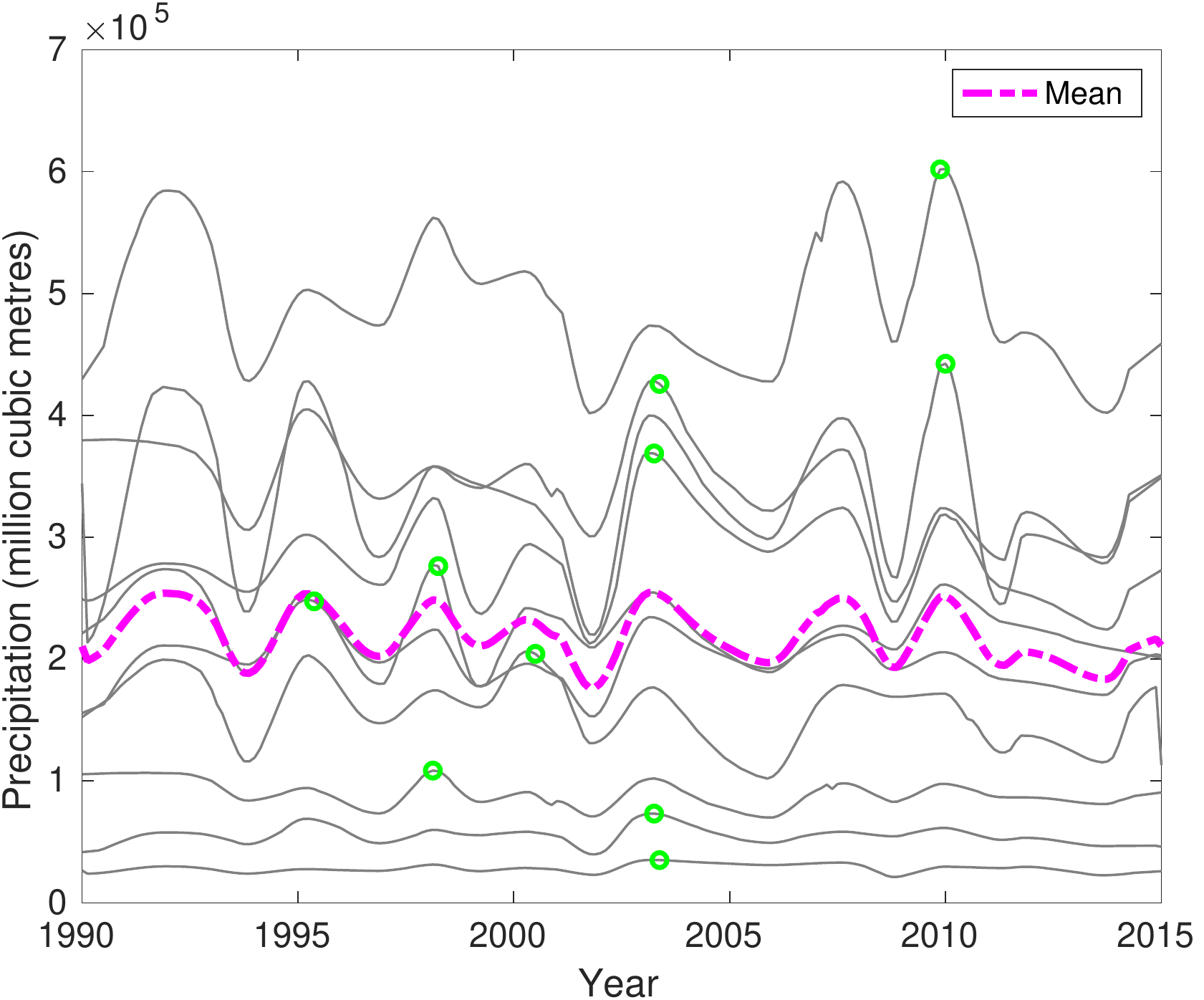}
& \includegraphics[width=0.25\textwidth]{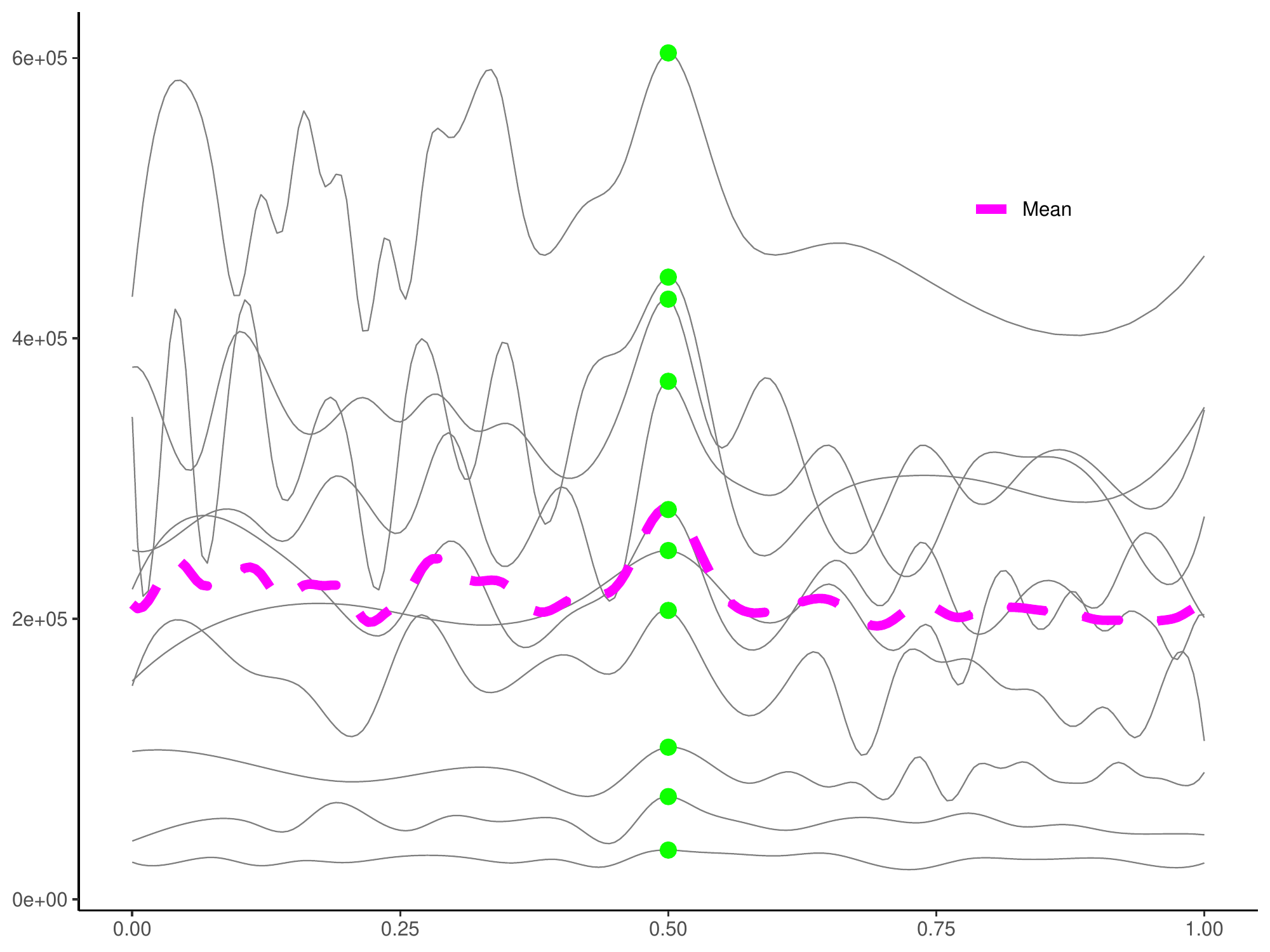}
\\
\tiny{(a) Raw  Data}
& \tiny{(b) Soft Alignment, $\lambda = 736840$}
& \tiny{(c) Unconstrained Alignment}
& \tiny{(d) \cite{ramsay2021fda}}
\\
\includegraphics[width=0.23\textwidth]{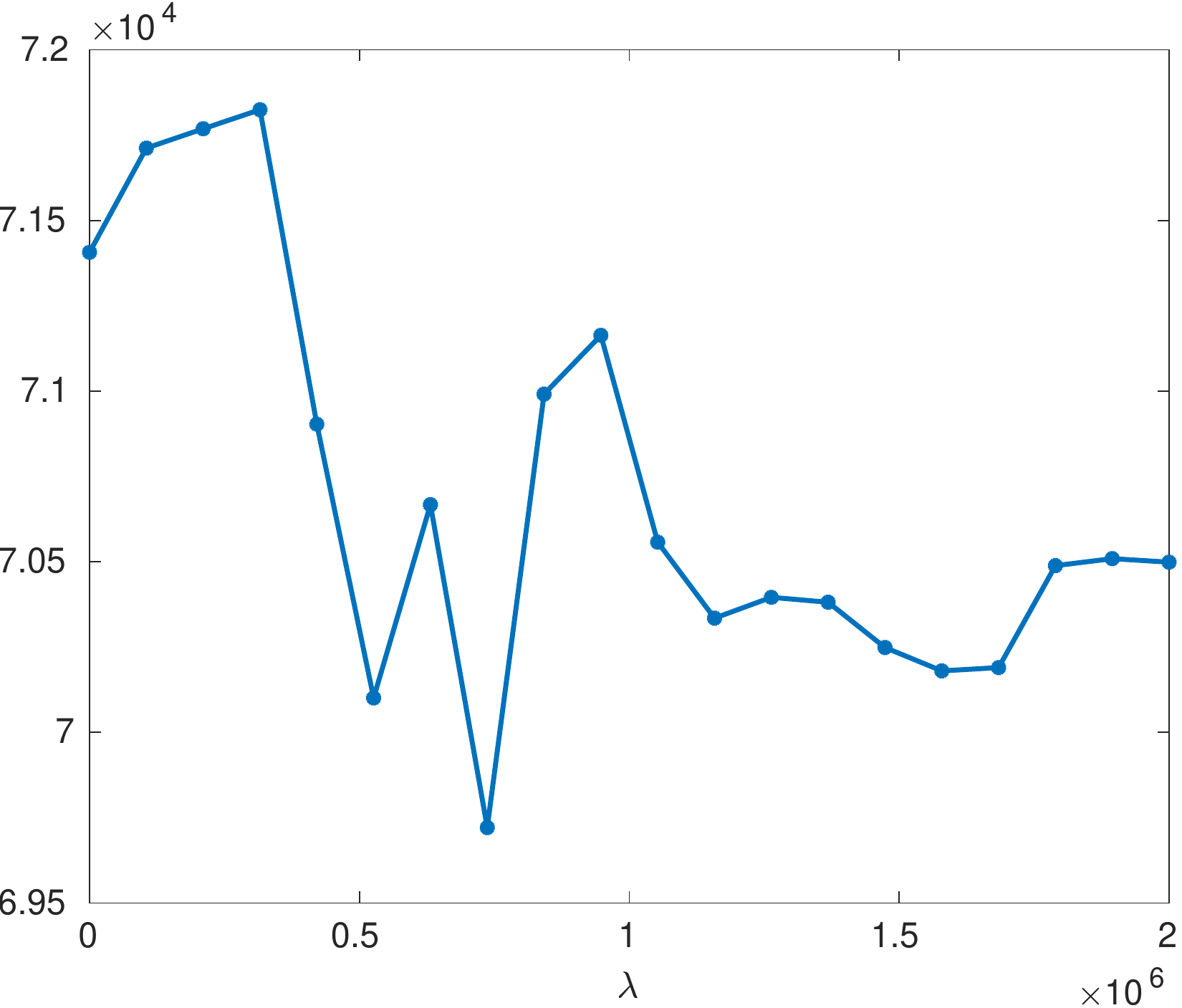}
& \includegraphics[width=0.25\textwidth]{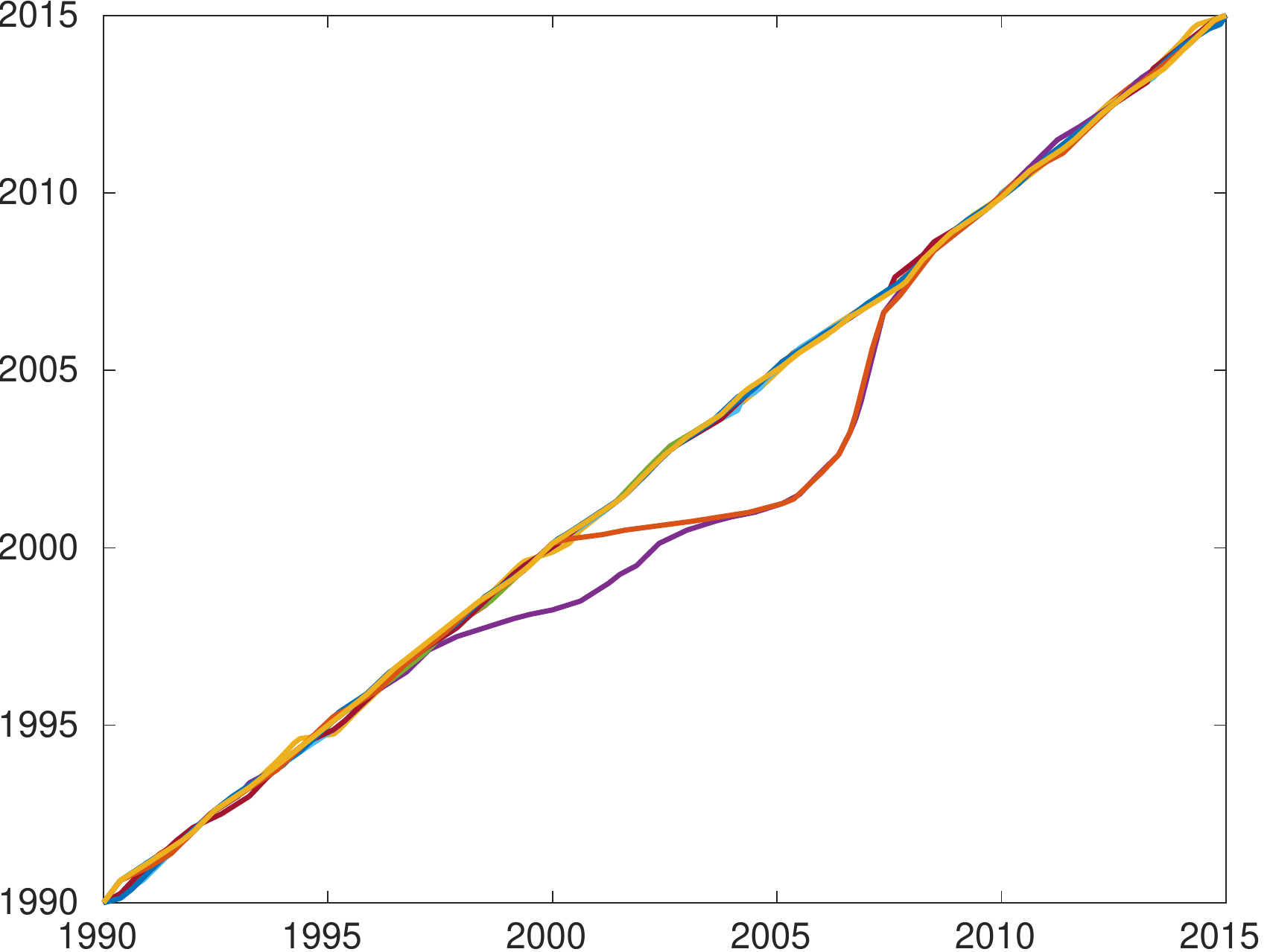}
& \includegraphics[width=0.25\textwidth]{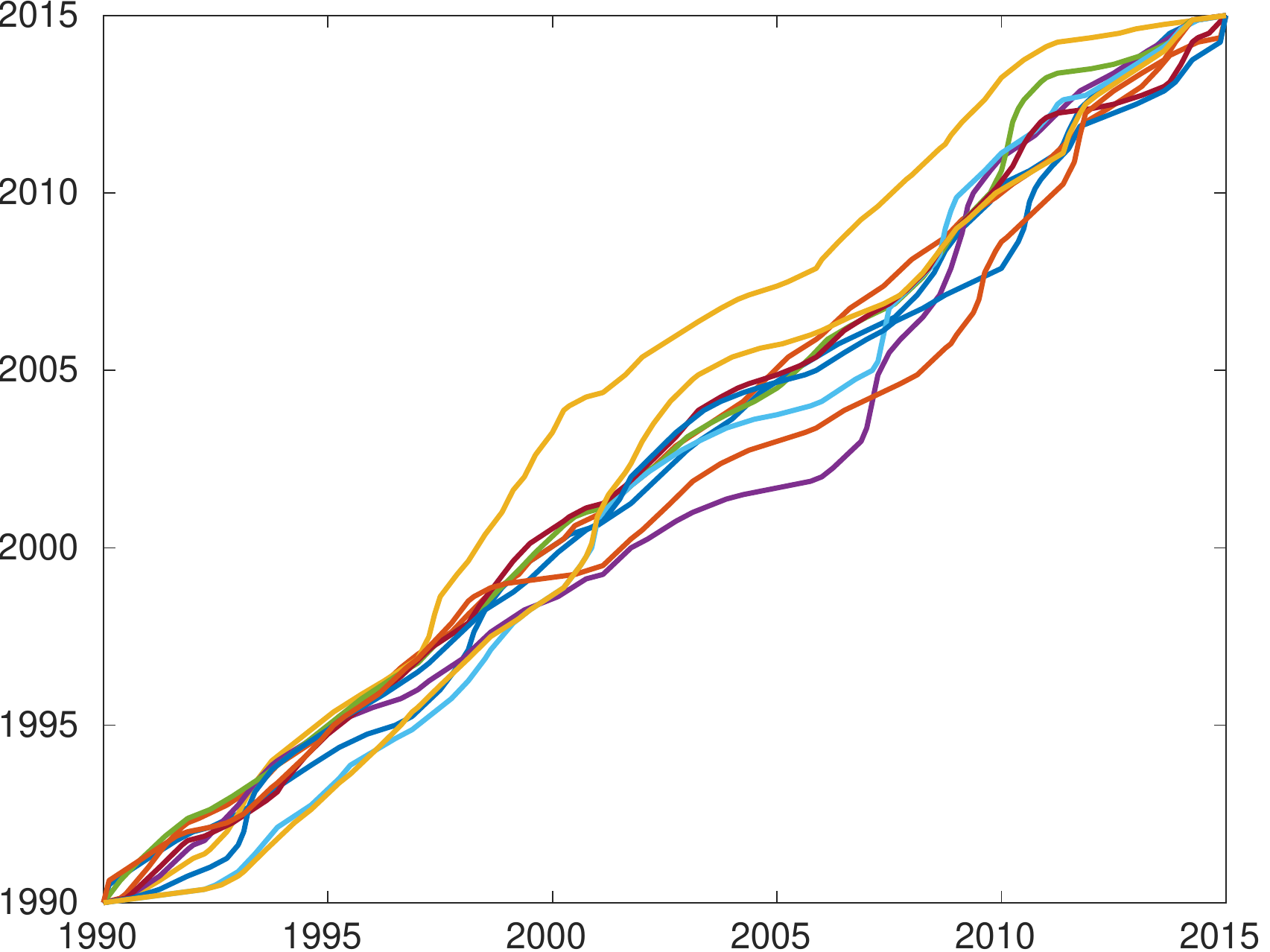}
& \includegraphics[width=0.25\textwidth]{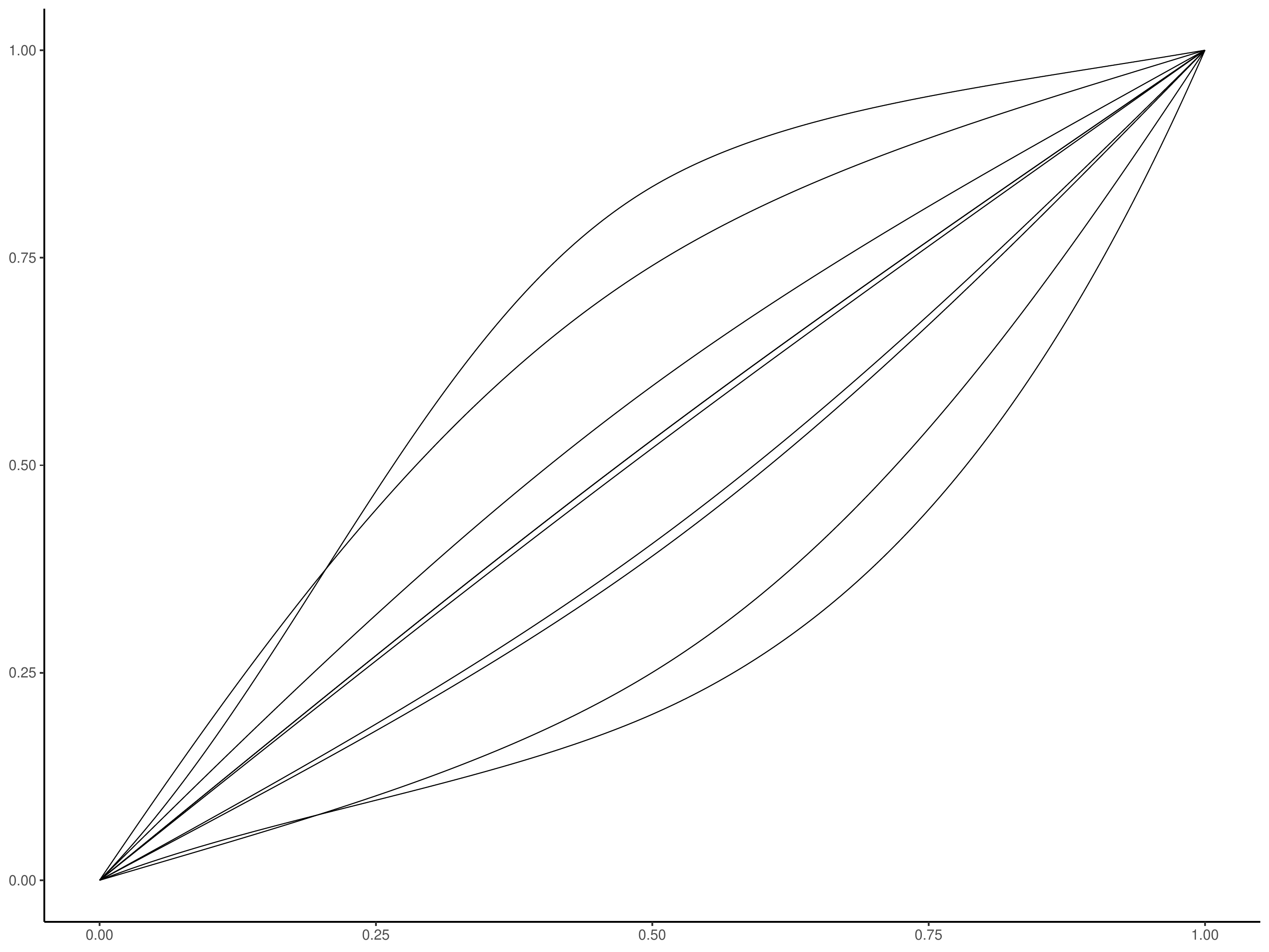}
\\
\tiny{(e) LOOCV Shape Error}
& \tiny{(f) Additional Warpings, $\lambda = 736840$ }
& \tiny{(g) Unconstrained Warpings}
& \tiny{(h) \cite{ramsay2021fda}}
\end{tabular}
\caption{Soft alignment for precipitation data. }
\label{fig:precip}
\end{figure}

\noindent {\bf Precipitation Data}: The last example relates to annual precipitation of 10 European countries.
As we know, the peaks and valleys of precipitation will not be temporally synchronized across different countries. 
To study the common structure of the precipitation for those countries, we need to remove or reduce the phase variability through alignment.
We select the largest peak as landmarks, represented by green circles in Figure \ref{fig:precip}.
As before, we balance the result between unconstrained alignment and hard registration with $\lambda =  736840$ (from LOOCV) and get the soft alignment. This is shown in (b) of Figure \ref{fig:precip}. {Comparing with the cross-sectional mean, mean pattern from both soft alignment and unconstrained alignment displays more prominent peaks and valleys. In addition, soft alignment preserves the largest peak not captured in the unconstrained alignment. }

\section{Summary}\label{sec:con_sfa}

In this work, we have presented a novel approach
 to incorporate landmarks into elastic functional alignment. The strength of this data-driven approach is 
 that it allows one to balance the contributions of landmarks 
 and function shapes in reaching an optimal registration. This is useful when alignment based purely on the geometry of functional data 
is not reliable, and some additional information in form of landmarks is present. 
In those cases, soft alignment can help provide a more meaningful solution. 

The objective function for soft alignment is actually a pseudometric and 
lends itself nice theoretical properties -- non-negativity, symmetry and invariance -- 
to the alignment solution. One can use this pseudometric for subsequent data analysis, 
such as clustering, principal component analysis, modeling functional data, and functional 
linear regression. 
We also note that although this work is limited to alignment of real-valued functions, 
this framework is directly applicable to curves in any Euclidean space.

\section{Appendix}
Proof of lemma \ref{lem:pseudom}: 
We need to establish the following properties of $d_{\lambda}$: 
\begin{enumerate}
\item {\bf Non-negativity:} $d_{\lambda}(q_1,q_2) \geq 0$. This holds because each of the components in $d_{\lambda}$ is 
non-negative.
\item {\bf Identity of indiscernibles:} $d_{\lambda}(q,q) = 0$; Unlike a metric space, points in a pseudometric space need not be distinguishable. If $d_{\lambda}(q_1,q_2) = 0$ , then $(q_1*\gamma_1) = (q_2 * \gamma_2) $, 
and this implies $q_1$ and $q_2$ are in the same orbit. 
In this situation, there will be two different scenarios. 
One the one hand, if $\boldsymbol{\tau}^{(1)} = \boldsymbol{\tau}^{(2)}$, this implies $q_1 = q_2$.
One the other hand, if at least one landmarks is different, then $q_1 \neq q_2$. 
However, in terms of alignment, $(q_1*\gamma_1) = (q_2 * \gamma_2) $ means alignment is done by the pre-alignment. 
\item {\bf Symmetry:} $d_{\lambda}(q_1,q_2)=d_{\lambda}(q_2,q_1)$. 
The proof uses the isometric property, as shown in Lemma \ref{lem:isometry}. 
\begin{align*}
&d^2_{\lambda}(q_1,q_2)\\
&= \inf\limits_{\gamma} \{\| (q_1*\gamma_1)-((q_2*\gamma_2) *\gamma) \|^2+\lambda  \| (\bf{1}*\gamma)-\bf{1} \|^2\}  \\
&= \inf\limits_{\gamma} \{\| ((q_1*\gamma_1)*\gamma^{-1})-(q_2*\gamma_2) \|^2+\lambda  \|((\bf{1}*\gamma)*\gamma^{-1})-(\bf{1}*\gamma^{-1}) \|^2\} \\
&=\inf\limits_{\gamma} \{\| ((q_1*\gamma_1)*\gamma^{-1})-(q_2*\gamma_2) \|^2+\lambda  \|\bf{1}-(\bf{1}*\gamma^{-1}) \|^2\} \\
&= d^2_{\lambda}(q_2,q_1)
\end{align*} 
\item {\bf Triangle Inequality:} $d_{\lambda}(q_1,q_3) \leq d_{\lambda}(q_1,q_2)+d_{\lambda}(q_1,q_3)$. The proof also uses Lemma \ref{lem:isometry}.
{
\begin{align*}
&\| (q_1*\gamma_1)-((q_3* \gamma_3)* \gamma_{31}) \|^2+\lambda  \|({\bf 1}*\gamma_{31})-\bf{1} \|^2 \\
&= \| \tilde{q}_1-(\tilde{q}_2 * \gamma_{21}) +(\tilde{q}_2 * \gamma_{21}) -(\tilde{q}_3 * \gamma_{31}) \|^2+\lambda  \|({\bf 1} *\gamma_{31})-({\bf 1} *\gamma_{21})+({\bf 1} *\gamma_{21})-\bf{1} \|^2\\
&\leq(\|\tilde{q}_1-(\tilde{q}_2* \gamma_{21}) \|+\|(\tilde{q}_2* \gamma_{21}) -(\tilde{q}_3* \gamma_{31}) \|)^2+\lambda(\|({\bf 1}*\gamma_{31})-({\bf 1}*\gamma_{21})\| +\|({\bf 1}*\gamma_{21})-{\bf 1} \|)^2\\
&\leq \|\tilde{q}_1-(\tilde{q}_2* \gamma_{21})\|^2+\lambda \|({\bf 1}*\gamma_{21})-{\bf 1} \|^2+\|(\tilde{q}_2* \gamma_{21}) -(\tilde{q}_3*\gamma_{31}) \|^2+\lambda\|({\bf 1}*\gamma_{31})-(\bf{1}*\gamma_{21})\|^2\\
&+2\|\tilde{q}_1-(\tilde{q}_2* \gamma_{21})\|\|(\tilde{q}_2* \gamma_{21}) -(\tilde{q}_3* \gamma^{31}) \|+2\lambda\|({\bf 1}*\gamma_{31})-({\bf 1}*\gamma_{21})\| \|({\bf 1}*\gamma_{21})-{\bf 1} \|\\
&\leq \|\tilde{q}_1-(\tilde{q}_2* \gamma_{21})\|^2+\lambda \|({\bf 1}*\gamma_{21})-{\bf 1} \|^2+\|\tilde{q}_2 -((\tilde{q}_3* \gamma_{31})*\gamma^{-1}_{21}) \|^2+\lambda\|(({\bf 1}*\gamma_{31})*\gamma^{-1}_{21})-{\bf 1}\|^2\\
&+2\|\tilde{q}_1-(\tilde{q}_2* \gamma_{21})\|\|\tilde{q}_2 -((\tilde{q}_3* \gamma_{31})*\gamma^{-1}_{21})\|+2\lambda\|(({\bf 1}*\gamma_{31}),\gamma^{-1}_{21})-{\bf 1}\| \|({\bf 1}*\gamma_{21})-{\bf 1} \|\\
&\leq \|\tilde{q}_1-(\tilde{q}_2* \gamma_{21})\|^2+\lambda \|({\bf 1}*\gamma_{21})-{\bf 1} \|^2+\|\tilde{q}_2 -((\tilde{q}_3* \gamma_{31})*\gamma^{-1}_{21}) \|^2+\lambda\|(({\bf 1}*\gamma_{31})*\gamma^{-1}_{21})-{\bf 1}\|^2\\
&+2\sqrt{(\|\tilde{q}_1-(\tilde{q}_2* \gamma_{21})\|^2+\lambda \|({\bf 1}*\gamma_{21})-{\bf 1} \|^2)(\|\tilde{q}_2 -((\tilde{q}_3* \gamma_{31})*\gamma^{-1}_{21}) \|^2+\lambda\|(({\bf 1}*\gamma_{31})*\gamma^{-1}_{21})-{\bf 1}\|^2)}\\
&\leq (\sqrt{\|\tilde{q}_1-(\tilde{q}_2* \gamma_{21})\|^2+\lambda \|({\bf 1}*\gamma_{21})-{\bf 1} \|^2}+\sqrt{\|\tilde{q}_2 -((\tilde{q}_3* \gamma_{31}),\gamma^{-1}_{21}) \|^2+\lambda\|(({\bf 1}*\gamma_{31})*\gamma^{-1}_{21})-{\bf 1}\|^2})^2
\end{align*}
}
The LHS is less than or equal to the RHS for all $\gamma_{31} \in \Gamma$. Then $d(q_1,q_3) \leq LHS \leq RHS$. Besides, the above equation is true for all $\gamma_{21} \in \Gamma$. So 
{\small
\begin{equation*} 
d^2(q_1,q_3) \leq (d(q_1,q_2)+d(q_2,q_3))^2  \Leftrightarrow d(q_1,q_3) \leq d(q_1,q_2)+d(q_2,q_3)
\end{equation*}
}
~~~~~~~~~~ $\Box$
\end{enumerate}

\bibliographystyle{apalike}
\bibliography{bib}

\end{document}